\documentclass[11pt,a4paper]{article}

\usepackage[T1]{fontenc}
\usepackage[utf8]{inputenc}
\usepackage{lmodern}
\usepackage[english]{babel}
\usepackage[margin=2.6cm]{geometry}
\usepackage{amsmath,amssymb,amsthm,mathtools}
\usepackage{stmaryrd}
\usepackage{booktabs,array}
\usepackage{enumitem}
\usepackage{xcolor}
\usepackage{tikz}
\usetikzlibrary{arrows.meta,positioning,calc}
\usepackage{authblk}
\usepackage{microtype}
\usepackage[colorlinks=true,linkcolor=blue!45!black,citecolor=blue!45!black,urlcolor=blue!45!black]{hyperref}
\usepackage[capitalise,noabbrev]{cleveref}
\hypersetup{
  pdftitle={From State to Link-Register Model: A Transformer for Self-Stabilizing Distributed Algorithms (full version)},
  pdfauthor={Johanne Cohen, George Manoussakis, Laurence Pilard}}

\usepackage[appendix=append,bibliography=common]{apxproof}
\makeatletter
\newcommand{\onlywithappendix}[1]{\ifdefstring{\axp@appendix}{append}{#1}{}}
\makeatother

\newtheoremrep{theorem}{Theorem}[section]
\newtheoremrep{lemma}[theorem]{Lemma}
\newtheoremrep{proposition}[theorem]{Proposition}
\newtheoremrep{corollary}[theorem]{Corollary}
\theoremstyle{definition}
\newtheorem{definition}[theorem]{Definition}
\newtheorem{example}[theorem]{Example}
\theoremstyle{remark}
\newtheorem{remark}[theorem]{Remark}

\newcommand{\A}{\mathcal{A}_{SM}}
\newcommand{\ASLRM}{\mathcal{A}_{SLRM}}
\newcommand{\ALRM}{\mathcal{A}_{LRM}}
\newcommand{\ACLRM}{\mathcal{A}_{CLRM}}
\newcommand{\TA}{\textnormal{\textsc{S2SLR}}}
\newcommand{\TB}{\textnormal{\textsc{S2LR}}}
\newcommand{\TC}{\textnormal{\textsc{S2LR-C}}}
\newcommand{\Coloring}{\mathcal{C}\mathit{oloring}}

\newcommand{\rulename}[1]{\textsf{#1}}
\newcommand{\Write}{\rulename{Write}}
\newcommand{\Read}{\rulename{Read}}
\newcommand{\WantToAct}{\rulename{WantToAct}}
\newcommand{\ModifyMyState}{\rulename{ModifyMyState}}
\newcommand{\toWAIT}{\rulename{toWAIT}}
\newcommand{\ExecuteAlgo}{\rulename{ExecuteAlgo}}
\newcommand{\StartAgain}{\rulename{StartAgain}}
\newcommand{\ResetRule}{\rulename{Reset}}
\newcommand{\ToTrue}{\rulename{ToTrue}}
\newcommand{\CycleLocal}{\rulename{Local}}
\newcommand{\CycleIdle}{\rulename{Idle}}
\newcommand{\ChangeColor}{\rulename{Change\_color}}
\newcommand{\state}{\mathit{state}}
\newcommand{\phase}{\mathit{phase}}
\newcommand{\VAR}{\mathit{VAR}}
\newcommand{\id}{\mathit{id}}
\newcommand{\cp}[2]{\mathit{copy}_{#1}[#2]}
\newcommand{\YS}{\mathit{YourState}}
\newcommand{\MS}{\mathit{MyState}}
\newcommand{\reg}[2]{r_{#1#2}}
\newcommand{\LV}[2]{\mathit{LocalValues}_{#1}(#2)}
\newcommand{\RR}[2]{\mathit{ReadRegister}_{#1#2}}
\newcommand{\CR}[2]{\mathit{CopyRegister}_{#1}(#2)}
\newcommand{\CBS}{\mathit{CBS}}
\newcommand{\waitfull}{\mathit{waitNeighborWithLowerId}}
\newcommand{\wait}{\mathit{wait}}
\newcommand{\isAct}{\mathit{isActivable}}
\newcommand{\Activate}{\mathit{Activate}}

\newcommand{\wflag}[2]{\mathit{write}_{#1}[#2]}
\newcommand{\true}{\mathit{True}}
\newcommand{\false}{\mathit{False}}
\newcommand{\proj}[1]{\llbracket #1\rrbracket}
\newcommand{\Ex}{\mathcal{E}}
\newcommand{\Act}{\mathit{Act}}
\newcommand{\figrule}{\par\noindent\rule{\linewidth}{0.4pt}\par}
\newcommand{\modif}[1]{#1}
\newcommand{\Faithful}{Faithful}
\newcommand{\faithful}{faithful}

\title{From State to Link-Register Model:\\
A Transformer for Self-Stabilizing Distributed Algorithms\\[1.2ex]
\large Research report (full version)\thanks{A preliminary version of this work
appeared in the proceedings of SYNASC 2023~\cite{CohenMP23}.
\Cref{sec:differences} lists the differences between the two versions.}}

\author[1]{Johanne Cohen}
\author[2]{George Manoussakis}
\author[2]{Laurence Pilard}
\affil[1]{LISN, CNRS, Universit\'e Paris-Saclay, Orsay, France\\
\texttt{Johanne.Cohen@lisn.fr}}
\affil[2]{LI-PaRAD, Universit\'e de Versailles-St Quentin, Guyancourt, France\\
\texttt{George.Manoussakis@uvsq.fr}, \texttt{Laurence.Pilard@uvsq.fr}}
\date{November 2026}

\begin{document}
\maketitle

\begin{abstract}
%
In the link-register model, two neighboring nodes communicate through
registers located on the link between them, and an atomic step
{writes a single register or reads registers}. There is thus a delay
between the time a node takes an
action and the time its neighbors are informed of it. This delay captures the
asynchronism induced by communications. In the state model, in contrast, a
node reads the states of all its neighbors and updates its own state in one
atomic step.

The link-register models that we consider differ on what a node can
read, and this difference decides what a transformer can achieve. A node that
can read its own registers knows what its neighbors read about it. A node
that cannot must rewrite its registers forever, since a corrupted register
would stay unnoticed. With strict read/write atomicity, a node moreover reads
at most one register per atomic step.

We present  {three} transformers that convert a self-stabilizing algorithm written
in the state model into an algorithm for a link-register model with read/write
atomicity, one for each of these models. The first transformer, $\TA$, targets the strong-link-register
model, in which a node can also read its own registers. It works under the
unfair distributed daemon, and the transformed algorithm can be
silent. We prove that an execution of the transformed
algorithm that simulates $M$ moves of the original algorithm contains
$O(\Delta^2 M+\Delta n)$ moves, where $\Delta$ is the maximum degree of the
network and $n$ its number of nodes. 
The second transformer, $\TB$,
targets the classical link-register model, in which a node cannot read its own
registers. It works under the weakly fair distributed daemon{, since
every node is always enabled}. We prove that
each simulated move costs $O(\Delta)$ rounds{.}
The third transformer, $\TC$, targets the link-register model with
strict read/write atomicity. Each node writes its registers and reads the
registers of its neighbors in a cyclic order, and each simulated move still
costs $O(\Delta)$ rounds.

This report is the full version of a paper presented at SYNASC 2023. It
contains complete proofs. 
\end{abstract}

\newpage
\section{Introduction}\label{sec:introduction}

Self-stabilization~\cite{Dijk74} deals with transient faults in distributed
systems. A fault, or a sequence of faults, can leave the system in an
arbitrary configuration of the nodes and of the communication registers.
Starting from this configuration, the execution must resume a correct behavior
in finite time, without external intervention. A complete study of this notion
can be found in~\cite{Dole00}.

The behavior of a self-stabilizing algorithm depends on two modeling choices:
how nodes communicate, and how their actions are scheduled.

\paragraph{Communication models.}
A network is modeled by a graph. Nodes represent processors, and there is a
link between two nodes if and only if they can communicate with each other.
The simplest way to communicate in self-stabilization is the \emph{state
model}. Each node has a set of local variables, which make up its local state.
A node can read its own variables and the variables of all its neighbors, but
it can only write its own variables. Another way is to communicate through
\emph{locally shared registers}: neighboring nodes communicate by reading and
writing registers shared between them. The registers are either
single-writer/multi-reader and located at the nodes, which is the state model
again, or single-writer/single-reader and located on the links. The latter
case is the \emph{link-register model}.

These communication models come with several atomicities. With
\emph{composite atomicity}, a node reads the registers of all its neighbors
and writes its own registers in one atomic step. With \emph{read/write
atomicity}, an atomic step consists of a single read operation or a single
write operation on one register. Other atomicity conditions are used in the
link-register model; see~\cite{JohnenH08,JohnenH09} for a survey and for
results on their relative strength.

Combining the type of registers with the atomicity leads to three main
models~\cite{HJ06}, summarized in \cref{tab:models}. We use the names
of~\cite{HJ06}. The state model with composite atomicity is the
\emph{composite state} model. Dijkstra introduced it in his seminal paper on
self-stabilization~\cite{Dijk74}. It does not capture the asynchronous
phenomena that occur in many real distributed systems. This led Dolev, Israeli
and Moran to introduce the link-register model with read/write
atomicity~\cite{DIM93}, called the \emph{atomic register} model. The last
model is the state model with read/write atomicity, called the \emph{atomic
state} model. It is used, for instance, in~\cite{chatto02}.

\begin{table}[ht]
\centering
\small
\begin{tabular}{@{}lll@{}}
\toprule
Atomicity & State model & Link-register model\\
 & (single-writer/multi-reader) & (single-writer/single-reader)\\
\midrule
Composite  & composite state model & none\\
Read/write & atomic state model & atomic register model\\
\bottomrule
\end{tabular}
\caption{The three main communication models.}
\label{tab:models}
\end{table}

\paragraph{What changes from one model to another.}
This work is about the difference between these models. In the state model,
a node that executes a rule sees the current variables of all its
neighbors. In a link-register model, it only sees copies. A value goes from a
node to its neighbor in two atomic steps, a write and a read, and the
scheduler decides when each of them takes place. A node can thus act on a
value that its neighbor has already changed.

A second difference separates the two link-register models that we consider.
If a node cannot read its own registers, it cannot know what its neighbors
read about it, and a corrupted register can stay unnoticed forever. These two
differences explain the design of our transformers and the assumptions of our
results. \Cref{sec:models} puts the three models side by side.

\paragraph{Daemons.}
The environment of a self-stabilizing algorithm is modeled by a \emph{daemon},
which chooses the nodes that act at each step. A daemon is \emph{sequential}
(or central) if exactly one enabled node acts at each step, and
\emph{distributed} if any nonempty set of enabled nodes may act
simultaneously. Independently, a daemon is \emph{fair} if every node that
remains enabled is eventually chosen, and \emph{unfair} if it only guarantees
global progress:
at each step, at least one enabled node acts, if such a node exists. The
unfair distributed daemon is the most powerful adversary, since every
execution allowed by another daemon is also allowed by this one.
See~\cite{DubTix11} for a taxonomy of daemons.

\paragraph{From the state model to the link-register model.}
It is easier to design and to prove an algorithm in the composite state model
than in the atomic register model. The atomic register model is more
realistic. A natural approach is therefore to design the algorithm in the
state model and to convert it automatically. This is the purpose of this
work: we present self-stabilizing \emph{transformers} that convert an
algorithm in the state model into an algorithm in a link-register model with
read/write atomicity.

In a link-register model, a node must keep copies of the variables of its
neighbors, and these copies may be out of date. To simulate one atomic step of
the state model, a node $u$ must execute a sequence of read and write steps
during which none of its neighbors changes its variables. Our transformers
enforce this with a local mutual exclusion, in the spirit of the
distributed constructions of a \emph{maximal independent set} (MIS). An
independent set is a set of nodes no two of which are neighbors. It is maximal
if it is maximal for inclusion. This problem has been studied a lot in
parallel and distributed settings, following the seminal works
of~\cite{Alon1986,linial1987,luby1986}. A node joins the set under
construction if none of its neighbors is in the set, and identifiers break
the symmetry between neighbors. MIS constructions are used as a black box to
organize communications in wireless networks~\cite{liu2017cooperative,gao2017novel},
and they are a basic tool to transform self-stabilizing algorithms from one
model to another~\cite{grati2007,turau2006randomized}. In our transformers,
the nodes that execute a rule of the original algorithm at a given time form
an independent set, once the effect of the initial corruption has
disappeared. A node enters this set only after it has checked, through its
registers, that its neighbors know it.

\paragraph{The two target models.}
In the atomic register model, a node cannot read its own registers. This is
close to message passing: a node cannot read again the last message it has
sent. As a consequence, a node cannot check whether the content of one of its
registers is corrupted. In the transformer that we propose for this model, a
node therefore rewrites its registers again and again. Every node is then
enabled in every configuration, and an unfair daemon could starve any of
them. For this reason, this transformer is analyzed under the weakly fair
distributed daemon, and its time complexity is measured in rounds.

To obtain a result under the unfair distributed daemon, with a complexity in
moves, we also consider a slightly stronger model, which we call the
\emph{strong-link-register model}. In this model, a node is allowed to read
its own registers. This is the only difference with the atomic register model.
A node then writes a register only when the register does not hold the right
value, and the transformed algorithm can be silent.

\paragraph{Names and notation.}
We give a name to each transformer. The transformer for the
strong-link-register model is called $\TA$, for ``state to
strong-link-register''. The transformer for the link-register model is called
$\TB$, for ``state to link-register''. They are the
transformers $\mathcal{T}_1$ and $\mathcal{T}_2$ of the conference
version~\cite{CohenMP23}. As in~\cite{CohenMP23}, $\A$ denotes the algorithm
of the state model that is transformed, and we write $\ASLRM=\TA(\A)$ and
$\ALRM=\TB(\A)$ for its two transformed versions. Our running example is the
algorithm $\Coloring$ of~\cite{CohenMP23}.

\subsection{Contributions}

We put the state model, the
link-register model and the strong-link-register model side by side: what a
node can read, what it does in one atomic step, and what follows for the
transformers. {We identify three differences, from the state model to
the most constrained model. Each of them takes some information away from the
nodes, and each transformer below rebuilds this information.}
\begin{enumerate}
\item {\emph{Copies instead of variables.} In the state model, a node
reads the current variables of all its neighbors and updates its own
variables in one atomic step. In a link-register model, it reads registers. A
value goes from a node to its neighbor in two atomic steps, a write and a
read. A node thus only holds copies, and they can be out of date.}
\item {\emph{Reading one's own registers.} In the strong-link-register
model, a node can read the registers that it writes. In the link-register
model, it cannot. It does not know what its neighbors read about it, and it
cannot detect that one of its registers is corrupted.}
\item {\emph{One register per move.} With strict read/write atomicity,
a move reads at most one register. This variant of the link-register model is
defined in \cref{sec:t3}.}
\end{enumerate}

 {We give three transformers. Each of them turns an algorithm $\A$ of
the state model into an algorithm for a link-register model, on any network
with $n$ nodes and maximum degree $\Delta$. A node simulates a move of $\A$
by executing a rule called $\ExecuteAlgo$. We write $M$ for the number of
simulated moves, that is, the number of executions of $\ExecuteAlgo$.}

\begin{description}[leftmargin=1.2em]

\item[Transformer $\TA$ (\cref{sec:t1,sec:t1-analysis}).]  {The algorithm
$\TA(\A)$ runs in the strong-link-register model under the unfair
distributed daemon. Acknowledgments and a local mutual exclusion ensure that
a node simulates a move of $\A$ only when its copies are up to date. A node
reads its own registers, so it writes a register only when needed. The cost
is therefore counted in moves, and the transformed algorithm can be silent.}
\begin{itemize}
\item \emph{Simulation.}  {Every execution of $\ExecuteAlgo$ is a legal
move of $\A$, except for at most one per node
(\cref{lem:faithful,cor:unfaithful,thm:simulation}). An exception requires a
corrupted start: the variable $\state$ of the node is initially between $3$
and $6$.}
\item \emph{Move complexity.}  {One simulated move costs $O(\Delta^2)$
moves: an execution with $M$ simulated moves contains
$O(\Delta^2M+\Delta n)$ moves (\cref{thm:moves}).}
\item \emph{Silent algorithms.}  {If every execution of $\A$ has at most
$F$ moves, then $\TA(\A)$ is silent, and when it stops the variables of $\A$
form a terminal configuration of $\A$. It simulates at most $(n+1)F+n$
moves: each of the at most $n$ exceptions acts as one more transient fault,
from which $\A$ recovers in at most $F$ moves. Without exception, $M\le F$
and the execution contains $O(\Delta^2F+\Delta n)$ moves
(\cref{thm:t1-silent}).}
\end{itemize}

\item[Transformer $\TB$ (\cref{sec:t2,sec:t2-analysis}).]  {The algorithm
$\TB(\A)$ runs in the link-register model (the atomic register model) under
the weakly fair distributed daemon. A node cannot read its own registers, so
it cannot detect a corrupted one. Flags replace this test, and every node
rewrites its registers forever. Every node is thus always enabled: the
daemon must be weakly fair, time is counted in rounds, and the executions
are infinite.}
\begin{itemize}
\item \emph{Simulation.}  {Forget the flags: the moves of $\TB(\A)$ that
change a register or a variable form a sequence of steps of $\TA(\A)$
(\cref{lem:projection}). The simulation result above carries over.}
\item \emph{Round complexity.}  {One simulated move costs $O(\Delta)$
rounds: if $M$ is finite, after $O(\Delta(M+1))$ rounds the registers and
the variables of $\A$ never change again, and the variables of $\A$ form a
terminal configuration of $\A$ (\cref{thm:rounds,cor:t2-silent}).}
\end{itemize}

\item[Transformer $\TC$ (\cref{sec:t3,sec:t3-analysis}).]  {The algorithm
$\TC(\A)$ runs in the link-register model with strict read/write atomicity,
under the weakly fair distributed daemon: a move accesses at most one
register. In $\TB$, one rule still reads the registers of all the neighbors
of a node. $\TC$ removes it. Each node writes its registers, reads the
registers of its neighbors and checks its local rules in a fixed cyclic
order. The cost stays the same: if $M$ is finite, after $O(\Delta(M+1))$
rounds the registers and the variables of $\A$ never change again
(\cref{thm:t3-rounds}).}

\end{description}

\paragraph{Organization}

\Cref{sec:related} reviews related work. \Cref{sec:model} defines the models,
and \cref{sec:models} compares them.
\Cref{sec:t1} presents the transformer $\TA$ and \cref{sec:t1-analysis}
analyzes it. \Cref{sec:t2} presents the transformer $\TB$ and
\cref{sec:t2-analysis} analyzes it.  {\Cref{sec:t3} presents the
transformer $\TC$ and \cref{sec:t3-analysis} analyzes it.}
\Cref{sec:conclusion} concludes.
\onlywithappendix{{The proofs of the technical lemmas are given in
\cref{app:t1,app:t2,app:t3}.}}
\section{Related work}\label{sec:related}

Self-stabilizing algorithms are written in several models, and many works
relate these models to one another. The usual goal is to design an algorithm
in a comfortable model and to convert it automatically into an algorithm for a
more realistic one. This section reviews these conversions. We start with the
models themselves (\cref{sec:rw-models}). We then consider the conversions
that stay within the state model (\cref{sec:rw-state}), those that go from the
state model to registers on links (\cref{sec:rw-links}), and those that target
message passing and wireless networks (\cref{sec:rw-messages}).
\Cref{sec:rw-other} mentions other transformations, and
\cref{sec:rw-position} places our results. \Cref{tab:related} lists the
transformers that are the closest to ours.

\subsection{Communication models}\label{sec:rw-models}

The state model with composite atomicity goes back to Dijkstra~\cite{Dijk74}.
Most self-stabilizing algorithms are written in this model. The book of
Altisen \emph{et al.}~\cite{ADDP19} is a recent introduction to
self-stabilization in this setting, and~\cite{DubTix11} classifies the
daemons. Dolev, Israeli and Moran introduced read/write atomicity and the
registers on links~\cite{DIM93}. Higham and Johnen~\cite{HJ06} give a common
framework for these models. Registers are located at the nodes in their
\emph{atomic-state} model and on the links in their \emph{atomic-link} model,
and each read or write is one atomic step. \Cref{tab:models} follows this
classification. Our atomic register model is their atomic-link model.

The terminology is not uniform. In recent papers such
as~\cite{DevismesIJM23,KakugawaKSO25}, the name atomic-state model denotes
the state model with composite atomicity, which we call the composite state
model. The state model is also called the state-reading model or the locally
shared memory model~\cite{KakugawaKSO25}.

\paragraph{Who reads a register on a link.}
Two variants of the registers on links coexist. In the atomic-link model, a
register is single-writer/single-reader: only the neighbor reads it. This is
our link-register model. In other works, the writer can also read its
register. This is the case in~\cite{johnen2010reliable}, where both endpoints
of a link can read a register, and in~\cite{CohenLMPS16}. This variant is our
strong-link-register model.

\paragraph{Weaker registers.}
Registers can be weaker than atomic. Dolev and Herman~\cite{DolevH01} revisit
Dijkstra's algorithm when registers are only regular or safe. Johnen and
Higham study the implementation of the atomic-state model in weaker
networks~\cite{JohnenH07}. They give a wait-free and self-stabilizing
implementation of regular registers by safe registers, with applications to
networks~\cite{JohnenH08,JohnenH09}. These works weaken the semantics of the
registers. They are complementary to ours, which keeps atomic registers and
changes their location and the atomicity of the steps.

\paragraph{Registers on links with larger steps.}
Registers on links are also used with steps that are not limited to one read
or one write. Hoepman~\cite{Hoepman98} compares the link-register model and
the state-reading model under the central daemon. He shows that the two
models are eventually equivalent on graphs with a suitable edge labelling,
which includes oriented rings. In one direction, a node simply writes its
state in all its link registers. With read/write atomicity, this simple
translation lets a node act on a value that is out of date, since a neighbor
may have changed its state without having written it yet.

\subsection{Refining atomicity and daemons within the state model}\label{sec:rw-state}

All executions that are possible with composite atomicity are possible with
read/write atomicity, but the converse is not true. It is thus easier to
design algorithms with composite atomicity, or even for a central daemon,
while weaker assumptions are more realistic. Many transformers go from the
former to the latter and stay within the state model. Huang, Wuu and
Tsai~\cite{HuangWT94} already study how to convert a system designed for the
serial model into a system for the distributed model.

A common approach is to implement a local mutual exclusion among neighboring
nodes. The algorithm $\A$, written for the composite state model, is composed
with a local mutual exclusion algorithm written for the atomic state model, so
that a node executes an action of $\A$ only when it has access to the critical
section~\cite{Bo04}. Beauquier \emph{et al.}~\cite{Bea00} build a daemon
refinement on a self-stabilizing local mutual exclusion algorithm, and
Antonoiu and Srimani~\cite{Anto99} solve local mutual exclusion with
read/write atomicity. A solution to the dining philosophers problem is also a
solution to the local mutual exclusion problem, and this problem has many
solutions~\cite{Canta03,dan06,Hu00,NE02,NES02}. In particular, Nesterenko and
Arora~\cite{NES02} go from a model where a process reads the states of all its
neighbors and updates its own state in one step to a model where a step
either reads the state of one neighbor or updates the state of the process.
Cantarell, Datta and Petit~\cite{Canta03} give a transformer that lets
neighbors execute concurrently and that preserves silence. Timestamps are
another solution~\cite{MizunoK96,MI98}. In~\cite{MI98}, they guarantee that
neighbors execute their guarded commands in mutual exclusion, and the
transformation preserves silence. Alternators are one more
solution~\cite{gou07,KU05}: they build an independent set of nodes, and only
the nodes of this set can execute a rule. Conflict
managers~\cite{grati2007} ask for less than local mutual exclusion. Two nodes
in conflict never execute simultaneously and at least one node makes
progress, but no fairness is required.

Our transformers rely on the same ingredient, a local mutual exclusion based
on identifiers. The difference lies in the target model. In the state model, a
node still reads the current state of a neighbor, even if it reads one
neighbor at a time. With registers on links, a node only reads what its
neighbor has written for it. This value can be out of date, or wrong without
its owner noticing (\cref{sec:one-link}).

\subsection{From the state model to registers on links}\label{sec:rw-links}

Fewer results address the transformation from the state model to a
link-register model. The closest work is that of Higham and
Johnen~\cite{HJ06}. They determine under which conditions, and with which
guarantees, a solution in the atomic-state model can be transformed into a
solution in the atomic-link model, for self-stabilization and for
wait-freedom. As far as we know, this is the only transformer from the atomic
state model to the atomic register model, and no time complexity is given.
Chapter~4 of Dolev's book~\cite{Dole00} is devoted to model conversions. It
presents a transformer from the composite state model to the atomic register
model under the fair sequential daemon, based on token passing. Its round
complexity can be exponential. Ioannidou~\cite{Ioannidou02} gives a technique
to prove that a transformation between two models preserves stabilization.
She applies it to the other direction, from message passing to shared memory
with bounded registers.

Our transformers behave differently: one move of the original algorithm costs
$O(\Delta^2)$ moves with $\TA$ and $O(\Delta)$ rounds with $\TB$. 

Algorithms have also been designed directly for a link-register model. This is
the case for maximal matching~\cite{CohPil17,CohenMPS18}, also in anonymous
networks~\cite{CohenLMPS16}. The analysis of~\cite{CohPil17,CohenMPS18} has
the same structure as the one of this report: a first algorithm for a relaxed
communication model under the unfair distributed daemon, then a second
algorithm for the link-register model under the fair distributed daemon.

\paragraph{Communication primitives.}
Another approach is to use communication primitives that give guarantees on
the read and write operations, and to simulate the composite state model on
top of them. The primitives of Johnen \emph{et al.}~\cite{johnen2010reliable}
are a natural candidate. They are designed for read/write atomicity and a
fair scheduler. They guarantee properties between two consecutive writes of
the same node. The strongest one states that between two writes, all
neighbors read the last written value exactly once. This is obtained by
blocking the write of a node until all its neighbors have read. In these
protocols, each node writes in its registers infinitely often.

These primitives cannot be used directly as the basis of a transformer. In
the composite state model, a node reads the states of all its neighbors in one
atomic action. Each value it reads is thus the last value written by the
corresponding neighbor. In particular, a node cannot act on the basis of the
last written value of a neighbor $a$ together with an older value of another
neighbor $b$. With the primitives alone, such a combination is possible, and
the node may then reach a state that is unreachable in the composite state
model. Our transformers exclude these combinations: when a node executes a
rule of the original algorithm, all its copies are up to date
(\cref{lem:faithful}).

\subsection{Message passing and wireless networks}\label{sec:rw-messages}

Registers on links are close to message passing, and the same questions arise
there. Dolev's book converts shared memory to message passing with
self-stabilizing data-link algorithms~\cite[Chapter~4]{Dole00}, and the
refinement of~\cite{NES02} can be continued down to a message-passing model.
Gouda and Multari~\cite{GoudaM91} show that a stabilizing communication
protocol is nonterminating and has timeout actions. The rule $\ToTrue$ of
$\TB$ plays the role of such a timeout: a node that cannot read its own
registers has to write them again from time to time.

In wireless sensor networks, a node broadcasts to all its neighbors, and
messages can be lost. The cached sensornet transform of
Herman~\cite{Herman03} runs an algorithm of the state model in such a network.
Each node periodically broadcasts its variables, keeps a cache of the
variables of each neighbor, and evaluates its guards on the caches. Its
analyses are probabilistic, and they first concerned silent algorithms.
Kakugawa \emph{et al.}~\cite{KakugawaYKM11} exhibit a non-silent algorithm
whose transform does not behave correctly, and they give a sufficient
condition for correctness. In our setting no value is lost, but the
execution is chosen by an adversarial daemon, and a node can still act on a
value that is out of date. Kulkarni and
Arumugam~\cite{KulkarniA06} introduce the write-all-with-collision model for
sensor networks and study the transformations between this model and the
read/write model. Turau and Weyer~\cite{turau2006randomized,TurauW09} propose
randomized transformations for sensor networks. The transformed algorithms
are probabilistically self-stabilizing.

\begin{table}[t]
\centering
\small
\begin{tabular}{@{}>{\raggedright\arraybackslash}p{3.1cm}>{\raggedright\arraybackslash}p{3.7cm}>{\raggedright\arraybackslash}p{3.7cm}>{\raggedright\arraybackslash}p{3.7cm}@{}}
\toprule
Reference & From & To & Tool or remark\\
\midrule
Huang \emph{et al.}~\cite{HuangWT94} & serial model & distributed model & \\
Mizuno, Nesterenko~\cite{MI98} & serial model & asynchronous shared memory &
timestamps, preserves silence\\
Gouda, Haddix~\cite{gou07} & serial execution & concurrent execution &
alternator\\
Beauquier \emph{et al.}~\cite{Bea00} & weaker daemon & stronger daemon &
local mutual exclusion\\
Nesterenko, Arora~\cite{NES02} & a step reads all neighbors and updates &
a step reads one neighbor or updates & dining philosophers\\
Cantarell \emph{et al.}~\cite{Canta03} & central daemon or composite
atomicity & distributed read/write atomicity & neighbors may execute
concurrently, preserves silence\\
Gradinariu, Tixeuil~\cite{grati2007} & sequential model & asynchronous
distributed model & conflict managers\\
\midrule
Higham, Johnen~\cite{HJ06} & atomic-state model & atomic-link model & under
conditions\\
Dolev~\cite{Dole00} & central daemon & read/write atomicity & token
passing\\
Herman~\cite{Herman03} & state model & sensor networks & caches and periodic
broadcasts\\
Turau, Weyer~\cite{turau2006randomized,TurauW09} & state model & sensor
networks & randomization\\
Turau~\cite{Turau12} & distance-two model & state model, distributed daemon &
slowdown of $O(m)$ moves\\
Kakugawa \emph{et al.}~\cite{KakugawaKSO25} & $R(1)W(1)$ model, central
daemon & synchronous message passing & randomization\\
\midrule
$\TA$ (\cite{CohenMP23},this report) & composite state model, unfair distributed daemon &
strong-link-register model, unfair distributed daemon & $O(\Delta^2)$ moves
per move\\
$\TB$ (\cite{CohenMP23}, this report) & composite state model, unfair distributed daemon &
link-register model, weakly fair distributed daemon & $O(\Delta)$ rounds per
move\\
\bottomrule
\end{tabular}
\caption{Transformers between models for self-stabilizing algorithms. The
first group stays within the state model. The second group changes the
communication model.}
\label{tab:related}
\end{table}

\subsection{Other transformations between models}\label{sec:rw-other}

\paragraph{Larger views.}
Some models give a node more than the state model does. In the distance-two
model, a node reads the variables of the nodes at distance at most two.
Gairing \emph{et al.}~\cite{GairingGHKM04} transform the algorithms of this
model into algorithms of the usual state model. Goddard
\emph{et al.}~\cite{GoddardHJT08} extend this approach to distance $k$, with
a slowdown of $n^{O(\log k)}$. Turau~\cite{Turau12} emulates the distance-two
model under the distributed daemon with a slowdown of $O(m)$ moves. Kakugawa
\emph{et al.}~\cite{KakugawaKSO25} recently proposed the $R(1)W(1)$ model, in
which a process reads and writes its own variables and those of its neighbors
in a single step, under the unfair central daemon. They give a randomized
transformer from this model to synchronous message passing with synchronized
clocks. They mention the conference version of this
report~\cite{CohenMP23} as a transformation from the state model to the
link-register model with read/write atomicity. A follow-up paper replaces
the writes on neighbors by reads at a larger
distance~\cite{KakugawaKSOW26}.

\paragraph{Adding self-stabilization.}
Other transformers start from an algorithm that is not self-stabilizing. Katz
and Perry~\cite{KatzP93} and Awerbuch and Varghese~\cite{AwerbuchV91} do so
in message-passing systems. Awerbuch \emph{et al.}~\cite{Var94} transform
locally checkable protocols into self-stabilizing ones with a
self-stabilizing reset algorithm. They assume that communications are
acknowledged. We make no such assumption in our communication model, and
acknowledgments are part of what our transformers implement. More recently,
several transformers turn a synchronous algorithm into an asynchronous
self-stabilizing algorithm of the state
model~\cite{CohenPRS23,DevismesIJM23}, and a self-stabilizing unison is used
to simulate synchronous self-stabilizing algorithms in an asynchronous
environment~\cite{DevismesIJM25}. The transformer of~\cite{DevismesIJM23}
produces silent algorithms with a polynomial number of moves under the unfair
distributed daemon. This is the class of algorithms covered by
\cref{thm:t1-silent,cor:t2-silent}, so our transformers can be applied to
its output.

\paragraph{Silence.}
We call an algorithm silent when all its executions are finite. For register
models, Dolev, Gouda and Schneider~\cite{DolevGS99} use a weaker definition:
the values stored in the communication registers eventually stop changing.
When $\A$ has bounded move complexity, $\TA(\A)$ is silent in the first
sense, and $\TB(\A)$ in the second one only (\cref{rem:silence}).

\subsection{Position of this work}\label{sec:rw-position}

\Cref{tab:related} summarizes the transformers discussed above that are the
closest to ours. To our knowledge, $\TA$ and $\TB$ are the only transformers
from the state model with composite atomicity to registers on links with
read/write atomicity that come with a cost per simulated move: $O(\Delta^2)$
moves under the unfair distributed daemon for $\TA$, and $O(\Delta)$ rounds
under the weakly fair distributed daemon for $\TB$.
\section{Model and definitions}\label{sec:model}

\subsection{Network}

The network is an undirected graph $G=(V,E)$ with $n=|V|$ nodes and $m=|E|$
links. Communications are bidirectional. Two nodes are \emph{neighbors} if
they are joined by a link. We denote by $N(u)$ the set of neighbors of a node
$u$, by $N[u]=N(u)\cup\{u\}$ its closed neighborhood, by $\deg(u)=|N(u)|$ its
degree, and by $\Delta$ the maximum degree of $G$. We assume $\Delta\ge 1$.

Each node $u$ has a unique identifier $\id_u$ taken from a totally ordered
set. Identifiers are constants: faults do not modify them. All nodes have the
same variables. If $x$ is a variable, $x_u$ denotes its instance at node $u$.

\subsection{The state model}

In the state model, an algorithm $\A$ gives each node $u$ a set of variables.
We denote by $\VAR_u$ the tuple of their current values. A node can read its
own variables and those of its neighbors, and it can only write its own
variables. A \emph{configuration} of $\A$ is the vector $(\VAR_u)_{u\in V}$.
The \emph{view} of $u$ in a configuration is the tuple
\[
\bigl(\id_u,\ \VAR_u,\ (\id_v,\VAR_v)_{v\in N(u)}\bigr).
\]

The algorithm consists of a finite number of \emph{guarded rules}
$\mathcal{G}_j\to\mathcal{I}_j$. The guard $\mathcal{G}_j$ is a predicate on
the view of the node. The action $\mathcal{I}_j$ computes a new value of
$\VAR_u$ from the view. A rule is \emph{enabled} at $u$ in a configuration if
its guard holds on the view of $u$. A node is enabled if at least one of its
rules is. A configuration is \emph{terminal} if no node is enabled.

A \emph{step} of $\A$ transforms a configuration $\gamma$ into a configuration
$\gamma'$ as follows: a nonempty set of nodes enabled in $\gamma$ is chosen,
and each chosen node executes the action of one of its enabled rules. All
guards and actions are evaluated in $\gamma$. This is composite atomicity. An
\emph{execution} of $\A$ is a maximal sequence of configurations
$\gamma_0,\gamma_1,\ldots$ such that each $\gamma_t\to\gamma_{t+1}$ is a step.
Maximal means that the sequence is infinite, or that it ends in a terminal
configuration.

The transformers use the algorithm $\A$ through two notations. Both apply to
any tuple $\nu$ that has the shape of a view of $u$, even if $\nu$ is not the
view of $u$ in the current configuration.
\begin{itemize}
\item $\isAct(\nu,\A)$ is true if and only if the guard of at least one rule
of $\A$ holds on $\nu$.
\item $\Activate(\nu,\A)$ is the value of $\VAR_u$ computed by the action of a
rule of $\A$ whose guard holds on $\nu$. If several guards hold, the rule is
chosen arbitrarily.
\end{itemize}

\subsection{Link-register models}\label{sec:registers}

In the link-register models, two neighbors $u$ and $v$ share two registers
$\reg uv$ and $\reg vu$. Node $u$ is the only node that writes in $\reg uv$.
We consider two models.
\begin{itemize}
\item In the \emph{link-register model}, $v$ is the only node that reads
$\reg uv$. With read/write atomicity, this is the atomic register model
of~\cite{DIM93,HJ06}.
\item In the \emph{strong-link-register model}, both $u$ and $v$ can read
$\reg uv$.
\end{itemize}
In both models a node has local variables that no other node can read. A
\emph{configuration} is the vector of the values of all variables and all
registers.

An algorithm is again a set of guarded rules. We express read/write atomicity
as a restriction on the rules, as in~\cite{CohenMP23}. A rule has one of the
two following forms.
\begin{itemize}
\item It writes one register of the node, and possibly some of its local
variables. Its guard depends only on the local variables of the node and, in
the strong-link-register model, on the register that is written.
\item It writes only local variables of the node. Its guard may read registers
written by the neighbors of the node.
\end{itemize}
Link-register algorithms are often presented as an infinite loop of read
operations, local computations and write operations at each node. The
atomicity then tells at which points a node can be suspended. Guarded rules
make these points explicit: each move of a node is one atomic step.

The moment when $u$ writes in $\reg uv$ is the moment from which the written
value is available to $v$. A write is thus similar to the reception of a
message by $v$ in a message-passing model. In the strong-link-register model,
$u$ can compare $\reg uv$ with the value it should hold, and it writes only
when they differ.

Enabled rules, enabled nodes, terminal configurations, steps and executions
are defined as in the state model: in a step, each chosen node executes one of
its enabled rules, and all guards and actions are evaluated in the
configuration at the beginning of the step. Since a node writes only its own
variables and registers, the result of a step is well defined.

\subsection{Daemons, moves and rounds}

A \emph{daemon} is a predicate on executions. It models the scheduler. At each
step, the daemon chooses the nodes that act and, for each of them, which of
its enabled rules is executed. We consider the \emph{distributed} daemon,
which can choose any nonempty set of enabled nodes. The distributed daemon is
\emph{unfair} if it has no other constraint. In particular, it can prevent a
node from acting forever, as long as another node is enabled. It is
\emph{weakly fair} if every node that is enabled continuously from some
configuration on is eventually chosen. The \emph{central} daemon chooses
exactly one node at each step, and the \emph{synchronous} daemon chooses all
enabled nodes at each step. Every execution under the central daemon or under
the weakly fair distributed daemon is an execution under the unfair
distributed daemon. Every execution under the synchronous daemon is an
execution under the weakly fair distributed daemon.

When a node executes a rule, we say that it performs a \emph{move}. The
number of moves of a step is the number of nodes chosen in this step. The
number of moves of an execution is its time complexity under the unfair
daemon.

Under a weakly fair daemon, time is measured in \emph{rounds}~\cite{DolevIM97}.
A node is \emph{neutralized} in a step $\gamma\to\gamma'$ if it is enabled in
$\gamma$, it is not chosen in this step, and it is not enabled in $\gamma'$.
The first round of an execution $e$ is the minimal prefix $e'$ of $e$ such
that every node enabled in the first configuration of $e$ performs a move or
is neutralized during $e'$. The second round of $e$ is the first round of the
execution obtained by removing $e'$ from $e$, and so on. A round measures the
time needed by the slowest node to act.

\subsection{Self-stabilization and silent algorithms}

An algorithm is \emph{self-stabilizing} for a specification if there is a set
$\mathcal{L}$ of configurations, called legitimate, such that every execution
starting in $\mathcal{L}$ satisfies the specification (\emph{correctness}) and
every execution, from any configuration, reaches a configuration of
$\mathcal{L}$ (\emph{convergence}).

An algorithm is \emph{silent} if all its executions are finite. A silent
algorithm is self-stabilizing for a predicate $P$ on configurations if all its
terminal configurations satisfy $P$. Most of our results concern algorithms of
the following kind.

\begin{definition}\label{def:F}
Let $F$ be an integer. An algorithm $\A$ in the state model has \emph{move
complexity at most $F$} if every execution of $\A$ under the unfair
distributed daemon, from any configuration, contains at most $F$ moves.
\end{definition}

Such an algorithm is silent, and each of its executions ends in a terminal
configuration after at most $F$ moves. Any prefix of an execution of $\A$,
and more generally any finite sequence of steps of $\A$, also contains at
most $F$ moves, since it can be extended into an execution.

\begin{remark}[Two notions of silence]\label{rem:silence}
For register models, Dolev, Gouda and Schneider~\cite{DolevGS99} call an
algorithm silent if it converges to a configuration after which the values
stored in the communication registers are fixed. Nodes may still read these
registers and write the same values again. This definition is weaker than
ours, which requires the execution to stop. The difference matters in the
link-register model. A node cannot check its own registers and has to rewrite
them forever, so the executions of $\TB(\A)$ are
infinite. When $\A$ has bounded move complexity, $\TA(\A)$ is silent in our
sense (\cref{thm:t1-silent}), and $\TB(\A)$ is silent in the sense
of~\cite{DolevGS99}: its registers eventually stop changing
(\cref{cor:t2-silent}).
\end{remark}
\section{The difference between the models}\label{sec:models}

The three models of \cref{sec:model} differ on two points: what a node can
read, and how much it does in one atomic step. This section makes these
differences concrete before we present the transformers. \Cref{sec:one-link}
puts the three models side by side. \Cref{sec:coloring} recalls the algorithm
$\Coloring$ of the conference version~\cite{CohenMP23}, which is our running
example. \Cref{sec:consequences} gives the consequences of these differences
for the transformers.

\subsection{Three models on one link}\label{sec:one-link}

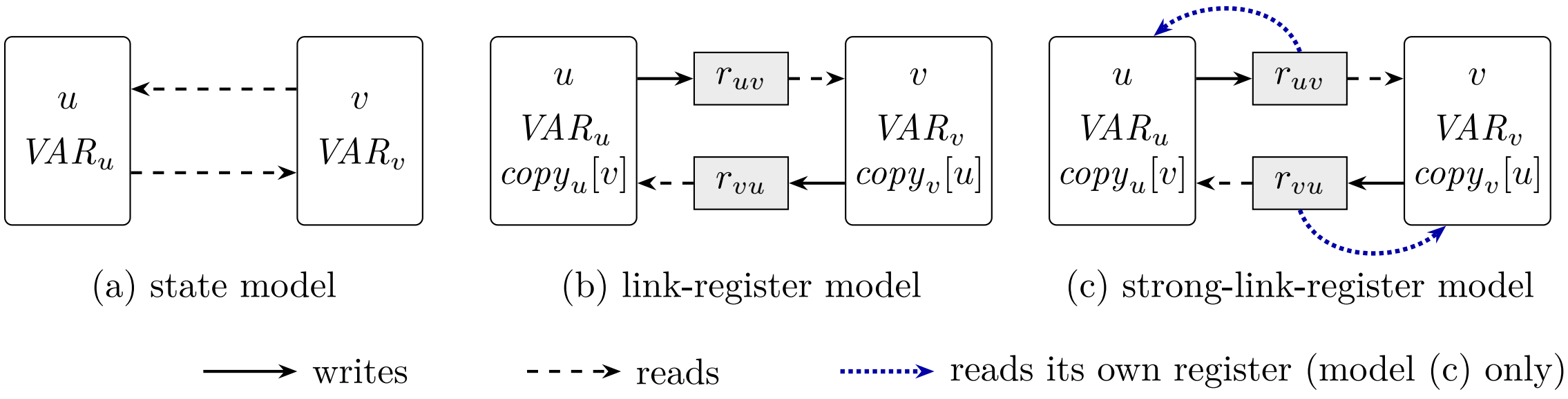
\begin{figure}[ht]
\centering
\begin{tikzpicture}[x=1cm,y=1cm,>={Stealth[length=1.8mm]},font=\small,
  proc/.style={draw,rounded corners=2pt,minimum width=1.2cm,minimum height=1.8cm,
               align=center,inner sep=2pt},
  regi/.style={draw,fill=gray!15,minimum width=0.9cm,minimum height=0.5cm,inner sep=1pt},
  wr/.style={->,thick},
  rd/.style={->,thick,dashed},
  own/.style={->,very thick,densely dotted,blue!65!black}]
  \begin{scope}[shift={(0,0)}]
    \node[proc] (u) at (0,0) {$u$\\[3pt]$\VAR_u$};
    \node[proc] (v) at (2.8,0) {$v$\\[3pt]$\VAR_v$};
    \draw[rd] ([yshift=4mm]v.west) -- ([yshift=4mm]u.east);
    \draw[rd] ([yshift=-4mm]u.east) -- ([yshift=-4mm]v.west);
    \node at (1.4,-1.5) {(a) state model};
  \end{scope}
  \begin{scope}[shift={(4.75,0)}]
    \node[proc] (u) at (0,0) {$u$\\[3pt]$\VAR_u$\\$\cp uv$};
    \node[proc] (v) at (3.4,0) {$v$\\[3pt]$\VAR_v$\\$\cp vu$};
    \node[regi] (ruv) at (1.7,0.5) {$\reg uv$};
    \node[regi] (rvu) at (1.7,-0.5) {$\reg vu$};
    \draw[wr] ([yshift=5mm]u.east) -- (ruv.west);
    \draw[rd] (ruv.east) -- ([yshift=5mm]v.west);
    \draw[wr] ([yshift=-5mm]v.west) -- (rvu.east);
    \draw[rd] (rvu.west) -- ([yshift=-5mm]u.east);
    \node at (1.7,-1.5) {(b) link-register model};
  \end{scope}
  \begin{scope}[shift={(10.1,0)}]
    \node[proc] (u) at (0,0) {$u$\\[3pt]$\VAR_u$\\$\cp uv$};
    \node[proc] (v) at (3.4,0) {$v$\\[3pt]$\VAR_v$\\$\cp vu$};
    \node[regi] (ruv) at (1.7,0.5) {$\reg uv$};
    \node[regi] (rvu) at (1.7,-0.5) {$\reg vu$};
    \draw[wr] ([yshift=5mm]u.east) -- (ruv.west);
    \draw[rd] (ruv.east) -- ([yshift=5mm]v.west);
    \draw[wr] ([yshift=-5mm]v.west) -- (rvu.east);
    \draw[rd] (rvu.west) -- ([yshift=-5mm]u.east);
    \draw[own] (ruv.north) to[out=110,in=40] ([xshift=3mm]u.north);
    \draw[own] (rvu.south) to[out=-70,in=-140] ([xshift=-3mm]v.south);
    \node at (1.7,-1.5) {(c) strong-link-register model};
  \end{scope}
  \draw[wr] (1.3,-2.3) -- (2.2,-2.3) node[right] {writes};
  \draw[rd] (4.4,-2.3) -- (5.3,-2.3) node[right] {reads};
  \draw[own] (7.4,-2.3) -- (8.3,-2.3) node[right,black] {reads its own register (model (c) only)};
\end{tikzpicture}
\caption{What two neighbors $u$ and $v$ can read and write. In the state
model, each node reads the variables of the other. In the link-register
models, the nodes only communicate through the registers $\reg uv$ and
$\reg vu$, and each of them keeps a copy of what it has read. In the
strong-link-register model, a node can also read the register that it
writes.}
\label{fig:models}
\end{figure}

Consider two neighbors $u$ and $v$. \Cref{fig:models} shows what each of them
can read and write in the three models.

\begin{itemize}
\item \emph{State model.} The variables of $v$ are visible to $u$. In one
atomic step, $u$ reads the variables of all its neighbors and updates its own
variables. What $u$ reads is the current configuration: if $v$ changes its
variables, the next step of $u$ sees the new values.
\item \emph{Link-register model.} Node $u$ cannot read the variables of $v$.
All it knows about $v$ comes from the register $\reg vu$, and it keeps what it
has read in local variables, the \emph{copies}. An atomic step is much
smaller: $u$ writes one register, or it reads registers and updates its local
variables. A new value of the variables of $v$ reaches $u$ in two atomic
steps: $v$ writes it in $\reg vu$, then $u$ reads $\reg vu$. The daemon
chooses when each of these steps takes place, and $v$ may change its
variables again in the meantime. The copies of $u$ can thus be out of date,
and $u$ has no means to know it. Moreover, $u$ cannot read $\reg uv$: it does
not know what $v$ can read about $u$.
\item \emph{Strong-link-register model.} The only change is that $u$ can also
read its own register $\reg uv$. It can compare it with its variables, and it
knows whether $v$ can read its current values.
\end{itemize}

\Cref{tab:compare} summarizes these points, together with their consequences
for the results of this report. The delay of the link-register models also
exists in the atomic state model of \cref{tab:models}, where a node reads the
state of one neighbor per atomic step.

\begin{table}[ht]
\centering
\small
\renewcommand{\arraystretch}{1.25}
\begin{tabular}{@{}>{\raggedright\arraybackslash}p{2.8cm}>{\raggedright\arraybackslash}p{3.1cm}>{\raggedright\arraybackslash}p{3.9cm}>{\raggedright\arraybackslash}p{3.9cm}@{}}
\toprule
 & State model & Link-register model & Strong-link-register model\\
\midrule
What $u$ can read &
the variables of all its neighbors &
the registers $\reg vu$ written by its neighbors &
the registers $\reg vu$ and its own registers $\reg uv$\\
One atomic step of $u$ &
reads all its neighbors and updates its variables &
writes one register, or reads registers and updates its local variables &
same as the link-register model\\
When $v$ learns a change of $u$ &
at its next step &
after a write by $u$ and a read by $v$, at times chosen by the daemon &
same as the link-register model\\
Can $u$ check what $v$ reads about $u$? &
nothing to check &
no &
yes\\
\midrule
In this report & source model & target of $\TB$ & target of $\TA$\\
Registers &
none &
rewritten forever, every node is always enabled &
written only when needed, the algorithm can be silent\\
Daemon and measure &
unfair distributed, moves &
weakly fair distributed, rounds &
unfair distributed, moves\\
Cost of a simulated move &
one move &
$O(\Delta)$ rounds &
$O(\Delta^2)$ moves\\
\bottomrule
\end{tabular}
\caption{The three models on a link $\{u,v\}$, and what each of them implies
in this report.}
\label{tab:compare}
\end{table}

\subsection{The algorithm $\Coloring$}\label{sec:coloring}

Our running example is the algorithm $\Coloring$ of~\cite{CohenMP23}, defined
in \cref{fig:coloring}. It colors the nodes with the colors
$1,\ldots,\Delta+1$ so that two neighbors have different colors. A node that
has the same color as a neighbor with a larger identifier takes the smallest
color that none of its neighbors uses. The node with the largest identifier
never moves, and an induction on the identifiers in decreasing order shows
that every execution is finite. In a terminal configuration the coloring is
proper. For instance, take two neighbors $s$ and $t$ with $\id_s<\id_t$ and
$c_s=c_t=1$. Only $s$ is enabled. After its move, $c_s=2$ and the
configuration is terminal.

\begin{figure}[ht]
\figrule
\small
\noindent\textbf{$\Coloring$ - Local Variables}
\begin{tabbing}
\hspace{1.0cm}\=\hspace{0.6cm}\=\kill
$\id_{u}$ \> : \> the identifier of $u$\\
$c_{u}$ \> : \> the local variable of $u$ corresponding to a color
\end{tabbing}

\noindent\textbf{$\Coloring$ - Rule for each node $u$}
\begin{tabbing}
\hspace{2.95cm}\=\kill
\textbf{\ChangeColor} \textbf{::} \> $\exists a\in N(u),\ (c_{a}=c_{u}\wedge \id_{u}<\id_{a})\ \to\ c_{u}:=\min\bigl([\Delta+1]\setminus\{c_{a}:a\in N(u)\}\bigr)$
\end{tabbing}
\figrule
\caption{Algorithm $\Coloring$ written in the state model~\cite{CohenMP23}.
Here $[\Delta+1]$ is the set of colors $\{1,\ldots,\Delta+1\}$.}
\label{fig:coloring}
\end{figure}

\subsection{Consequences for the transformers}\label{sec:consequences}

The two differences of \cref{sec:one-link} correspond to the two steps of our
construction.

\begin{itemize}
\item \emph{From the state model to registers.} Copies can be out of date: a
neighbor may have changed its variables without having written them yet. The
transformer $\TA$ adds acknowledgments and a local mutual exclusion to the
exchange of copies. With them, a node that executes a rule of $\A$ holds the
current variables of its neighbors (\cref{lem:faithful}). The price is a
handshake of four moves with each neighbor for each step of the protocol,
and the attempts that a node abandons when a neighbor moves first. This
leads to {$O(\Delta^2)$} moves per simulated move
(\cref{thm:moves}).
\item \emph{From the strong-link-register model to the link-register model.}
A node cannot check its own registers. The transformer
$\TB$ replaces the comparison between a register and the local variables by
flags, and adds the rule $\ToTrue$, which rewrites all registers again and
again. The algorithm is not silent anymore, except in the weaker sense of
\cref{rem:silence}. It needs the weakly fair daemon,
and one simulated move costs $O(\Delta)$ rounds
(\cref{thm:rounds}).
\end{itemize}
\section{The transformer \TA: from the state model to
the strong-link-register model}\label{sec:t1}

This section describes the transformer $\TA$, which is the transformer
$\mathcal{T}_1$ of the conference version~\cite{CohenMP23}. It takes an
algorithm $\A$ written in the state model and produces an algorithm
$\ASLRM=\TA(\A)$ for the strong-link-register model. The algorithm $\A$ is
given by the variables $\VAR_u$ of each node $u$ and by its guarded rules
$\mathcal{G}_j\to\mathcal{I}_j$.

In the transformed algorithm, nodes publish their variables in their
registers and keep copies of what their neighbors publish. Since these copies
can be out of date (\cref{sec:one-link}), the transformer adds acknowledgments
and a local mutual exclusion. We use the algorithm $\Coloring$ of
\cref{fig:coloring} as a running example.

\subsection{Principle of the transformation}

\paragraph{Copies.}
In the strong-link-register model, a node $u$ cannot read the variables of its
neighbors. For each neighbor $a$, it keeps a record $\cp ua$ that contains,
among other things, a copy $\cp ua.\id$ of the identifier of $a$ and a copy
$\cp ua.\VAR$ of the variables of $a$. Nodes exchange this information
through the registers. A rule of $\A$ is then evaluated on the copies: in the
guards and in the actions of $\A$, every occurrence of $\id_a$ or of a
variable of a neighbor $a$ is replaced by the corresponding field of
$\cp ua$. In other words, $u$ evaluates $\isAct$ and $\Activate$ on its
\emph{local view}
\[
\bigl(\id_u,\ \VAR_u,\ (\cp ua.\id,\ \cp ua.\VAR)_{a\in N(u)}\bigr).
\]
We write $\isAct(u,\A)$ and $\Activate(u,\A)$ for these two evaluations.

\begin{example}
For $\Coloring$, the guard of $\ChangeColor$ becomes
$\exists a\in N(u): (\cp ua.c=c_u\wedge \id_u<\cp ua.\id)$, and its action
becomes $c_u:=\min(\{1,\ldots,\Delta+1\}\setminus\{\cp ua.c : a\in N(u)\})$.
\end{example}

\paragraph{Three steps.}
Before it executes a rule of $\A$, a node $u$ must make sure that its copies
are correct, that is, $\cp ua.\VAR=\VAR_a$ for every neighbor $a$. Reading
the registers is not enough for that, since a neighbor may have changed its
variables without having written them yet.
To this end, $\TA$ replaces one move of $\A$ by a sequence of moves of
$\TA(\A)$, organized in three steps.
\begin{enumerate}
\item Node $u$ makes sure that it has stored the current values of the
variables of its neighbors.
\item Node $u$ informs its neighbors that it is about to execute a rule, so
that they do not change their own variables in the meantime. It then executes
the rule. This step relies on a local mutual exclusion based on identifiers,
as in the distributed constructions of a maximal independent set.
\item Node $u$ informs its neighbors of the new values of its variables, and
releases the lock only when it knows that all of them have received these
values.
\end{enumerate}

\paragraph{The variable $\state$.}
The lock is implemented with a variable $\state_u\in\{-1,0,\ldots,7\}$. Apart
from corrupted initial values, its meaning is the following.
\begin{itemize}
\item $\state_u=-1$: $u$ is idle. It has not started to execute a rule of
$\A$.
\item $\state_u=0$: $u$ intends to execute a rule of $\A$. It then goes
through the states $1,2,3,4$ and $5$. Their only role is to force several
complete exchanges with each neighbor, which remove corrupted values from the
registers and from the copies.
\item $\state_u=5$: $u$ is a candidate. It moves to state $6$ only if it has
priority over its neighbors.
\item $\state_u=6$: $u$ is about to execute a rule of $\A$, as soon as it has
priority over its neighbors that are also in state $6$.
\item $\state_u=7$: $u$ has executed a rule of $\A$ and informs its neighbors
of its new variables. It then returns to state $-1$.
\end{itemize}
Priorities are defined by the pair (state, identifier): a node gives way to a
neighbor that has a larger state, or the same state and a smaller identifier.

\paragraph{Acknowledgments.}
Node $u$ writes in the register $\reg ua$ all the information that it sends to
its neighbor $a$: its identifier, its state, its variables $\VAR_u$, and the
last value of $\state_a$ that it has read. This last field is an
acknowledgment: it tells $a$ that $u$ has received this value of $\state_a$.
In the record $\cp ua$, the field $\cp ua.\YS$ is the state of $a$ as $u$
knows it, and the field $\cp ua.\MS$ is the acknowledgment received from $a$,
that is, the value of $\state_u$ that $a$ has read.

A node changes its state only when all its neighbors have acknowledged its
current state. This is the predicate $\CBS(u)$ of \cref{fig:t1}. Thus, each
change of $\state_u$ is followed by an exchange of four moves with each
neighbor $a$: $u$ writes its new state in $\reg ua$, $a$ reads it, $a$ writes
the acknowledgment in $\reg au$, and $u$ reads it. We call this exchange a
\emph{handshake}. \Cref{lem:handshake} makes it precise.

\subsection{Formal definition}

The transformer $\TA$ is defined in \cref{fig:t1}. The figure uses the
presentation of the conference version~\cite{CohenMP23}. For each neighbor
$a$, node $u$ has a record $\cp ua$ with four fields, named $\id$, $\YS$,
$\MS$ and $\VAR$. It also has a register $\reg ua$ with four fields, named
$\id$, $\state$, $\YS$ and $\VAR$. This register is written by $u$, and it is
read by $a$ and by $u$.

The rules use three tuples. $\LV ua$ is what $u$ has to publish in $\reg ua$.
$\RR au$ is the content of the register $\reg au$, which $u$ reads.
$\CR ua$ is the content of the record $\cp ua$. Tuples are compared and
assigned component by component, and $\reg ua$ stands for the tuple of its
four fields. Thus, when $u$ executes $\Read(a)$,
\begin{align*}
\cp ua.\id&:=\reg au.\id, & \cp ua.\YS&:=\reg au.\state,\\
\cp ua.\MS&:=\reg au.\YS, & \cp ua.\VAR&:=\reg au.\VAR .
\end{align*}
The predicate $\isAct(u,\A)$ and the function $\Activate(u,\A)$ are evaluated
on the local view of $u$, which is the tuple
$\bigl(\id_u,\VAR_u,(\cp ua.\id,\cp ua.\VAR)_{a\in N(u)}\bigr)$. In the
action of $\ExecuteAlgo$, the statement $\Activate(u,\A)$ executes one rule
of $\A$ that is enabled on this local view: it stands for
$\VAR_u:=\Activate(u,\A)$.

\begin{figure}[ht]
\figrule
\footnotesize
\noindent\textbf{$\ASLRM$ - Local Variables}
\begin{tabbing}
\hspace{1.4cm}\=\hspace{0.55cm}\=\kill
$\id_{u}$ \> : \> the identifier of $u$\\
$\VAR_{u}$ \> : \> the local variables of $u$ of Algorithm $\A$\\
$\state_{u}$ \> $\in$ \> $\{-1, 0, \dots, 7\}$\\[5pt]
\hspace{2.05cm}\=\hspace{3.1cm}\=\kill
$\forall a \in N(u):$ \> $\cp ua.\id$ \> /* id of $a$ */\\
 \> $\cp ua.\YS$ \> /* local state of $a$ */\\
 \> $\cp ua.\MS$ \> /* the $a$'s view of the state of $u$ */\\
 \> $\cp ua.\VAR$ \> /* the local variables of $a$ in Algorithm $\A$ */
\end{tabbing}

\noindent\textbf{$\ASLRM$ - Predicates and functions}
\begin{tabbing}
$\LV ua\equiv(\id_{u},\state_{u},\cp ua.\YS,\VAR_{u})$\\[3pt]
$\RR au\equiv(\reg au.\id,\reg au.\state,\reg au.\YS,\reg au.\VAR)$\\[3pt]
$\CBS(u)\equiv\forall a\in N(u):\cp ua.\MS=\state_{u}$\\[3pt]
$\CR ua\equiv(\cp ua.\id,\cp ua.\YS,\cp ua.\MS,\cp ua.\VAR)$\\[3pt]
$\waitfull(u)\equiv\exists a\in N(u):(\cp ua.\YS>\state_{u})\ \vee$\\
\hspace{5.3cm}$(\cp ua.\YS=\state_{u}\wedge\cp ua.\id<\id_{u})$
\end{tabbing}

\noindent\textbf{$\ASLRM$ - Rules for each node $u$}
\begin{tabbing}
\hspace{1.75cm}\=\kill
$\forall a\in N(u),$ \> \textbf{\Write(a)} \textbf{::} $\reg ua\neq\LV ua\ \to\ \reg ua:=\LV ua$\\[3pt]
 \> \textbf{\Read(a)} \textbf{::} $\CR ua\neq\RR au\ \to\ \CR ua:=\RR au$\\[3pt]
\hspace{2.85cm}\=\hspace{0.8cm}\=\kill
\textbf{\WantToAct} \textbf{::} \> $\CBS(u)\wedge\state_u=-1\wedge\isAct(u,\A)\ \to\ \state_{u}:=\state_{u}+1$\\[3pt]
\textbf{\ModifyMyState} \textbf{::} \> $\CBS(u)\wedge 0\leq\state_u<5\ \to\ \state_{u}:=\state_{u}+1$\\[3pt]
\textbf{\ExecuteAlgo} \textbf{::} \> $\CBS(u)\wedge\neg\waitfull(u)\wedge\state_u=6\wedge\isAct(u,\A)\ \rightarrow$\\
 \>\> $\Activate(u,\A)$ ; $\state_{u}:=\state_{u}+1$\\[3pt]
\textbf{\toWAIT} \textbf{::} \> $\CBS(u)\wedge\isAct(u,\A)\wedge\neg\waitfull(u)\wedge\state_u=5$\\
 \>\> $\rightarrow\ \state_{u}:=\state_{u}+1$\\[3pt]
\textbf{\StartAgain} \textbf{::} \> $\CBS(u)\wedge(\state_{u}=7)\ \rightarrow\ \state_{u}:=-1$\\[3pt]
\textbf{\ResetRule} \textbf{::} \> $\CBS(u)\wedge\state_u\in\{5,6\}\wedge\neg\isAct(u,\A)\ \to\ \state_{u}:=-1$
\end{tabbing}
\figrule
\caption{Transformer $\TA$ transforms an algorithm $\A$ into an algorithm
$\ASLRM$ in the strong-link-register model. This is the transformer
$\mathcal{T}_1$ of~\cite{CohenMP23}, with the same presentation.}
\label{fig:t1}
\end{figure}

The rules $\Write(a)$ and $\Read(a)$ are the communication rules. A node
writes in a register only if the register does not hold the right value: this
is where it reads its own register. The six other rules are called the
\emph{local rules}. They read no register and modify only $\state_u$ and
$\VAR_u$. Their guards are mutually exclusive, so at most one local rule is
enabled at a node.

\begin{itemize}
\item $\WantToAct$ starts an attempt: an idle node whose local view enables a
rule of $\A$ moves to state $0$.
\item $\ModifyMyState$ moves from state $0$ to state $5$, one unit at a time.
\item $\toWAIT$ moves a candidate to state $6$ if a rule of $\A$ is still
enabled on its local view and if, according to its copies, no neighbor has a
larger state or the same state with a smaller identifier.
\item $\ExecuteAlgo$ executes a rule of $\A$ on the local view, under the same
priority condition in state $6$. This is the only rule that modifies $\VAR_u$.
\item $\StartAgain$ ends the attempt after the neighbors have acknowledged
state $7$. By then they have read the new value of $\VAR_u$.
\item $\ResetRule$ abandons the attempt when the local view no longer enables a
rule of $\A$. This happens when a neighbor has changed its variables in the
meantime.
\end{itemize}
In the text and in the proofs, we use two shorthands:
\begin{itemize}
\item $\wait(u)$ stands for $\waitfull(u)$;
\item $\cp ua=\reg au$ stands for $\CR ua=\RR au$.
\end{itemize}

\subsection{An execution on two nodes}\label{sec:two-nodes}

Consider again two neighbors $s$ and $t$ with $\id_s=1$, $\id_t=2$ and
$c_s=c_t=1$, and the algorithm $\TA(\Coloring)$. We start from the
configuration in which both states are $-1$, the registers hold the values
$\LV st$ and $\LV ts$, and the copies are equal to the registers. In every
configuration of the execution, exactly one rule is enabled, so the execution
is unique. It has $45$ moves. Node $s$ executes nine local rules: $\WantToAct$,
five times $\ModifyMyState$, $\toWAIT$, $\ExecuteAlgo$ and $\StartAgain$.
Each of them is followed by the same handshake: $s$ executes $\Write(t)$, $t$
executes $\Read(s)$, $t$ executes $\Write(s)$ and $s$ executes $\Read(t)$.
\section{Analysis of \TA}\label{sec:t1-analysis}
\begin{toappendix}
\label{app:t1}
\end{toappendix}

We study an execution of an algorithm obtained by transformer $\TA$ from an
algorithm $\A$ of the state model. The daemon is the unfair distributed
daemon, and the initial configuration is arbitrary. All the proofs of this section is given in \Cref{app:t1}

\subsection{Notation}\label{sec:notation}

Let $\Ex$ be an execution. The configurations of $\Ex$ are denoted by capital
letters, and a configuration of $\Ex$ is identified with its position in
$\Ex$. As in the conference version~\cite{CohenMP23}, a step of $\Ex$ from a
configuration $C$ to the next configuration $D$ is called a
\emph{transition} and is denoted by $C\mapsto D$. For two
configurations $C$ and $D$ of $\Ex$, $C<D$ means that $C$ is before $D$ in
$\Ex$, and $C\le D$ means that $C<D$ or $C=D$.
\begin{itemize}
\item A configuration $C$ is \emph{between} $A$ and $B$ if $A\le C\le B$.
\item A transition $C\mapsto D$ is \emph{between} $A$ and $B$ if $A\le C$ and
$D\le B$.
\item A node executes a rule \emph{between} $A$ and $B$ if it executes this
rule in a transition between $A$ and $B$.
\item A \emph{part} of $\Ex$ is a sequence of consecutive configurations of
$\Ex$.
\end{itemize}
For a variable, a register or a predicate $x$, we denote by $x(C)$ its value
in configuration $C$. For instance, $\state_u(C)$ is the value of $\state_u$
in $C$, and $\CBS(u)(C)$ is the value of predicate $\CBS(u)$ in $C$.

We denote by \emph{local rules} the rules in $\{\WantToAct$, $\ModifyMyState$,
$\toWAIT$, $\ExecuteAlgo$, $\StartAgain$, $\ResetRule\}$. Only these rules
modify the local variable $\state_u$.

We say that $\Ex$ contains $k$ \emph{consecutive transitions}
$C_i\mapsto D_i$, $1\le i\le k$, in which $u$ executes a local rule if
\begin{itemize}
\item $D_i\le C_{i+1}$ for $1\le i<k$;
\item $u$ executes a local rule in $C_i\mapsto D_i$ for $1\le i\le k$;
\item $u$ executes no local rule between $D_i$ and $C_{i+1}$ for $1\le i<k$.
\end{itemize}
We use the same definition with rule $\Read(v)$ or rule $\Write(v)$ instead
of the local rules.

$\proj C$ is said to be the \emph{projected configuration} of a configuration
$C$ if it is the vector of all the local values $\VAR_u$. It is a
configuration of algorithm $\A$.

The following properties are read on \cref{fig:t1}.

\begin{lemma}\label{lem:basic}
Let $(u,v)$ be an edge. Let $\Ex$ be an execution of algorithm $\TA(\A)$
which contains a transition $C\mapsto D$. We have
\begin{enumerate}
\item\label{b:one} $u$ executes at most one rule in $C\mapsto D$, and the
guard of this rule holds in $C$;
\item\label{b:write} if $\reg uv(D)\neq\reg uv(C)$, then $u$ executes rule
$\Write(v)$ in $C\mapsto D$. If $u$ executes rule $\Write(v)$ in
$C\mapsto D$, then
\[
\reg uv(D)=\LV uv(C)=\bigl(\id_u,\state_u(C),\cp uv.\YS(C),\VAR_u(C)\bigr);
\]
\item\label{b:read} if $\cp uv(D)\neq\cp uv(C)$, then $u$ executes rule
$\Read(v)$ in $C\mapsto D$. If $u$ executes rule $\Read(v)$ in $C\mapsto D$,
then $\cp uv(D)=\reg vu(C)$;
\item\label{b:state} $\state_u(D)\neq\state_u(C)$ if and only if $u$ executes
a local rule in $C\mapsto D$. In this case, $\CBS(u)(C)$ is true, and
\begin{itemize}
\item $\state_u(D)=\state_u(C)+1$ if the rule is $\WantToAct$,
$\ModifyMyState$, $\toWAIT$ or $\ExecuteAlgo$;
\item $\state_u(D)=-1$ and $\state_u(C)\in\{5,6,7\}$ if the rule is
$\ResetRule$ or $\StartAgain$;
\end{itemize}
\item\label{b:var} if $\VAR_u(D)\neq\VAR_u(C)$, then $u$ executes rule
$\ExecuteAlgo$ in $C\mapsto D$.
\end{enumerate}
\end{lemma}

By \cref{lem:basic}(\ref{b:state}), state $6$ is reached only by rule
$\toWAIT$, from state $5$, and state $7$ only by rule $\ExecuteAlgo$, from
state $6$.

\subsection{Properties on rules  $\Read$ and $\Write$ }

\begin{lemmarep}\label{lem:2:read}
Let $(u,v)$ be an edge. Let $\Ex$ be an execution of algorithm $\TA(\A)$
which contains two consecutive transitions $A\mapsto B$ and $C\mapsto D$ with
$B\le C$ in which $u$ executes rule $\Read(v)$. Then
\begin{itemize}
\item $\reg vu(A)\neq\reg vu(C)$;
\item $v$ executes rule $\Write(u)$ between $A$ and $C$.
\end{itemize}
\end{lemmarep}


\begin{proof}
By definition of rule $\Read$, $\cp uv(B)=\reg vu(A)$. Node $u$ does not
execute rule $\Read(v)$ between $B$ and $C$. Thus
$\cp uv(C)=\cp uv(B)=\reg vu(A)$. Since $u$ executes rule $\Read(v)$ in
$C\mapsto D$, we have $\cp uv(C)\neq\reg vu(C)$. Thus
$\reg vu(A)\neq\reg vu(C)$. Only rule $\Write(u)$ of $v$ modifies $\reg vu$.
Thus $v$ executes rule $\Write(u)$ between $A$ and $C$.
\end{proof}

\begin{lemmarep}\label{lem:2:write}
Let $(u,v)$ be an edge. Let $\Ex$ be an execution of algorithm $\TA(\A)$
which contains two consecutive transitions $A\mapsto B$ and $C\mapsto D$ with
$B\le C$ in which $u$ executes rule $\Write(v)$. Then
\begin{itemize}
\item $\LV uv(A)\neq\LV uv(C)$;
\item between $B$ and $C$, node $u$ executes a local rule, or it executes
rule $\Read(v)$ in a transition that modifies $\cp uv.\YS$.
\end{itemize}
\end{lemmarep}


\begin{proof}
By definition of rule $\Write$, $\reg uv(B)=\LV uv(A)$. Node $u$ does not
execute rule $\Write(v)$ between $B$ and $C$. Thus
$\reg uv(C)=\reg uv(B)=\LV uv(A)$. Since $u$ executes rule $\Write(v)$ in
$C\mapsto D$, we have $\reg uv(C)\neq\LV uv(C)$. Thus
$\LV uv(A)\neq\LV uv(C)$.

Recall that $\LV uv=(\id_u,\state_u,\cp uv.\YS,\VAR_u)$. Rule $\Write(v)$
does not modify these values, so $\LV uv(B)=\LV uv(A)$. Thus $\state_u$,
$\VAR_u$ or $\cp uv.\YS$ is modified in a transition between $B$ and $C$. By
\cref{lem:basic}, in this transition $u$ executes a local rule or rule
$\Read(v)$.
\end{proof}

\begin{lemmarep}\label{lem:3:read}
Let $(u,v)$ be an edge. Let $\Ex$ be an execution of algorithm $\TA(\A)$
which contains three transitions $A_j\mapsto B_j$, $1\le j\le 3$, with
$B_j\le A_{j+1}$ for $j\in\{1,2\}$, such that
\begin{itemize}
\item $u$ executes rule $\Read(v)$ and modifies $\cp uv.\YS$ in
$A_j\mapsto B_j$, for $1\le j\le 3$;
\item $\cp uv.\YS$ remains unchanged between $B_j$ and $A_{j+1}$, for
$j\in\{1,2\}$.
\end{itemize}
Then
\begin{itemize}
\item for $j\in\{2,3\}$, $v$ executes rule $\Write(u)$ and modifies
$\reg vu.\state$ in a transition between $A_{j-1}$ and $A_j$. Let
$Y_j\mapsto Z_j$ be the first such transition;
\item $v$ executes a local rule between $Z_2$ and $Y_3$.
\end{itemize}
\end{lemmarep}


\begin{proof}
Let $j\in\{2,3\}$. Under the assumption of the lemma and by definition of
rule $\Read$,
\begin{align*}
\cp uv.\YS(A_j)&=\cp uv.\YS(B_{j-1})=\reg vu.\state(A_{j-1}),\\
\cp uv.\YS(B_j)&=\reg vu.\state(A_j),
\end{align*}
and $\cp uv.\YS(A_j)\neq\cp uv.\YS(B_j)$. Thus
$\reg vu.\state(A_{j-1})\neq\reg vu.\state(A_j)$. Only rule $\Write(u)$ of
$v$ modifies $\reg vu$. This proves the first point.

By definition of transition $Y_3\mapsto Z_3$ and of rule $\Write$,
$\reg vu.\state(Y_3)\neq\reg vu.\state(Z_3)=\state_v(Y_3)$. Let $Y\mapsto Z$
be the last transition with $Z\le Y_3$ in which $v$ executes rule
$\Write(u)$. It exists and $Z_2\le Z$, since $Z_2\le A_2\le Y_3$. We have
$\reg vu.\state(Y_3)=\reg vu.\state(Z)=\state_v(Y)$. Thus
$\state_v(Y)\neq\state_v(Y_3)$, and $v$ executes a local rule between $Y$ and
$Y_3$. It does not execute it in $Y\mapsto Z$. Thus $v$ executes a local rule
between $Z$ and $Y_3$, and the lemma holds.
\end{proof}

\subsection{Properties on local variables {$\state$}{}}

\begin{lemmarep}\label{lem:0:local:state}
Let $u$ be a node. Let $s$ be an integer in $\{0,\ldots,7\}$. Let $\Ex$ be an
execution of algorithm $\TA(\A)$ which contains two configurations $S_1$ and
$S_2$ with $S_1<S_2$ such that $\state_u(S_1)=-1$ and $\state_u(S_2)=s$.
Then $u$ executes at least $s+1$ local rules between $S_1$ and $S_2$.
\end{lemmarep}

\begin{proof}
Let $C$ be the last configuration between $S_1$ and $S_2$ such that
$\state_u(C)=-1$. By definition of configuration $C$, node $u$ executes
neither rule $\StartAgain$ nor rule $\ResetRule$ between $C$ and $S_2$. By
\cref{lem:basic}(\ref{b:state}), each local rule executed by $u$ between $C$
and $S_2$ increases $\state_u$ by $1$. Thus $u$ executes $s+1$ local rules
between $C$ and $S_2$.
\end{proof}

\begin{lemmarep}\label{lem:5:state}
Let $u$ be a node. Let $\Ex$ be an execution of algorithm $\TA(\A)$ which
contains five consecutive transitions $C_i\mapsto D_i$, $1\le i\le 5$, with
$D_i\le C_{i+1}$ for $1\le i<5$, in which $u$ executes a local rule. We have
\begin{enumerate}
\item\label{s:const} the local variables $\state_u$ and $\VAR_u$ remain
unchanged between $D_i$ and $C_{i+1}$, for $1\le i<5$;
\item\label{s:distinct} $\state_u(C_i)\neq\state_u(C_j)$ for
$1\le i<j\le 5$.
\end{enumerate}
\end{lemmarep}

\begin{proof}
Only local rules modify the local variables $\state_u$ and $\VAR_u$
(\cref{lem:basic}), and the first point can be deduced. In particular,
$\state_u(C_{i+1})=\state_u(D_i)$ for $1\le i<5$.

Now, we focus on the second point. We prove it by contradiction. We assume
that $\state_u(C_i)=\state_u(C_j)$ with $i<j$. Rules that increase $\state_u$
by $1$ cannot bring it back to the same value. Thus $u$ executes rule
$\ResetRule$ or rule $\StartAgain$ in a transition $C_r\mapsto D_r$ with
$i\le r<j$. Let $r'$ be the smallest such index and let $r$ be the largest
one.
\begin{itemize}
\item The local rules executed in the transitions $C_\ell\mapsto D_\ell$ with
$i\le\ell<r'$ increase $\state_u$ by $1$. Thus
$\state_u(C_{r'})=\state_u(C_i)+(r'-i)$. By \cref{lem:basic}(\ref{b:state}),
$\state_u(C_{r'})\ge 5$. Thus $\state_u(C_i)\ge 5-(r'-i)$.
\item We have $\state_u(D_r)=-1$, and the local rules executed in the
transitions $C_\ell\mapsto D_\ell$ with $r<\ell<j$ increase $\state_u$ by
$1$. Thus $\state_u(C_j)=-1+(j-r-1)$.
\end{itemize}
Since $r'-i\le 3$ and $j-r-1\le 3$, we obtain
$\state_u(C_i)\ge 2\ge\state_u(C_j)$. The equality $\state_u(C_i)=2$ requires
$r'=i+3$. Then $r'=r=j-1$ and $\state_u(C_j)=-1$. In all cases
$\state_u(C_i)\neq\state_u(C_j)$. This leads to a contradiction.
\end{proof}

\begin{remark}
\Cref{lem:5:state}(\ref{s:distinct}) is the place where the intermediate
states $0,\ldots,4$ are used. After rule $\ResetRule$ or rule $\StartAgain$,
a node needs several local rules before its state can take again a value
that it had shortly before. This prevents an old acknowledgment from being
mistaken for a recent one.
\end{remark}

\subsection{Properties on consecutive local rules}\label{sec:handshake}

\begin{figure}[ht]
\centering
\begin{tikzpicture}[x=1cm,y=1cm,>={Stealth[length=2mm]},font=\small]
  \draw[thick] (0,3.0) -- (13.4,3.0);
  \draw[thick] (0,0) -- (13.4,0);
  \draw[gray!60,dashed] (0,2.0) -- (13.4,2.0);
  \draw[gray!60,dashed] (0,1.0) -- (13.4,1.0);
  \node[left] at (0,3.0) {$u$};
  \node[left] at (0,0) {$v$};
  \node[left,gray!70!black] at (0,2.0) {$\reg uv$};
  \node[left,gray!70!black] at (0,1.0) {$\reg vu$};
  \foreach \x in {0.9,12.6} { \draw[very thick] (\x,2.82) -- (\x,3.18); }
  \node at (0.9,3.45) {$C_4\mapsto D_4$};
  \node at (0.9,3.95) {local rule};
  \node at (12.6,3.45) {$C_5\mapsto D_5$};
  \node at (12.6,3.95) {local rule};
  \fill[blue!60!black] (3.2,3.0) circle (2pt);
  \node at (3.2,3.45) {$\alpha_5\mapsto\beta_5$};
  \node[blue!60!black] at (3.2,3.95) {$\Write(v)$};
  \fill[blue!60!black] (10.2,3.0) circle (2pt);
  \node at (10.2,3.45) {$A_5\mapsto B_5$};
  \node[blue!60!black] at (10.2,3.95) {$\Read(v)$};
  \fill[blue!60!black] (5.6,0) circle (2pt);
  \node at (5.6,-0.4) {$W_5\mapsto X_5$};
  \node[blue!60!black] at (5.6,-0.9) {$\Read(u)$};
  \fill[blue!60!black] (8.0,0) circle (2pt);
  \node at (8.0,-0.4) {$Y_5\mapsto Z_5$};
  \node[blue!60!black] at (8.0,-0.9) {$\Write(u)$};
  \draw[->,blue!60!black,thick] (3.2,3.0) -- (3.2,2.0);
  \draw[->,blue!60!black,thick] (3.2,2.0) -- (5.6,2.0) -- (5.6,0);
  \draw[->,blue!60!black,thick] (8.0,0) -- (8.0,1.0);
  \draw[->,blue!60!black,thick] (8.0,1.0) -- (10.2,1.0) -- (10.2,3.0);
  \node[above,font=\footnotesize] at (4.4,2.0) {$\reg uv.\state=s$};
  \node[right,font=\footnotesize] at (8.05,0.7) {$\reg vu.\YS=s$};
\end{tikzpicture}
\caption{The handshake between a node $u$ and a neighbor $v$
(\cref{lem:handshake}), with $s=\state_u(C_5)$. Time goes from left to right.
Between its local rules of $C_4\mapsto D_4$ and $C_5\mapsto D_5$, node $u$
writes $s$ in $\reg uv$ in transition $\alpha_5\mapsto\beta_5$, node $v$
reads it in $W_5\mapsto X_5$ and acknowledges it in $\reg vu$ in
$Y_5\mapsto Z_5$, and $u$ reads the acknowledgment in $A_5\mapsto B_5$.}
\label{fig:handshake}
\end{figure}
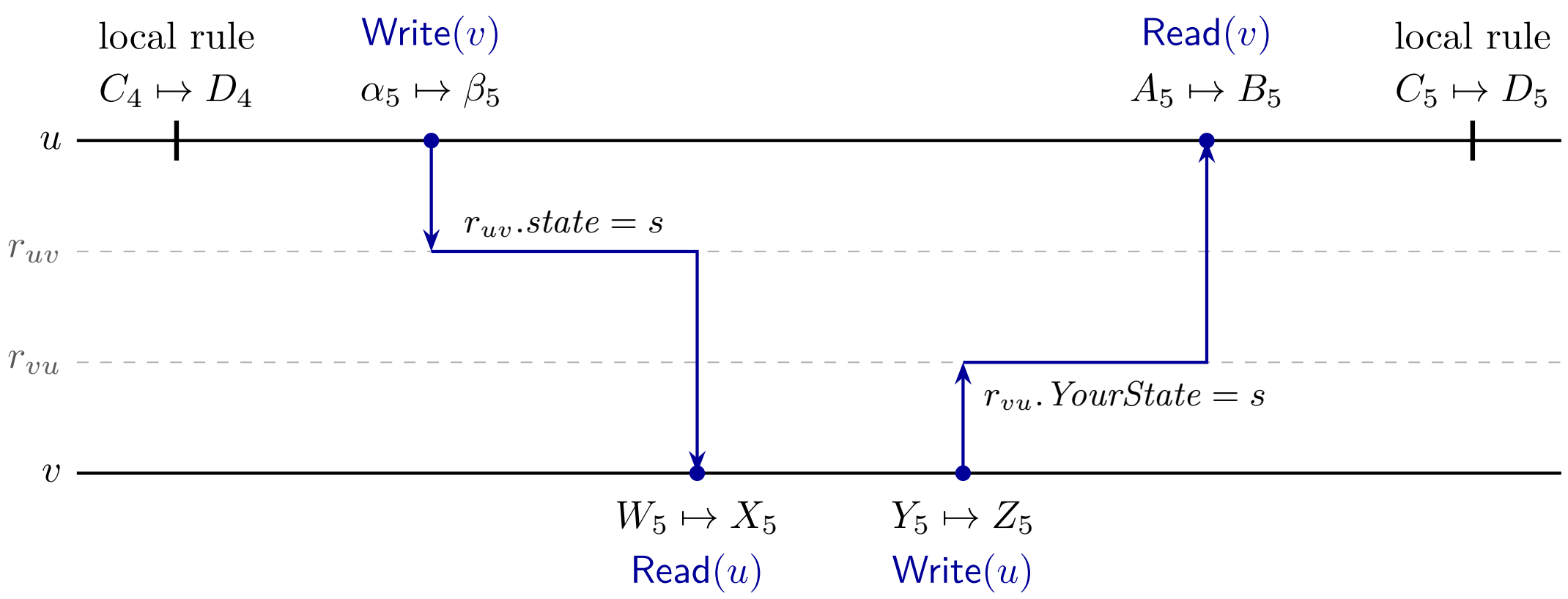

The lemmas of this subsection describe the communications that take place
between consecutive local rules of a node. They are illustrated in
\cref{fig:handshake}.

In \cref{lem:2:local:rules,lem:3:local:rules,lem:4:local:rules,lem:handshake},
$(u,v)$ is an edge and $\Ex$ contains $k$ consecutive transitions
$C_i\mapsto D_i$, $1\le i\le k$, in which $u$ executes a local rule. We use
the following transitions. The lemmas show that they exist.
\begin{itemize}
\item For $2\le i\le k$, $A_i\mapsto B_i$ is the last transition between
$D_{i-1}$ and $C_i$ in which $u$ executes rule $\Read(v)$.
\item For $3\le i\le k$, $Y_i\mapsto Z_i$ is the last transition between
$A_{i-1}$ and $A_i$ in which $v$ executes rule $\Write(u)$.
\item For $4\le i\le k$, $W_i\mapsto X_i$ is the last transition between
$Z_{i-1}$ and $Y_i$ in which $v$ executes rule $\Read(u)$.
\item For $5\le i\le k$, $\alpha_i\mapsto\beta_i$ is the last transition
between $W_{i-1}$ and $W_i$ in which $u$ executes rule $\Write(v)$.
\end{itemize}

\begin{lemmarep}\label{lem:2:local:rules}
Let $(u,v)$ be an edge. Let $\Ex$ be an execution of algorithm $\TA(\A)$
which contains two consecutive transitions $C_1\mapsto D_1$ and
$C_2\mapsto D_2$ with $D_1\le C_2$ in which $u$ executes a local rule. Then
\begin{itemize}
\item $\state_u(C_1)\neq\state_u(C_2)$;
\item $u$ executes rule $\Read(v)$ between $D_1$ and $C_2$;
\item $\cp uv(C_2)=\reg vu(A_2)$ and $\reg vu.\YS(A_2)=\state_u(C_2)$.
\end{itemize}
\end{lemmarep}

\begin{proof}
Since $u$ executes a local rule in transition $C_1\mapsto D_1$,
$\state_u(C_1)\neq\state_u(D_1)$ by \cref{lem:basic}(\ref{b:state}).
According to the assumption of this lemma, the local variable $\state_u$
remains unchanged between $D_1$ and $C_2$, and the first point is proven.

Since $u$ executes a local rule in $C_1\mapsto D_1$ and in $C_2\mapsto D_2$,
$\CBS(u)$ is true in $C_1$ and in $C_2$. Thus
$\cp uv.\MS(C_1)=\state_u(C_1)$ and $\cp uv.\MS(C_2)=\state_u(C_2)$. We
deduce that
\[
\cp uv.\MS(C_1)\neq\cp uv.\MS(C_2).
\]
The only way to modify $\cp uv.\MS$ is that $u$ executes rule $\Read(v)$.
Node $u$ does not execute it in $C_1\mapsto D_1$
(\cref{lem:basic}(\ref{b:one})). Thus $u$ executes rule $\Read(v)$ between
$D_1$ and $C_2$.

By definition of transition $A_2\mapsto B_2$ and of rule $\Read$,
$\cp uv(C_2)=\cp uv(B_2)=\reg vu(A_2)$. The third field gives
$\reg vu.\YS(A_2)=\cp uv.\MS(C_2)=\state_u(C_2)$.
\end{proof}

\begin{lemmarep}\label{lem:3:local:rules}
Let $(u,v)$ be an edge. Let $\Ex$ be an execution of algorithm $\TA(\A)$
which contains three consecutive transitions $C_i\mapsto D_i$,
$1\le i\le 3$, with $D_i\le C_{i+1}$ for $1\le i<3$, in which $u$ executes a
local rule. Then
\begin{itemize}
\item $v$ executes rule $\Write(u)$ between $A_2$ and $A_3$;
\item $\cp uv(C_3)=\bigl(\id_v,\state_v(Y_3),\state_u(C_3),\VAR_v(Y_3)\bigr)$;
\item $\cp vu.\YS(Y_3)=\state_u(C_3)$.
\end{itemize}
\end{lemmarep}

\begin{proof}
First, we apply \cref{lem:2:local:rules} to the execution between $C_1$ and
$D_2$ and to the execution between $C_2$ and $D_3$. We obtain
\[
\reg vu.\YS(A_2)=\state_u(C_2)\quad\text{and}\quad
\reg vu.\YS(A_3)=\state_u(C_3),
\]
with $\state_u(C_2)\neq\state_u(C_3)$. Thus
$\reg vu(A_2)\neq\reg vu(A_3)$, with $A_2<C_2<A_3$. Only rule
$\Write(u)$ of $v$ modifies $\reg vu$. Thus $v$ executes rule $\Write(u)$
between $A_2$ and $A_3$.

By definition of transition $Y_3\mapsto Z_3$ and of rule $\Write$,
\begin{align*}
\reg vu(A_3)&=\reg vu(Z_3)=\LV vu(Y_3)\\
&=\bigl(\id_v,\state_v(Y_3),\cp vu.\YS(Y_3),\VAR_v(Y_3)\bigr).
\end{align*}
By \cref{lem:2:local:rules}, $\cp uv(C_3)=\reg vu(A_3)$ and
$\reg vu.\YS(A_3)=\state_u(C_3)$. Thus $\cp vu.\YS(Y_3)=\state_u(C_3)$, and
the lemma holds.
\end{proof}

\begin{lemmarep}\label{lem:4:local:rules}
Let $(u,v)$ be an edge. Let $\Ex$ be an execution of algorithm $\TA(\A)$
which contains four consecutive transitions $C_i\mapsto D_i$, $1\le i\le 4$,
with $D_i\le C_{i+1}$ for $1\le i<4$, in which $u$ executes a local rule.
Then
\begin{itemize}
\item $v$ executes rule $\Read(u)$ between $Z_3$ and $Y_4$;
\item $\cp vu(Y_4)=\reg uv(W_4)$ and $\reg uv.\state(W_4)=\state_u(C_4)$.
\end{itemize}
\end{lemmarep}

 \begin{proof}
We apply \cref{lem:3:local:rules} to the execution between $C_1$ and $D_3$
and to the execution between $C_2$ and $D_4$. We obtain
$\cp vu.\YS(Y_3)=\state_u(C_3)$ and $\cp vu.\YS(Y_4)=\state_u(C_4)$. By
\cref{lem:2:local:rules}, $\state_u(C_3)\neq\state_u(C_4)$. Thus
\[
\cp vu.\YS(Y_3)\neq\cp vu.\YS(Y_4),\qquad\text{with } Y_3<Z_3\le A_3\le Y_4 .
\]
The only way to modify $\cp vu.\YS$ is that $v$ executes rule $\Read(u)$.
Node $v$ does not execute it in $Y_3\mapsto Z_3$. Thus $v$ executes rule
$\Read(u)$ between $Z_3$ and $Y_4$.

By definition of transition $W_4\mapsto X_4$ and of rule $\Read$,
$\cp vu(Y_4)=\cp vu(X_4)=\reg uv(W_4)$. The second field gives
$\reg uv.\state(W_4)=\cp vu.\YS(Y_4)=\state_u(C_4)$.
\end{proof}

\begin{lemmarep}\label{lem:handshake}\label{lem:5:local:rules}
Let $(u,v)$ be an edge. Let $\Ex$ be an execution of algorithm $\TA(\A)$
which contains five consecutive transitions $C_i\mapsto D_i$, $1\le i\le 5$,
with $D_i\le C_{i+1}$ for $1\le i<5$, in which $u$ executes a local rule.
Then
\begin{enumerate}
\item\label{h:b} $u$ executes rule $\Write(v)$ between $W_4$ and $W_5$;
\item\label{h:order}
$D_4\le\alpha_5<\beta_5\le W_5<X_5\le Y_5<Z_5\le A_5<B_5\le C_5$;
\item\label{h:eq}
$\state_u(\alpha_5)=\reg uv.\state(W_5)=\cp vu.\YS(Y_5)=\reg vu.\YS(A_5)
=\state_u(C_5)$;
\item\label{h:y}
$\cp uv(C_5)=\bigl(\id_v,\state_v(Y_5),\state_u(C_5),\VAR_v(Y_5)\bigr)$;
\item\label{h:w} $\cp vu.\id(C)=\id_u$, $\cp vu.\YS(C)=\state_u(C_5)$ and
$\cp vu.\VAR(C)=\VAR_u(C_5)$, for every configuration $C$ between $X_5$ and
$C_5$.
\end{enumerate}
\end{lemmarep}

\begin{proof}
We apply \cref{lem:4:local:rules} to the execution between $C_1$ and $D_4$
and to the execution between $C_2$ and $D_5$. We obtain
$\reg uv.\state(W_4)=\state_u(C_4)$ and $\reg uv.\state(W_5)=\state_u(C_5)$.
By \cref{lem:2:local:rules}, $\state_u(C_4)\neq\state_u(C_5)$. Thus
\[
\reg uv(W_4)\neq\reg uv(W_5),\qquad\text{with } W_4<X_4\le Y_4<Z_4\le W_5 .
\]
Only rule $\Write(v)$ of $u$ modifies $\reg uv$. Thus $u$ executes rule
$\Write(v)$ between $W_4$ and $W_5$, and the first point is proven.

By definition of transition $\alpha_5\mapsto\beta_5$ and of rule $\Write$,
\[
\reg uv(W_5)=\reg uv(\beta_5)=\LV uv(\alpha_5).
\]
The second field gives $\state_u(\alpha_5)=\reg uv.\state(W_5)$, which is
equal to $\state_u(C_5)$.
The other equalities of the third point, and the fourth point, are given by
\cref{lem:2:local:rules,lem:3:local:rules,lem:4:local:rules}, applied to
the last transitions.

Now, we prove the second point. By definition of the transitions,
\[
D_1\le A_2\le Y_3<Z_3\le W_4\le\alpha_5<\beta_5\le W_5<X_5\le Y_5<Z_5\le A_5<B_5\le C_5 .
\]
Thus $\alpha_5$ is between $D_{i-1}$ and $C_i$ for some $i\in\{2,3,4,5\}$,
and $\state_u(\alpha_5)=\state_u(C_i)$ by \cref{lem:5:state}(\ref{s:const}).
\Cref{lem:5:state}(\ref{s:distinct}) claims that
$\state_u(C_5)\neq\state_u(C_i)$ for $i\in\{2,3,4\}$. Since
$\state_u(\alpha_5)=\state_u(C_5)$, we have $i=5$, that is,
$D_4\le\alpha_5$.

It remains to prove the last point. By
\cref{lem:5:state}(\ref{s:const}), $\state_u(C)=\state_u(C_5)$ and
$\VAR_u(C)=\VAR_u(C_5)$ for every configuration $C$ between $D_4$ and $C_5$.
Thus every execution of rule $\Write(v)$ by $u$ between $D_4$ and $C_5$
writes $\id_u$, $\state_u(C_5)$ and $\VAR_u(C_5)$ in the fields $\id$,
$\state$ and $\VAR$ of $\reg uv$. Since $u$ executes rule $\Write(v)$ in
$\alpha_5\mapsto\beta_5$ and since only $u$ writes in $\reg uv$, for every
configuration $C'$ between $\beta_5$ and $C_5$ we have
\begin{equation}\label{eq:reg-stable}
\reg uv.\id(C')=\id_u,\qquad \reg uv.\state(C')=\state_u(C_5),\qquad
\reg uv.\VAR(C')=\VAR_u(C_5).
\end{equation}
Let $C$ be a configuration between $X_5$ and $C_5$. Let $C'\mapsto C''$ be
the last transition with $C''\le C$ in which $v$ executes rule $\Read(u)$.
It exists and $W_5\le C'$, since $X_5\le C$. Thus $C'$ is between $\beta_5$
and $C_5$, and $\cp vu(C)=\cp vu(C'')=\reg uv(C')$. The last point follows
from \eqref{eq:reg-stable}.
\end{proof}

\begin{remark}\label{rem:scope}
The proofs of
\cref{lem:2:read,lem:3:read,lem:0:local:state,lem:5:state,lem:2:local:rules,lem:3:local:rules,lem:4:local:rules,lem:handshake}
use only \cref{lem:basic} and the guards of rule $\Read$ and of the local
rules. They do not use the guard of rule $\Write$. The same holds for
\cref{lem:faithful,cor:un\faithful,thm:simulation,cor:M,lem:ten} below. This
will allow us to reuse these statements for transformer $\TB$.
\end{remark}

\subsection{\Faithful{} executions of rule {$\ExecuteAlgo$}{}}

\begin{definition}\label{def:faithful}
Let $\Ex$ be an execution of algorithm $\TA(\A)$ which contains a transition
$C\mapsto D$ in which a node $u$ executes rule $\ExecuteAlgo$. This execution
of rule $\ExecuteAlgo$ is said to be \emph{\faithful{}} if, for every
$v\in N(u)$,
\[
\cp uv.\id(C)=\id_v\qquad\text{and}\qquad\cp uv.\VAR(C)=\VAR_v(C).
\]
\end{definition}

In a \faithful{} execution of rule $\ExecuteAlgo$ in $C\mapsto D$, node $u$
evaluates $\isAct(u,\A)$ and $\Activate(u,\A)$ on its view in the projected
configuration $\proj C$.

\begin{lemmarep}\label{lem:faithful}
Let $u$ be a node. Let $\Ex$ be an execution of algorithm $\TA(\A)$ which
contains five consecutive transitions $C_i\mapsto D_i$, $1\le i\le 5$, with
$D_i\le C_{i+1}$ for $1\le i<5$, in which $u$ executes a local rule. Assume
that $u$ executes rule $\ExecuteAlgo$ in $C_5\mapsto D_5$. Then, for every
$v\in N(u)$,
\begin{itemize}
\item $\cp uv.\id(C_5)=\id_v$ and $\cp uv.\VAR(C_5)=\VAR_v(C_5)$;
\item $v$ does not execute rule $\ExecuteAlgo$ in $C_5\mapsto D_5$.
\end{itemize}
\end{lemmarep}

\begin{proof}
By definition of the guard of rule $\ExecuteAlgo$, $\state_u(C_5)=6$. Let
$v\in N(u)$. Let $W\mapsto X$ and $Y\mapsto Z$ be the transitions
$W_5\mapsto X_5$ and $Y_5\mapsto Z_5$ of \cref{lem:handshake}. We have
$X\le Y<Z\le C_5$ and
\begin{enumerate}[label=(\roman*)]
\item\label{f:i} $\cp uv(C_5)=\bigl(\id_v,\state_v(Y),6,\VAR_v(Y)\bigr)$;
\item\label{f:ii} $\cp vu.\id(C)=\id_u$ and $\cp vu.\YS(C)=6$ for every
configuration $C$ between $X$ and $C_5$.
\end{enumerate}
By~\ref{f:i}, $\cp uv.\id(C_5)=\id_v$ and $\cp uv.\VAR(C_5)=\VAR_v(Y)$. We
prove that $v$ does not execute rule $\ExecuteAlgo$ in a transition
$E\mapsto E'$ with $Y\le E\le C_5$. This proves the second point. By
\cref{lem:basic}(\ref{b:var}), it also gives $\VAR_v(Y)=\VAR_v(C_5)$, which
proves the first point.

We prove it by contradiction. We assume that $v$ executes rule
$\ExecuteAlgo$ in such a transition $E\mapsto E'$. Since $v$ executes rule
$\Write(u)$ in $Y\mapsto Z$, we have $Y<E\le C_5$. By definition of the guard
of rule $\ExecuteAlgo$, $\state_v(E)=6$ and $\wait(v)(E)$ is false.
By~\ref{f:ii}, $\cp vu.\YS(E)=6=\state_v(E)$ and $\cp vu.\id(E)=\id_u$. Since
$\wait(v)(E)$ is false, $\id_u>\id_v$.

By definition of the guard of rule $\ExecuteAlgo$ at $u$, $\wait(u)(C_5)$ is
false. By~\ref{f:i}, $\cp uv.\YS(C_5)=\state_v(Y)$ and
$\cp uv.\id(C_5)=\id_v<\id_u$. Since $\wait(u)(C_5)$ is false and
$\state_u(C_5)=6$, we have $\state_v(Y)\le 6$ and $\state_v(Y)\neq 6$. Thus
$\state_v(Y)\le 5$.

So $\state_v(Y)\le 5$ and $\state_v(E)=6$. Let $G\mapsto G'$ be the last
transition between $Y$ and $E$ such that $\state_v(G)\neq 6$. Then
$\state_v(G')=6$. Thus $v$ executes rule $\toWAIT$ in $G\mapsto G'$, and
$\state_v(G)=5$. Since $v$ executes rule $\Write(u)$ in $Y\mapsto Z$, we have
$Y<G<E\le C_5$. By~\ref{f:ii}, $\cp vu.\YS(G)=6>5=\state_v(G)$. Thus
$\wait(v)(G)$ is true. This leads to a contradiction with the guard of rule
$\toWAIT$.
\end{proof}

\begin{corollaryrep}\label{cor:unfaithful}
Let $u$ be a node. Let $\Ex$ be an execution of algorithm $\TA(\A)$ with
initial configuration $C_0$. We have
\begin{enumerate}
\item\label{u:one} at most one execution of rule $\ExecuteAlgo$ by $u$ in
$\Ex$ is not \faithful{};
\item\label{u:init} if an execution of rule $\ExecuteAlgo$ by $u$ is not
\faithful{}, then $\state_u(C_0)\in\{3,4,5,6\}$;
\item\label{u:correct} if $\Ex$ contains a configuration $S$ such that
$\state_u(S)=-1$, then every execution of rule $\ExecuteAlgo$ by $u$ after
$S$ is \faithful{}.
\end{enumerate}
Consequently, the number $k$ of executions of rule $\ExecuteAlgo$ in $\Ex$
that are not \faithful{} satisfies
\[
k\ \le\ \bigl|\{u\in V : \state_u(C_0)\in\{3,4,5,6\}\}\bigr|\ \le\ n .
\]
\end{corollaryrep}

\begin{proof}
By \cref{lem:faithful}, if an execution of rule $\ExecuteAlgo$ by $u$ in a
transition $C\mapsto D$ is not \faithful{}, then $u$ executes at most three
local rules before $C$. Moreover $\state_u(C)=6$.

We prove the second point. Let $j\le 3$ be the number of local rules executed
by $u$ before $C$. If one of them were rule $\ResetRule$ or rule
$\StartAgain$, the state of $u$ would be $-1$ after it. At most two local
rules could follow, and we would have $\state_u(C)\le 1$. Thus these $j$
local rules increase $\state_u$ by $1$, and
$\state_u(C_0)=6-j\in\{3,4,5,6\}$.

We prove the first point. After an execution of rule $\ExecuteAlgo$, the
state of $u$ is $7$. Before another execution of rule $\ExecuteAlgo$, node
$u$ must execute at least eight local rules: $\StartAgain$, $\WantToAct$,
five times $\ModifyMyState$ and $\toWAIT$. Thus every execution of rule
$\ExecuteAlgo$ by $u$, except the first one, is preceded by at least four
local rules of $u$, and it is \modif{\faithful}.

We prove the third point. By \cref{lem:0:local:state}, node $u$ executes at
least seven local rules between $S$ and a configuration $C$ after $S$ such
that $\state_u(C)=6$. Thus every execution of rule $\ExecuteAlgo$ by $u$
after $S$ is preceded by at least four local rules of $u$, and it is
\modif{\faithful}.
\end{proof}

Executions of rule $\ExecuteAlgo$ that are not \faithful{} do occur: a node may
start in state $6$ with corrupted copies that satisfy the guard of rule
$\ExecuteAlgo$, and the daemon may choose it in the first transition. By
\cref{cor:unfaithful}(\ref{u:correct}), this cannot happen to a node that has
been in state $-1$.

\subsection{Projected executions}

Let $\Ex'$ be a part of an execution of algorithm $\TA(\A)$. We write
\[
\Ex'=(\Ex_0\,C_1C'_1\,\Ex_1\,C_2C'_2\,\Ex_2\ldots)
\]
where
\begin{itemize}
\item for all $j$, at least one node executes rule $\ExecuteAlgo$ in
transition $C_j\mapsto C'_j$;
\item no node executes rule $\ExecuteAlgo$ in the other transitions of
$\Ex'$.
\end{itemize}
$\proj{\Ex'}$ is said to be the \emph{projected execution} of $\Ex'$ if
\begin{itemize}
\item $\proj{\Ex'}=(\proj{C_1}\proj{C_2}\ldots)$ when $\Ex'$ contains
infinitely many transitions $C_j\mapsto C'_j$;
\item $\proj{\Ex'}=(\proj{C_1}\proj{C_2}\ldots\proj{C_p}\proj{C'_p})$ when
$\Ex'$ contains $p\ge 1$ transitions $C_j\mapsto C'_j$;
\item $\proj{\Ex'}=(\proj{S})$, where $S$ is the first configuration of
$\Ex'$, when no node executes rule $\ExecuteAlgo$ in $\Ex'$.
\end{itemize}

\begin{theorem}[Simulation]\label{thm:simulation}
Let $\Ex'$ be a part of an execution of algorithm $\TA(\A)$ such that all the
executions of rule $\ExecuteAlgo$ in $\Ex'$ are  {\faithful}. Then
\begin{enumerate}
\item the projected execution $\proj{\Ex'}$ is a sequence of transitions of
algorithm $\A$ in the state model;
\item the number of executions of rule $\ExecuteAlgo$ in $\Ex'$ is the number
of moves of algorithm $\A$ in $\proj{\Ex'}$.
\end{enumerate}
\end{theorem}

\begin{proof}
We write $\Ex'=(\Ex_0\,C_1C'_1\,\Ex_1\,C_2C'_2\,\Ex_2\ldots)$ as above. Let
$j$ be an integer.

No node executes rule $\ExecuteAlgo$ between $C'_j$ and $C_{j+1}$. By
\cref{lem:basic}(\ref{b:var}), no node modifies its local variable $\VAR$. So
$\proj{C'_j}=\proj{C_{j+1}}$.

Now, we focus on transition $C_j\mapsto C'_j$. Let $u$ be a node executing
rule $\ExecuteAlgo$ in this transition. Since this execution is \faithful,
$\cp uv.\id(C_j)=\id_v$ and $\cp uv.\VAR(C_j)=\VAR_v(C_j)$ for every neighbor
$v$ of $u$. Since $\isAct(u,\A)$ is true in $C_j$, node $u$ is enabled in
the projected configuration $\proj{C_j}$, and $\VAR_u(C'_j)$ is the result of
a rule of $\A$ enabled at $u$ in $\proj{C_j}$. The other nodes do not modify
their local variable $\VAR$. So $\proj{C_j}\mapsto\proj{C'_j}$ is a
transition of algorithm $\A$ in the state model, in which only the nodes
executing rule $\ExecuteAlgo$ in $C_j\mapsto C'_j$ execute a rule of $\A$.

Since $\proj{C'_j}=\proj{C_{j+1}}$, the theorem holds.
\end{proof}

An execution of rule $\ExecuteAlgo$ that is not \faithful{} replaces $\VAR_u$ by
a value computed from corrupted copies. For algorithm $\A$, this is a
transient fault that hits one node. The next corollary bounds the number of
executions of rule $\ExecuteAlgo$.

\begin{corollaryrep}\label{cor:M}
Let $\A$ be an algorithm in the state model with move complexity at most $F$
(\cref{def:F}). Let $\Ex$ be an execution of algorithm $\TA(\A)$ which
contains $M$ executions of rule $\ExecuteAlgo$, $k$ of which are not
{\faithful}. Then
\[
M\le (k+1)F+k .
\]
\end{corollaryrep}

\begin{proof}
Let $T_1,\ldots,T_p$ be the transitions of $\Ex$ which contain an execution
of rule $\ExecuteAlgo$ that is not \modif{\faithful}. We have $p\le k$. These
transitions split $\Ex$ into at most $p+1$ parts
$\Ex'_0,\ldots,\Ex'_p$ in which all the executions of rule $\ExecuteAlgo$ are
\modif{\faithful}. By \cref{thm:simulation}, $\proj{\Ex'_j}$ is a sequence of
transitions of algorithm $\A$, whose moves are the executions of rule
$\ExecuteAlgo$ in $\Ex'_j$.

Consider a transition $T_j=(C\mapsto D)$ and the set $U$ of the nodes whose
execution of rule $\ExecuteAlgo$ in $T_j$ is {\faithful}. Assume that
$U\neq\emptyset$. As in the proof of \cref{thm:simulation}, the nodes of $U$
are enabled in $\proj C$, and their new values $\VAR$ are the result of a
transition of algorithm $\A$ from $\proj C$ in which only the nodes of $U$
execute a rule. This transition extends the sequence $\proj{\Ex'_{j-1}}$,
which ends in $\proj C$.

Thus the \faithful{} executions of rule $\ExecuteAlgo$ are partitioned into at
most $p+1$ groups, and each group is the set of moves of a sequence of
transitions of algorithm $\A$. Each group has at most $F$ elements. Hence at
most $(p+1)F\le(k+1)F$ executions of rule $\ExecuteAlgo$ are \faithful, and
$k$ are not.
\end{proof}

\subsection{Move complexity}

\begin{lemmarep}\label{lem:ten}
Let $x$ be a node. Let $\Ex$ be an execution of algorithm $\TA(\A)$ which
contains two configurations $S$ and $T$ with $S\le T$ such that no node of
$N[x]$ executes rule $\ExecuteAlgo$ between $S$ and $T$. Then $x$ executes at
most $10$ local rules between $S$ and $T$.
\end{lemmarep}

\begin{proof}
Let $C_i\mapsto D_i$, $1\le i\le k$, be the transitions between $S$ and $T$
in which $x$ executes a local rule. They are consecutive. By
\cref{lem:basic}(\ref{b:var}), the local variable $\VAR_v$ remains unchanged
between $S$ and $T$, for every $v\in N[x]$.

Let $3\le i\le k$ and let $v\in N(x)$. We apply \cref{lem:3:local:rules} to
the transitions $C_{i-2}\mapsto D_{i-2}$, $C_{i-1}\mapsto D_{i-1}$ and
$C_i\mapsto D_i$. There exists a configuration $Y$ with
$S\le C_{i-2}<Y<C_i\le T$ such that $\cp xv.\id(C_i)=\id_v$ and
$\cp xv.\VAR(C_i)=\VAR_v(Y)=\VAR_v(S)$. Thus, in each configuration $C_i$
with $i\ge 3$, $\isAct(x,\A)$ is evaluated on the same tuple
\[
\bigl(\id_x,\ \VAR_x(S),\ (\id_v,\VAR_v(S))_{v\in N(x)}\bigr),
\]
and it has the same value. We split the proof into two cases according to
this value.

Case $1$. Assume that this value is false. The rules executed in the
transitions $C_i\mapsto D_i$ with $i\ge 3$ are in
$\{\ModifyMyState,\ResetRule,\StartAgain\}$. Starting from any state, $x$
executes at most five times rule $\ModifyMyState$, then rule $\ResetRule$.
Then $\state_x=-1$ and none of these rules is enabled. So $k-2\le 6$.

Case $2$. Assume that this value is true. Since $x$ does not execute rule
$\ExecuteAlgo$ between $S$ and $T$, the rules executed in the transitions
$C_i\mapsto D_i$ with $i\ge 3$ are in
$\{\StartAgain,\WantToAct,\ModifyMyState,\toWAIT\}$. Starting from any state,
$x$ executes at most rule $\StartAgain$, rule $\WantToAct$, five times rule
$\ModifyMyState$ and rule $\toWAIT$. Then $\state_x=6$ and none of these
rules is enabled. So $k-2\le 8$.

In both cases $k\le 10$.
\end{proof}

Let $\Ex'$ be a finite prefix of an execution of algorithm $\TA(\A)$. For an
edge $(u,v)$, we denote by
\begin{itemize}
\item $L_u$ the number of local rules executed by $u$ in $\Ex'$;
\item $M_u$ the number of executions of rule $\ExecuteAlgo$ by $u$ in
$\Ex'$;
\item $\mathit{Wr}_{uv}$ the number of executions of rule $\Write(v)$ by $u$
in $\Ex'$;
\item $\mathit{Rd}_{uv}$ the number of executions of rule $\Read(v)$ by $u$
in $\Ex'$.
\end{itemize}
Let $L=\sum_{u\in V}L_u$ and $M=\sum_{u\in V}M_u$.

\begin{lemmarep}\label{lem:counting}
Let $(u,v)$ be an edge. Let $\Ex'$ be a finite prefix of an execution of
algorithm $\TA(\A)$. We have
\begin{enumerate}
\item\label{c:L} $L_u\le 10+11\sum_{x\in N[u]}M_x$;
\item\label{c:R} $\mathit{Rd}_{uv}\le 1+\mathit{Wr}_{vu}$;
\item\label{c:W} $\mathit{Wr}_{uv}\le L_u+L_v+3$.
\end{enumerate}
\end{lemmarep}

\begin{proof}
We prove the first point. Let $q=\sum_{x\in N[u]}M_x$. At most $q$
transitions of $\Ex'$ contain an execution of rule $\ExecuteAlgo$ by a node
of $N[u]$. They split $\Ex'$ into at most $q+1$ parts in which no node of
$N[u]$ executes rule $\ExecuteAlgo$. By \cref{lem:ten}, $u$ executes at most
$10$ local rules in each part. It executes at most one local rule in each of
the $q$ transitions. Thus $L_u\le 10(q+1)+q$.

We prove the second point. By \cref{lem:2:read}, if $A\mapsto B$ and
$C\mapsto D$ are two consecutive transitions in which $u$ executes rule
$\Read(v)$, then $v$ executes rule $\Write(u)$ between $A$ and $C$. These
executions of rule $\Write(u)$ are different for different pairs of
consecutive transitions. Thus $\mathit{Rd}_{uv}-1\le\mathit{Wr}_{vu}$.

We prove the third point. Let $\mathit{Rd}^{\star}_{uv}$ be the number of
transitions of $\Ex'$ in which $u$ executes rule $\Read(v)$ and modifies
$\cp uv.\YS$. By \cref{lem:2:write}, between two consecutive transitions in
which $u$ executes rule $\Write(v)$, node $u$ executes a local rule, or it
executes rule $\Read(v)$ in a transition that modifies $\cp uv.\YS$. Thus
\begin{equation}\label{eq:W}
\mathit{Wr}_{uv}\le 1+L_u+\mathit{Rd}^{\star}_{uv}.
\end{equation}
Let $A_j\mapsto B_j$, $1\le j\le p$, be the transitions of $\Ex'$ in which
$u$ executes rule $\Read(v)$ and modifies $\cp uv.\YS$, with
$B_j\le A_{j+1}$. Let $3\le j\le p$. We apply \cref{lem:3:read} to the
transitions $A_{j-2}\mapsto B_{j-2}$, $A_{j-1}\mapsto B_{j-1}$ and
$A_j\mapsto B_j$. Node $v$ executes a local rule between $Z_{j-1}$ and $Y_j$,
where $Y_\ell\mapsto Z_\ell$ is the first transition between $A_{\ell-1}$ and
$A_\ell$ in which $v$ executes rule $\Write(u)$ and modifies
$\reg vu.\state$. These local rules are different for different values of
$j$. Thus $p-2\le L_v$, that is,
\begin{equation}\label{eq:Y}
\mathit{Rd}^{\star}_{uv}\le 2+L_v .
\end{equation}
The third point follows from \eqref{eq:W} and \eqref{eq:Y}.
\end{proof}

\begin{theorem}[Move complexity of $\TA$]\label{thm:moves}
Let $\A$ be an algorithm in the state model. Let $\Ex'$ be a finite prefix of
an execution of algorithm $\TA(\A)$ under the unfair distributed daemon, in
which rule $\ExecuteAlgo$ is executed $M$ times. Then $\Ex'$ contains
$O(\Delta^2M+\Delta n)$ moves. In particular, an execution of algorithm
$\TA(\A)$ in which rule $\ExecuteAlgo$ is executed finitely many times is
finite.
\end{theorem}

\begin{proof}
Let $L$ be the number of local rules executed in $\Ex'$. We prove that
$\Ex'$ contains at most
\begin{equation}\label{eq:moves}
(4\Delta+1)\,L+14m\ \le\ (4\Delta+1)\bigl(11(\Delta+1)M+10n\bigr)+14m
\end{equation}
moves. Since $m\le n\Delta/2$, this is $O(\Delta^2M+\Delta n)$.

Summing \cref{lem:counting}(\ref{c:L}) over all nodes, each $M_x$ is counted
$\deg(x)+1$ times:
\[
L\le 10n+11\sum_{x\in V}(\deg(x)+1)M_x\le 10n+11(\Delta+1)M .
\]
The number of moves of $\Ex'$ is
$L+\sum_{(u,v)}(\mathit{Rd}_{uv}+\mathit{Wr}_{uv})$, where the sum ranges
over the $2m$ ordered pairs of neighbors. By
\cref{lem:counting}(\ref{c:R},\ref{c:W}),
\begin{align*}
\sum_{(u,v)}(\mathit{Rd}_{uv}+\mathit{Wr}_{uv})
&\le 2m+2\sum_{(u,v)}\mathit{Wr}_{uv}
\le 2m+2\sum_{(u,v)}(L_u+L_v+3)\\
&= 14m+4\sum_{u\in V}\deg(u)L_u\le 14m+4\Delta L .
\end{align*}
This gives the first bound of \eqref{eq:moves}, and the second one follows
from the bound on $L$. The bound holds for every finite prefix, which gives
the last claim.
\end{proof}

\subsection{Terminal configurations and silent algorithms}

A configuration of algorithm $\TA(\A)$ is said to be \emph{quiescent} if no
rule $\Write$ and no rule $\Read$ is enabled in it.

\begin{lemmarep}\label{lem:quiescent}
Let $C$ be a quiescent configuration of algorithm $\TA(\A)$. For every edge
$(u,v)$, we have
\begin{itemize}
\item $\reg uv(C)=\bigl(\id_u,\state_u(C),\state_v(C),\VAR_u(C)\bigr)$;
\item $\cp vu(C)=\bigl(\id_u,\state_u(C),\state_v(C),\VAR_u(C)\bigr)$.
\end{itemize}
In particular, for every node $u$, $\CBS(u)(C)$ is true, and $\wait(u)(C)$ is
true if and only if $u$ has a neighbor $v$ such that
$\state_v(C)>\state_u(C)$, or $\state_v(C)=\state_u(C)$ and $\id_v<\id_u$.
\end{lemmarep}


\begin{proof}
Since no rule $\Write$ is enabled in $C$,
\[
\reg uv=\LV uv=\bigl(\id_u,\state_u,\cp uv.\YS,\VAR_u\bigr).
\]
Since no rule $\Read$ is enabled in $C$, $\cp vu=\reg uv$ and
$\cp uv=\reg vu$. The second field of the last equality gives
$\cp uv.\YS=\reg vu.\state=\state_v$. Thus
$\reg uv=(\id_u,\state_u,\state_v,\VAR_u)$, and $\cp vu$ has the same
fields. The third field of $\cp vu$ is $\cp vu.\MS=\state_v$, which gives
$\CBS(v)$. The last claim follows from the definition of $\wait(u)$.
\end{proof}

\begin{lemmarep}\label{lem:terminal}
A configuration $C$ of algorithm $\TA(\A)$ is terminal if and only if
\begin{itemize}
\item $C$ is quiescent;
\item $\state_u(C)=-1$ for every node $u$;
\item $\proj C$ is a terminal configuration of algorithm $\A$.
\end{itemize}
\end{lemmarep}


\begin{proof}
Assume that $C$ is terminal. It is quiescent, and we use
\cref{lem:quiescent}. No node has a state in $\{0,\ldots,4\}$, since rule
$\ModifyMyState$ would be enabled. No node is in state $7$, since rule
$\StartAgain$ would be enabled. Let $U$ be the set of nodes in state $5$ or
$6$. We assume that $U\neq\emptyset$. Let $u$ be the node of $U$ with the
largest state and, among those, the smallest identifier. Then $\wait(u)$ is
false. If $\isAct(u,\A)$ is false, rule $\ResetRule$ is enabled at $u$.
Otherwise, rule $\toWAIT$ or rule $\ExecuteAlgo$ is enabled at $u$. This
leads to a contradiction. So $U=\emptyset$ and all states are $-1$. Since
rule $\WantToAct$ is not enabled, $\isAct(u,\A)$ is false for every node
$u$. By \cref{lem:quiescent}, $\isAct(u,\A)$ is evaluated on the view of $u$
in $\proj C$. So no node is enabled in $\proj C$.

Conversely, in a quiescent configuration where all states are $-1$ and whose
projected configuration is terminal, the only local rule whose guard could
hold is $\WantToAct$, and $\isAct(u,\A)$ is false for every node $u$.
\end{proof}

We can now state the main result on $\TA$.

\begin{theorem}[Silent algorithms]\label{thm:t1-silent}
Let $\A$ be an algorithm in the state model with move complexity at most $F$.
Let $\Ex$ be an execution of algorithm $\TA(\A)$ in the strong-link-register
model under the unfair distributed daemon, with initial configuration $C_0$.
Let $M$ be the number of executions of rule $\ExecuteAlgo$ in $\Ex$\modif{.}
Then
\begin{enumerate}
\item {$M\le(n+1)F+n$;}
\item $\Ex$ is finite and contains $O(\Delta^2M+\Delta n)$ moves;
\item the last configuration $T$ of $\Ex$ is quiescent, $\state_u(T)=-1$ for
every node $u$, and $\proj T$ is a terminal configuration of algorithm $\A$;
\item {if $\state_u(C_0)\notin\{3,4,5,6\}$ for every node $u$, then
$M\le F$,} $\Ex$ contains $O(\Delta^2F+\Delta n)$ moves, and the
projected execution $\proj{\Ex}$ is an execution of algorithm $\A$.
\end{enumerate}
\end{theorem}

\begin{proof}
 {By \cref{cor:unfaithful}, at most $n$ executions of rule
$\ExecuteAlgo$ are not \faithful. The first point then follows from
\cref{cor:M}.} Since $M$ is finite, the
second point is \cref{thm:moves}. The last configuration of a finite
execution is terminal, and the third point is \cref{lem:terminal}.
 {Assume that $\state_u(C_0)\notin\{3,4,5,6\}$ for every node $u$. By
\cref{cor:unfaithful}(\ref{u:init}), all the executions of rule
$\ExecuteAlgo$ are \faithful. Then $M\le F$ by \cref{cor:M}, the bound on
the number of moves follows from the second point,} and
\cref{thm:simulation} applies to $\Ex$. The projected
execution $\proj{\Ex}$ is a sequence of transitions of algorithm $\A$ that
ends in $\proj T$, which is terminal. So it is an execution of algorithm
$\A$.
\end{proof}
In particular, algorithm $\TA(\A)$ is silent. If $\A$ is self-stabilizing for
a predicate $P$ on its terminal configurations, then $\proj T$ satisfies $P$.
By the first point, $k=0$ if no node is in a state of $\{3,4,5,6\}$ in
$C_0$.

\begin{remark}[On the factor  {$n+1$}]\label{rem:k}
An execution of rule $\ExecuteAlgo$ that is not \faithful{} is the trace, in the
simulated algorithm, of a corrupted initial configuration of the
transformer. It modifies the variables of $\A$ at one node in an arbitrary
way, and $\A$ may need up to $F$ further moves to recover. Since the daemon
is unfair, it can delay such an execution for a long time: the node concerned
is simply not chosen. This is why $F$ is multiplied by  {$n+1$} in the
worst case.  {More precisely, by \cref{cor:M}, $n$ can be replaced by
the number of executions of rule $\ExecuteAlgo$ that are not \faithful. By
\cref{cor:unfaithful}, this number is at most the number of nodes that start
in the middle of an attempt.} We do not know whether the factor
\modif{$n+1$} can be reduced for the
transformer $\TA$. Such executions cannot be avoided completely by a local
test: the initial configuration may be the one that immediately precedes an
execution of rule $\ExecuteAlgo$ in a regular attempt, with corrupted
copies.
\end{remark}

For an algorithm that is not silent, the same arguments give the following
statement.

\begin{theorem}[Arbitrary algorithms]\label{thm:t1-general}
Let $\A$ be an algorithm in the state model. Let $\Ex$ be an execution of
algorithm $\TA(\A)$ under the unfair distributed daemon. Then $\Ex$ contains
a configuration $S$ such that the projected execution of the part of $\Ex$
that starts in $S$ is an execution of algorithm $\A$ under the unfair
distributed daemon.
\end{theorem}

\begin{proof}
By \cref{cor:unfaithful}, at most $n$ executions of rule $\ExecuteAlgo$ are
not \faithful. Let $S$ be a configuration after the last one, and let $\Ex'$
be the part of $\Ex$ that starts in $S$. By \cref{thm:simulation},
$\proj{\Ex'}$ is a sequence of transitions of algorithm $\A$. It remains to
show that it is maximal. If $\Ex$ is finite, its last configuration $T$ is
terminal, $\proj{\Ex'}$ ends in $\proj T$, and $\proj T$ is terminal for
$\A$ by \cref{lem:terminal}. If $\Ex$ is infinite, then rule $\ExecuteAlgo$
is executed infinitely often by \cref{thm:moves}, and $\proj{\Ex'}$ is
infinite.
\end{proof}

\begin{remark}
\Cref{thm:t1-general} shows that $\TA(\A)$ eventually behaves like $\A$. For a
non-silent algorithm under the unfair daemon, this does not give a closed set
of legitimate configurations in general. A node that is ready to perform an
execution of rule $\ExecuteAlgo$ that is not \faithful{} may be ignored by the
daemon for an arbitrarily long time, while other nodes keep moving.
\end{remark}

\section{The transformer \texorpdfstring{\TB}{S2LR}: from the state model to
the link-register model}\label{sec:t2}

In the link-register model, a node cannot read its own registers. The rule
$\Write(a)$ of $\TA$ compares $\reg ua$ with $\LV ua$, so $\TA(\A)$ cannot be
used as it is. The transformer $\TB$ is a modification of $\TA$ that does not
need this comparison. It is the transformer $\mathcal{T}_2$ of the conference
version~\cite{CohenMP23}. We write $\ALRM=\TB(\A)$ for the transformed
algorithm.

Each node $u$ has an additional boolean array $\mathit{write}_u$, indexed by
its neighbors. The flag $\wflag ua$ means that $u$ must write in $\reg ua$.
\begin{itemize}
\item The guard of $\Write(a)$ is the flag $\wflag ua$. The rule writes
$\LV ua$ in $\reg ua$ and sets the flag to $\false$.
\item $\Read(a)$ sets $\wflag ua$ to $\true$: the acknowledgment sent to $a$
may have to be updated.
\item Each local rule sets all the flags of $u$ to $\true$: the new state of
$u$ must be sent to all its neighbors.
\item A new rule $\ToTrue$ sets all the flags of $u$ to $\true$. It is
enabled only when no other rule of $u$ is enabled. This rule makes the
algorithm self-stabilizing: a register may initially hold a corrupted value
while the corresponding flag is $\false$. Thanks to $\ToTrue$, each node
rewrites all its registers again and again.
\end{itemize}
The other variables, the predicates and the guards of the local rules are
those of $\TA$. \Cref{fig:t2} gives the formal definition, with the
presentation of the conference version. The register $\reg ua$ is written by
$u$ and read by $a$ only.

\begin{figure}[tp]
\figrule
\footnotesize
\noindent\textbf{$\ALRM$ - Local Variables}
\begin{tabbing}
\hspace{1.4cm}\=\hspace{0.55cm}\=\kill
$\id_{u}$ \> : \> the identifier of $u$\\
$\VAR_{u}$ \> : \> the local variables of $u$ in Algorithm $\A$\\
$\state_{u}$ \> $\in$ \> $\{-1, 0, \dots, 7\}$\\
$\mathit{write}_u$ \> : \> a boolean array of size $deg(u)$\\[5pt]
\hspace{2.05cm}\=\hspace{3.1cm}\=\kill
$\forall a \in N(u):$ \> $\cp ua.\id$ \> /* id of $a$ */\\
 \> $\cp ua.\YS$ \> /* state of $a$ */\\
 \> $\cp ua.\MS$ \> /* the $a$'s view of the state of $u$ */\\
 \> $\cp ua.\VAR$ \> /* the local variables of $a$ in Algorithm $\A$ */
\end{tabbing}

\noindent\textbf{$\ALRM$ - Predicates and functions}
\begin{tabbing}
$\LV ua\equiv(\id_{u},\state_{u},\cp ua.\YS,\VAR_{u})$\\[3pt]
$\RR au\equiv(\reg au.\id,\reg au.\state,\reg au.\YS,\reg au.\VAR)$\\[3pt]
$\CBS(u)\equiv\forall a\in N(u):\cp ua.\MS=\state_{u}$\\[3pt]
$\CR ua\equiv(\cp ua.\id,\cp ua.\YS,\cp ua.\MS,\cp ua.\VAR)$\\[3pt]
$\waitfull(u)\equiv\exists a\in N(u):(\cp ua.\YS>\state_{u})\ \vee$\\
\hspace{5.3cm}$(\cp ua.\YS=\state_{u}\wedge\cp ua.\id<\id_{u})$
\end{tabbing}

\noindent\textbf{$\ALRM$ - Rules for each node $u$}
\begin{tabbing}
\hspace{1.75cm}\=\kill
$\forall a\in N(u),$ \> \textbf{\Write(a)} \textbf{::} $\wflag ua\ \to\ \wflag ua:=\false$ ; $\reg ua:=\LV ua$\\[3pt]
 \> \textbf{\Read(a)} \textbf{::} $\CR ua\neq\RR au\ \to$\\
 \> \hspace{1.35cm}$\CR ua:=\RR au$ ; $\wflag ua:=\true$\\[3pt]
\hspace{2.85cm}\=\hspace{0.8cm}\=\kill
\textbf{\WantToAct} \textbf{::} \> $\CBS(u)\wedge\state_u=-1\wedge\isAct(u,\A)\ \to$\\
 \>\> $\state_{u}:=\state_{u}+1$ ; $\forall a\in N(u),\wflag ua:=\true$\\[3pt]
\textbf{\ModifyMyState} \textbf{::} \> $\CBS(u)\wedge 0\leq\state_u<5\ \to\ \state_{u}:=\state_{u}+1$ ; $\forall a\in N(u),\wflag ua:=\true$\\[3pt]
\textbf{\ExecuteAlgo} \textbf{::} \> $\CBS(u)\wedge\neg\waitfull(u)\wedge\state_u=6\wedge\isAct(u,\A)\ \rightarrow$\\
 \>\> $\Activate(u,\A)$ ; $\state_{u}:=\state_{u}+1$ ; $\forall a\in N(u),\wflag ua:=\true$\\[3pt]
\textbf{\toWAIT} \textbf{::} \> $\CBS(u)\wedge\isAct(u,\A)\wedge\neg\waitfull(u)\wedge\state_u=5$\\
 \>\> $\rightarrow\ \state_{u}:=\state_{u}+1$ ; $\forall a\in N(u),\wflag ua:=\true$\\[3pt]
\textbf{\StartAgain} \textbf{::} \> $\CBS(u)\wedge(\state_{u}=7)\ \rightarrow\ \state_{u}:=-1$ ; $\forall a\in N(u),\wflag ua:=\true$\\[3pt]
\textbf{\ResetRule} \textbf{::} \> $\CBS(u)\wedge\state_u\in\{5,6\}\wedge\neg\isAct(u,\A)\ \to$\\
 \>\> $\state_{u}:=-1$ ; $\forall a\in N(u),\wflag ua:=\true$\\[8pt]
\textbf{\ToTrue} \textbf{::} \> $(\forall a\in N(u):\neg\wflag ua)$\\
 \> $\wedge\ (\forall a\in N(u)^{\star}:\neg(\CR ua\neq\RR au))$\\
 \> $\wedge\ \neg(\CBS(u)\wedge\state_u=-1\wedge\isAct(u,\A))$\\
 \> $\wedge\ \neg(\CBS(u)\wedge 0\leq\state_u<5)^{\star}$\\
 \> $\wedge\ \neg(\CBS(u)\wedge\neg\waitfull(u)\wedge\state_u=6\wedge\isAct(u,\A))$\\
 \> $\wedge\ \neg(\CBS(u)\wedge\isAct(u,\A)\wedge\neg\waitfull(u)\wedge\state_u=5)$\\
 \> $\wedge\ \neg(\CBS(u)\wedge(\state_{u}=7))$\\
 \> $\wedge\ \neg(\CBS(u)\wedge\state_u\in\{5,6\}\wedge\neg\isAct(u,\A))$\\
 \> $\rightarrow\ \forall a\in N(u):\wflag ua:=\true$
\end{tabbing}
\smallskip
\noindent$^{\star}$ Not in the figure of~\cite{CohenMP23}: the quantifier of
the second line, and the fourth line, which is the negation of the guard of
$\ModifyMyState$ (see \cref{rem:totrue}).
\figrule
\caption{Transformer $\TB$ transforms an algorithm $\A$ into an algorithm
$\ALRM$ in the link-register model. This is the transformer $\mathcal{T}_2$
of~\cite{CohenMP23}, with the same presentation and the two additions marked
with a star in the guard of $\ToTrue$.}
\label{fig:t2}
\end{figure}

The rules respect the restriction of \cref{sec:registers}. $\Write(a)$ writes
one register and its guard is a local variable. The other rules write only
local variables. $\Read(a)$ reads one register, and the guard of $\ToTrue$
reads the registers written by the neighbors of $u$.

\begin{remark}[Guard of $\ToTrue$]\label{rem:totrue}
The guard of $\ToTrue$ in \cref{fig:t2} is the conjunction of
$\forall a\in N(u):\neg\wflag ua$, of the negation of the guard of
$\Read(a)$ for every neighbor $a$, and of the negations of the guards of the
six local rules. In other words, $\ToTrue$ is enabled if and only if no other
rule of $u$ is enabled. This is the definition given in the text of the
conference version~\cite{CohenMP23}. The figure of the conference version
differs from \cref{fig:t2} on the two points marked with a star. The
condition on $\Read(a)$ has no quantifier, and the negation of the guard of
$\ModifyMyState$ is missing.

The second point matters. Without it, $\ToTrue$ and $\ModifyMyState$ can be
enabled together at a node $u$. A weakly fair daemon only guarantees that $u$
is chosen. It can let $u$ alternate forever between $\ToTrue$ and its rules
$\Write$, and $u$ never changes its state. This happens with $\Coloring$ on
two neighbors $s$ and $t$ with $\id_s<\id_t$ and $c_s=c_t=1$, from the
configuration where all states are $-1$, all flags are $\false$ and all
registers and copies are up to date. Under the synchronous daemon, $s$
reaches state $0$ and $t$ acknowledges it. From this point on,
$\ModifyMyState$ is enabled at $s$, but $s$ can alternate between $\ToTrue$
and $\Write(t)$ forever. Its color never changes. With the guard of
\cref{fig:t2}, $s$ changes its color in every execution
(\cref{cor:t2-silent}). Both behaviors have been checked with a simulator.
\end{remark}

By definition of $\ToTrue$, every node has an enabled rule in every
configuration of $\TB(\A)$. Hence $\TB(\A)$ has no terminal configuration,
all its executions are infinite, and an unfair daemon could prevent any node
from ever moving. We therefore assume the weakly fair distributed daemon:
every node performs infinitely many moves. Since no node is ever neutralized,
a round is a minimal sequence of consecutive steps in which every node
performs at least one move. Under the weakly fair daemon, every round is
finite and an execution is an infinite sequence of rounds.

In a configuration that $\TB(\A)$ is expected to reach, the registers and the
copies hold the right values, all states are $-1$ and no rule of $\A$ is
enabled. The nodes then only execute $\ToTrue$ and rules $\Write$ that
rewrite the value already present in the register. We say that the algorithm
has converged when it has reached such a configuration. The next section
makes this precise.
\section{Analysis of \texorpdfstring{\TB}{S2LR}}\label{sec:t2-analysis}
\begin{toappendix}
\label{app:t2}
\end{toappendix}

We study an execution of an algorithm obtained by transformer $\TB$ from an
algorithm $\A$ of the state model. The daemon is the weakly fair distributed
daemon, and the initial configuration is arbitrary. We use the notation of
\cref{sec:notation}. All the proofs omitted of this section are given in \Cref{app:t2}.

\subsection{Relation with \TA}

Let $C$ be a configuration of algorithm $\TB(\A)$. We denote by $\pi(C)$ the
configuration of algorithm $\TA(\A)$ obtained by removing from $C$ the arrays
$\mathit{write}$. The projected configuration $\proj C$ is defined as in
\cref{sec:notation}.

A move of algorithm $\TB(\A)$ is said to be a \emph{stutter} if it is
\begin{itemize}
\item an execution of rule $\ToTrue$, or
\item an execution of rule $\Write(a)$ by a node $u$ in a configuration where
$\reg ua=\LV ua$.
\end{itemize}
A stutter modifies only the arrays $\mathit{write}$. The other moves are said
to be \emph{effective}.

\begin{lemma}\label{lem:projection}
Let $C\mapsto D$ be a transition of algorithm $\TB(\A)$. We have
\begin{itemize}
\item if all the moves of $C\mapsto D$ are stutters, then $\pi(D)=\pi(C)$;
\item otherwise, $\pi(C)\mapsto\pi(D)$ is a transition of algorithm
$\TA(\A)$, whose moves are the effective moves of $C\mapsto D$.
\end{itemize}
\end{lemma}

\begin{proof}
An effective move is of one of three kinds.
\begin{itemize}
\item It is an execution of rule $\Write(a)$ by $u$ with
$\reg ua(C)\neq\LV ua(C)$. Then the guard of rule $\Write(a)$ of $\TA(\A)$
holds in $\pi(C)$.
\item It is an execution of rule $\Read(a)$. This rule has the same guard in
$\TA(\A)$.
\item It is an execution of a local rule. This rule has the same guard in
$\TA(\A)$.
\end{itemize}
In the three cases, the rule has the same effect on the registers and on the
variables of $\TA(\A)$ in both algorithms. Stutters do not modify $\pi(C)$.
\end{proof}

Thus an execution of algorithm $\TB(\A)$ is an execution of algorithm
$\TA(\A)$ with additional moves that only rewrite registers and modify the
arrays $\mathit{write}$.

\begin{lemma}\label{lem:transfer}
\Cref{lem:basic,lem:2:read,lem:3:read,lem:0:local:state,lem:5:state,lem:2:local:rules,lem:3:local:rules,lem:4:local:rules,lem:handshake,lem:faithful,cor:unfaithful,thm:simulation,cor:M,lem:ten}
hold for the executions of algorithm $\TB(\A)$.
\end{lemma}

\begin{proof}
\Cref{lem:basic} holds for algorithm $\TB(\A)$: this is read on
\cref{fig:t2}, since the arrays $\mathit{write}$ appear in none of its
points. In particular, if $u$ executes rule $\Write(v)$ in a transition
$C\mapsto D$, then $\reg uv(D)=\LV uv(C)$, whether this move is a stutter or
not. The proofs of the other statements use only \cref{lem:basic} and the
guards of rule $\Read$ and of the local rules (\cref{rem:scope}). These
guards are the same in both algorithms.
\end{proof}

In particular, at most one execution of rule $\ExecuteAlgo$ per node is not
\faithful.

\begin{corollary}\label{cor:effective}
Let $\Ex'$ be a finite prefix of an execution of algorithm $\TB(\A)$ in
which rule $\ExecuteAlgo$ is executed $M$ times. Then $\Ex'$ contains
$O(\Delta^{2}M+\Delta n)$ effective moves.
\end{corollary}

\begin{proof}
By \cref{lem:projection}, the effective moves of $\Ex'$ are the moves of a
finite sequence of transitions of algorithm $\TA(\A)$, which contains the
same executions of rule $\ExecuteAlgo$. The bound is \eqref{eq:moves},
established in the proof of \cref{thm:moves}.
\end{proof}

The total number of moves of algorithm $\TB(\A)$ cannot be bounded in this
way: a round may contain arbitrarily many stutters of a node that the daemon
chooses often. This is why the complexity of $\TB(\A)$ is measured in rounds.

\subsection{Properties on the arrays \texorpdfstring{$\mathit{write}$}{write}}

For an edge $(u,a)$, let $P(u,a)$ be the predicate
\(
P(u,a)\ \equiv\ \wflag ua\ \vee\ \reg ua=\LV ua .
\)

\begin{lemmarep}\label{lem:flags}
Let $u$ be a node. Let $\Ex$ be an execution of algorithm $\TB(\A)$ which
contains a transition $C\mapsto D$. We have
\begin{enumerate}
\item\label{fl:closed} for every $a\in N(u)$, if $P(u,a)(C)$ is true, then
$P(u,a)(D)$ is true;
\item\label{fl:est} if $u$ executes rule $\ToTrue$ or a local rule in
$C\mapsto D$, then $P(u,a)(D)$ is true for every $a\in N(u)$;
\item\label{fl:totrue} if $u$ executes rule $\ToTrue$ in $C\mapsto D$ and
$P(u,a)(C)$ is true for every $a\in N(u)$, then
\begin{itemize}
\item $\reg ua(C)=\LV ua(C)$ and $\cp ua(C)=\reg au(C)$ for every
$a\in N(u)$;
\item no local rule is enabled at $u$ in $C$.
\end{itemize}
\end{enumerate}
\end{lemmarep}

\begin{proof}
Only the moves of $u$ modify $\wflag ua$, $\reg ua$ and $\LV ua$. We focus on
transition $C\mapsto D$.
\begin{itemize}
\item If $u$ executes rule $\Write(a)$, then $\reg ua(D)=\LV ua(C)=\LV ua(D)$.
\item If $u$ executes rule $\Read(a)$, a local rule or rule $\ToTrue$, then
$\wflag ua(D)=\true$.
\item If $u$ executes rule $\Write(b)$ or rule $\Read(b)$ with $b\neq a$, or
if $u$ executes no rule, then $\wflag ua$, $\reg ua$ and $\LV ua$ are not
modified.
\end{itemize}
This proves the first two points. By definition of the guard of rule
$\ToTrue$, $\wflag ua(C)=\false$ for every $a\in N(u)$, no rule $\Read$ is
enabled at $u$ in $C$, and no local rule is enabled at $u$ in $C$. Since
$P(u,a)(C)$ is true and $\wflag ua(C)=\false$, we have
$\reg ua(C)=\LV ua(C)$. This proves the third point.
\end{proof}

\subsection{Properties on windows}

Let $\Ex$ be an execution of algorithm $\TB(\A)$. A \emph{window} of $\Ex$ is
a sequence $J$ of consecutive rounds of $\Ex$ such that no node executes rule
$\ExecuteAlgo$ during $J$. The rounds of $J$ are numbered from $1$ to $q$,
where $q$ may be infinite. We use the following notation.
\begin{itemize}
\item $J_0$ is the first configuration of $J$. For $1\le r\le q$, $J_r$ is
the last configuration of the $r$th round of $J$. Thus the $r$th round of $J$
consists of the transitions between $J_{r-1}$ and $J_r$.
\item A configuration $C$ of $\Ex$ is a \emph{configuration of $J$} if
$J_0\le C$, and $C\le J_q$ when $q$ is finite.
\item $\Act$ is the set of the nodes that are enabled in the projected
configuration $\proj{J_0}$ for algorithm $\A$. By
\cref{lem:basic}(\ref{b:var}), $\proj C=\proj{J_0}$ for every configuration
$C$ of $J$.
\item $B=59\Delta+11$. This value is such that a node performs at most $B-1$
moves in a window without executing rule $\ToTrue$ (\cref{lem:win-count}).
\end{itemize}

{\Cref{fig:window} illustrates these definitions.
\Cref{fig:window-thresholds} shows the properties that this subsection
establishes in a window with at least $48B$ rounds.}

\begin{figure}[ht]
\centering
\begin{tikzpicture}[x=1cm,y=1cm,font=\small]
  \foreach \a/\b in {0/1.7,2.05/3.05,3.4/5.2} {
    \draw[fill=blue!8] (\a,3.6) rectangle (\b,4.2);
  }
  \foreach \a in {1.7,3.05,5.2,13.0} {
    \fill[black!75] (\a,3.6) rectangle (\a+0.35,4.2);
  }
  \draw[thick,fill=blue!8] (5.55,3.6) rectangle (13.0,4.2);
  \node at (9.275,3.9) {window $J$};
  \draw[gray!60,dashed] (5.55,3.6) -- (0,2.3);
  \draw[gray!60,dashed] (13.0,3.6) -- (13.35,2.3);
  \foreach \a/\b/\t in {0/2.1/{round $1$},2.1/4.2/{round $2$},4.2/5.9/{$\cdots$},
      5.9/8.0/{round $r$},8.0/11.25/{$\cdots$},11.25/13.35/{round $q$}} {
    \draw[fill=blue!8] (\a,1.7) rectangle (\b,2.3);
    \node at ({(\a+\b)/2},2.0) {\t};
  }
  \foreach \x/\t in {0/{J_0},2.1/{J_1},4.2/{J_2},5.9/{J_{r-1}},8.0/{J_r},
      11.25/{J_{q-1}},13.35/{J_q}} {
    \draw (\x,1.7) -- (\x,1.55);
    \node[anchor=north] at (\x,1.55) {$\t$};
  }
  \fill[black!75] (0,0.2) rectangle (0.35,0.55);
  \node[anchor=west] at (0.5,0.375) {round that contains an execution of $\ExecuteAlgo$};
  \draw[fill=blue!8] (8.6,0.2) rectangle (9.3,0.55);
  \node[anchor=west] at (9.45,0.375) {window};
\end{tikzpicture}
\caption{{The rounds of an execution of $\TB(\A)$. A window is a
sequence of consecutive rounds in which no node executes rule $\ExecuteAlgo$.
If rule $\ExecuteAlgo$ is executed $M$ times, at most $M$ rounds contain an
execution of this rule. Bottom: a window $J$ with $q$ rounds. The
configuration $J_0$ is the first configuration of $J$, and $J_r$ is the last
configuration of its $r$th round.}}
\label{fig:window}
\end{figure}

\begin{figure}[ht]
\centering
\begin{tikzpicture}[x=1cm,y=1cm,font=\small]
  \draw[fill=blue!8] (0,3.0) rectangle (13.35,3.5);
  \node at (7.0,3.25) {$\cdots$};
  \foreach \x/\t in {0/{J_0},1.9/{J_B},3.8/{J_{2B}},5.7/{J_{3B}},8.3/{J_{33B}},
      10.4/{J_{37B}},13.35/{J_{48B}}} {
    \draw (\x,3.0) -- (\x,3.5);
    \node[anchor=south] at (\x,3.55) {$\t$};
  }
  \foreach \x/\y/\t in {
      8.3/1.5/{all the states are $-1$},
      5.7/0.8/{the fields $\id$ and $\VAR$ of the copies are correct},
      3.8/0.1/{the fields $\id$ and $\VAR$ of the registers are correct},
      1.9/-0.6/{$P(u,a)$ holds for every edge $(u,a)$}} {
    \draw[gray!60,dashed] (\x,3.0) -- (\x,\y+0.5);
    \draw[fill=black!6] (\x,\y) rectangle (13.35,\y+0.5);
    \node[anchor=west] at (\x+0.1,\y+0.25) {\t};
  }
  \draw[gray!60,dashed] (10.4,3.0) -- (10.4,2.7);
  \draw[thick,fill=blue!8] (10.4,2.2) rectangle (13.35,2.7);
  \node[anchor=west] at (10.5,2.45) {$\pi(C)$ is terminal};
\end{tikzpicture}
\caption{{A window $J$ with at least $48B$ rounds, where
$B=59\Delta+11$. Each bar starts at the
configuration from which the property holds, and the property then holds in
every configuration of $J$. The first three properties are those of
\cref{lem:cleaning}. The last two are established in the proof of
\cref{lem:window}, where $\pi(C)$ is terminal for algorithm $\TA(\A)$.}}
\label{fig:window-thresholds}
\end{figure}
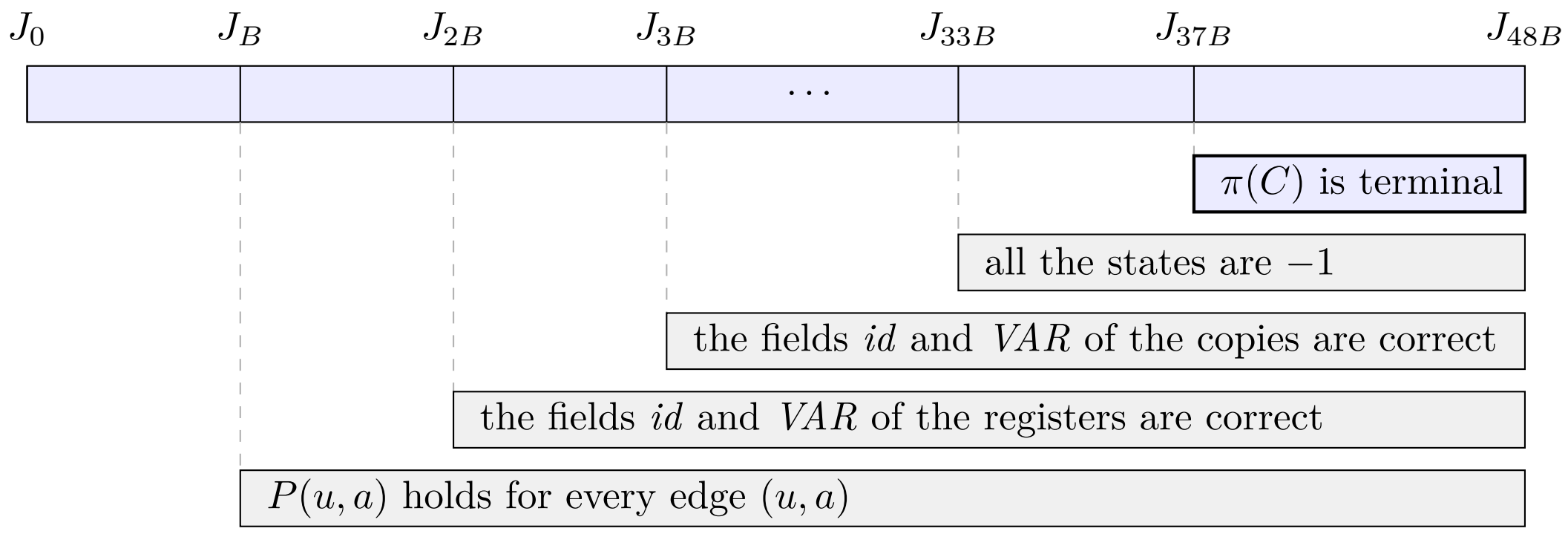
\begin{lemmarep}\label{lem:win-count}
Let $(u,v)$ be an edge. Let $\Ex$ be an execution of algorithm $\TB(\A)$
which contains a window $J$. We have
\begin{enumerate}
\item\label{w:L} $u$ executes at most $10$ local rules during $J$;
\item\label{w:R} $u$ executes rule $\Read(v)$ at most $24$ times during $J$;
\item\label{w:W} if $S$ and $T$ are two configurations of $J$ with $S\le T$
such that $u$ does not execute rule $\ToTrue$ between $S$ and $T$, then,
between $S$ and $T$, node $u$ executes rule $\Write(v)$ at most $35$ times
and it performs at most $59\deg(u)+10\le B-1$ moves.
\end{enumerate}
\end{lemmarep}

\begin{proof}
The first point is \cref{lem:ten}, which applies by \cref{lem:transfer}.

We prove the second point. We count the moves performed during $J$. We
denote by
\begin{itemize}
\item $L_u$ the number of local rules executed by $u$;
\item $\mathit{Rd}_{uv}$ the number of executions of rule $\Read(v)$ by $u$;
\item $\mathit{Wr}^{\star}_{vu}$ the number of executions of rule $\Write(u)$
by $v$ that modify $\reg vu$;
\item $\mathit{Rd}^{\star}_{vu}$ the number of executions of rule $\Read(u)$
by $v$ that modify $\cp vu.\YS$.
\end{itemize}
By \cref{lem:2:read}, $\reg vu$ is modified between two consecutive
executions of rule $\Read(v)$ by $u$. Thus
$\mathit{Rd}_{uv}\le 1+\mathit{Wr}^{\star}_{vu}$.

Let $Y\mapsto Z$ and $Y'\mapsto Z'$ be two transitions with $Z\le Y'$ in
which $v$ executes rule $\Write(u)$ and modifies $\reg vu$, such that
$\reg vu$ remains unchanged between $Z$ and $Y'$. Then
$\reg vu(Y')=\reg vu(Z)=\LV vu(Y)$ and
$\reg vu(Y')\neq\reg vu(Z')=\LV vu(Y')$. Thus $\LV vu(Y)\neq\LV vu(Y')$. So
$\LV vu$ is modified between $Z$ and $Y'$, by a local rule of $v$ or by an
execution of rule $\Read(u)$ by $v$ that modifies $\cp vu.\YS$. Thus
$\mathit{Wr}^{\star}_{vu}\le 1+L_v+\mathit{Rd}^{\star}_{vu}$.

Finally $\mathit{Rd}^{\star}_{vu}\le 2+L_u$: this is \eqref{eq:Y} with $u$
and $v$ exchanged, and its proof uses only \cref{lem:3:read}. With the first
point, we obtain
$\mathit{Rd}^{\star}_{vu}\le 12$, $\mathit{Wr}^{\star}_{vu}\le 23$ and
$\mathit{Rd}_{uv}\le 24$.

We prove the third point. By definition of the guard of rule $\Write$, an
execution of rule $\Write(v)$ by $u$ requires $\wflag uv=\true$, and it sets
$\wflag uv$ to $\false$. Between two consecutive executions of rule
$\Write(v)$ by $u$, $\wflag uv$ is set to $\true$ by an execution of rule
$\Read(v)$, of a local rule or of rule $\ToTrue$ by $u$. Between $S$ and $T$,
node $u$ does not execute rule $\ToTrue$. Thus it executes rule $\Write(v)$
at most $1+24+10=35$ times, and it performs at most
$10+24\deg(u)+35\deg(u)$ moves.
\end{proof}

Rule $\ToTrue$ is enabled at a node only if no other rule is enabled at this
node. So, when a node $u$ executes rule $\ToTrue$, its flags are $\false$,
its copies are equal to the registers of its neighbors, and none of its local
rules is enabled. The next lemma shows that this happens at least once every
$B$ rounds in a window. All the bounds of this section in number of rounds
come from this fact.

\begin{lemma}\label{lem:totrue}
Let $u$ be a node. Let $\Ex$ be an execution of algorithm $\TB(\A)$ which
contains a window $J$. Then $u$ executes rule $\ToTrue$ at least once during
any $B$ consecutive rounds of $J$.
\end{lemma}

\begin{proof}
Every node is enabled in every configuration. So $u$ performs at least one
move in each round, and at least $B$ moves during $B$ consecutive rounds. By
\cref{lem:win-count}(\ref{w:W}), if $u$ does not execute rule $\ToTrue$
during these rounds, it performs at most $B-1$ moves.
\end{proof}

Recall that rule $\ToTrue$ is enabled at a node only if no other rule is
enabled at this node. By \cref{lem:totrue}, in a window, no rule other than
$\ToTrue$ remains enabled at a node during $B$ consecutive rounds.

\begin{lemmarep}\label{lem:cleaning}
Let $(u,a)$ be an edge. Let $\Ex$ be an execution of algorithm $\TB(\A)$
which contains a window $J$. We have
\begin{enumerate}
\item\label{cl:P} $P(u,a)(C)$ is true for every configuration $C$ of $J$
with $J_B\le C$;
\item\label{cl:reg} $\reg ua.\id(C)=\id_u$ and $\reg ua.\VAR(C)=\VAR_u(C)$
for every configuration $C$ of $J$ with $J_{2B}\le C$;
\item\label{cl:copy} $\cp ua.\id(C)=\id_a$ and $\cp ua.\VAR(C)=\VAR_a(C)$
for every configuration $C$ of $J$ with $J_{3B}\le C$.
\end{enumerate}
In particular, for every configuration $C$ of $J$ with $J_{3B}\le C$,
$\isAct(u,\A)(C)$ is true if and only if $u\in\Act$.
\end{lemmarep}

\begin{proof}
We prove the first point. By \cref{lem:totrue}, $u$ executes rule $\ToTrue$
between $J_0$ and $J_B$. The first point follows from
\cref{lem:flags}(\ref{fl:closed},\ref{fl:est}).

We prove the second point. By \cref{lem:totrue}, $u$ executes rule $\ToTrue$
in a transition $S\mapsto S'$ between $J_B$ and $J_{2B}$. By the first point
and \cref{lem:flags}(\ref{fl:totrue}), $\reg ua(S)=\LV ua(S)$. Thus
$\reg ua.\id(S)=\id_u$ and $\reg ua.\VAR(S)=\VAR_u(S)$. After $S$, the
register $\reg ua$ is modified only by executions of rule $\Write(a)$ by
$u$, which write $\id_u$ and $\VAR_u$ in these fields. The local variable
$\VAR_u$ remains unchanged during $J$.

We prove the third point. By \cref{lem:totrue}, $u$ executes rule $\ToTrue$
in a transition $S\mapsto S'$ between $J_{2B}$ and $J_{3B}$. By
\cref{lem:flags}(\ref{fl:totrue}), $\cp ua(S)=\reg au(S)$. By the second
point, $\cp ua.\id(S)=\id_a$ and $\cp ua.\VAR(S)=\VAR_a(S)$. After $S$,
$\cp ua$ is modified only by executions of rule $\Read(a)$ by $u$, which copy
a register whose fields $\id$ and $\VAR$ are $\id_a$ and $\VAR_a$.

By the third point, in every configuration $C$ of $J$ with $J_{3B}\le C$,
$\isAct(u,\A)$ is evaluated on the view of $u$ in $\proj C=\proj{J_0}$.
\end{proof}

The next lemma bounds the duration of a handshake in a window. In its
statement, no configuration is between $J_{r'}$ and $T$ if $J$ has fewer than
$r'$ rounds or if $T<J_{r'}$.

\begin{lemmarep}\label{lem:propagation}
Let $(x,v)$ be an edge. Let $\Ex$ be an execution of algorithm $\TB(\A)$
which contains a window $J$. Let $\Sigma$ be a set of states, let $r\ge B$ be
an integer and let $T$ be a configuration of $J$ such that
$\state_x(C)\in\Sigma$ for every configuration $C$ between $J_r$ and $T$. We
have
\begin{enumerate}
\item\label{pr:1} $\reg xv.\state(C)\in\Sigma$ for every configuration $C$
between $J_{r+B}$ and $T$;
\item\label{pr:2} $\cp vx.\YS(C)\in\Sigma$ for every configuration $C$
between $J_{r+2B}$ and $T$.
\end{enumerate}
If moreover $\Sigma=\{s\}$, we have
\begin{enumerate}[resume]
\item\label{pr:3} $\reg vx.\YS(C)=s$ for every configuration $C$ between
$J_{r+3B}$ and $T$;
\item\label{pr:4} $\cp xv.\MS(C)=s$ for every configuration $C$ between
$J_{r+4B}$ and $T$.
\end{enumerate}
\end{lemmarep}

\begin{proof}
In each point, we assume that there is a configuration between the given
configuration $J_{r'}$ and $T$. All the configurations considered are after
$J_B$. So \cref{lem:cleaning}(\ref{cl:P}) and
\cref{lem:flags}(\ref{fl:totrue}) apply to every execution of rule $\ToTrue$
below. These executions exist by \cref{lem:totrue}.

We prove the first point. Node $x$ executes rule $\ToTrue$ in a transition
$S_1\mapsto S'_1$ between $J_r$ and $J_{r+B}$. We have
$\reg xv(S_1)=\LV xv(S_1)$, so $\reg xv.\state(S_1)=\state_x(S_1)\in\Sigma$.
After $S_1$, an execution of rule $\Write(v)$ by $x$ in a transition
$C\mapsto C'$ with $C'\le T$ writes $\state_x(C)\in\Sigma$ in this field.
Thus $\reg xv.\state(C)\in\Sigma$ for every configuration $C$ between $S_1$
and $T$.

We prove the second point. Node $v$ executes rule $\ToTrue$ in a transition
$S_2\mapsto S'_2$ between $J_{r+B}$ and $J_{r+2B}$. We have
$\cp vx(S_2)=\reg xv(S_2)$, so
$\cp vx.\YS(S_2)=\reg xv.\state(S_2)\in\Sigma$ by the first point. After
$S_2$, an execution of rule $\Read(x)$ by $v$ in a transition $C\mapsto C'$
with $C'\le T$ copies $\reg xv.\state(C)\in\Sigma$ in this field.

We prove the third point. Node $v$ executes rule $\ToTrue$ in a transition
$S_3\mapsto S'_3$ between $J_{r+2B}$ and $J_{r+3B}$. We have
$\reg vx(S_3)=\LV vx(S_3)$. By the second point,
$\reg vx.\YS(S_3)=\cp vx.\YS(S_3)=s$. After $S_3$, an execution of rule
$\Write(x)$ by $v$ in a transition $C\mapsto C'$ with $C'\le T$ writes
$\cp vx.\YS(C)=s$ in this field.

We prove the fourth point. Node $x$ executes rule $\ToTrue$ in a transition
$S_4\mapsto S'_4$ between $J_{r+3B}$ and $J_{r+4B}$. We have
$\cp xv(S_4)=\reg vx(S_4)$, so $\cp xv.\MS(S_4)=\reg vx.\YS(S_4)=s$ by the
third point. After $S_4$, an execution of rule $\Read(v)$ by $x$ in a
transition $C\mapsto C'$ with $C'\le T$ copies $\reg vx.\YS(C)=s$ in this
field.
\end{proof}

\begin{lemmarep}\label{lem:free}
Let $x$ be a node. Let $\Ex$ be an execution of algorithm $\TB(\A)$ which
contains a window $J$ with at least $r+5B$ rounds, where $r\ge 3B$. Assume
that one of the following conditions holds:
\begin{itemize}
\item $\state_x(J_r)\in\{0,\ldots,4,7\}$;
\item $\state_x(J_r)=-1$ and $x\in\Act$;
\item $\state_x(J_r)\in\{5,6\}$ and $x\notin\Act$.
\end{itemize}
Then $x$ executes a local rule between $J_r$ and $J_{r+5B}$.
\end{lemmarep}

\begin{proof}
We prove this lemma by contradiction. We assume that $x$ executes no local
rule between $J_r$ and $J_{r+5B}$. Then $\state_x(C)=\state_x(J_r)$ for every
configuration $C$ between $J_r$ and $J_{r+5B}$.

Let $C$ be a configuration between $J_{r+4B}$ and $J_{r+5B}$. By
\cref{lem:propagation}(\ref{pr:4}), applied to every neighbor of $x$,
$\CBS(x)(C)$ is true. By \cref{lem:cleaning}, $\isAct(x,\A)(C)$ is true if
and only if $x\in\Act$. Thus the guard of rule $\ModifyMyState$,
$\StartAgain$, $\WantToAct$ or $\ResetRule$, according to the case, holds at
$x$ in $C$. So rule $\ToTrue$ is not enabled at $x$ in $C$.

Thus $x$ does not execute rule $\ToTrue$ between $J_{r+4B}$ and $J_{r+5B}$.
This leads to a contradiction with \cref{lem:totrue}.
\end{proof}

\begin{lemmarep}\label{lem:window}
Let $\Ex$ be an execution of algorithm $\TB(\A)$ which contains a window $J$
with at least $48B$ rounds. Then, for every configuration $C$ of $J$ with
$J_{37B}\le C$, $\pi(C)$ is a terminal configuration of algorithm $\TA(\A)$.
\end{lemmarep}

\begin{proof}
In this proof, all the configurations are configurations of $J$.

\emph{Step 1: the nodes that are not in $\Act$.} Let $x\notin\Act$. By
\cref{lem:cleaning}, $\isAct(x,\A)(C)$ is false for every configuration $C$
with $J_{3B}\le C$. After $J_{3B}$, node $x$ can only execute rules
$\ModifyMyState$, $\ResetRule$ and $\StartAgain$. In state $-1$, it executes
no local rule. From any other state, it reaches state $-1$ after at most six
local rules: at most five executions of rule $\ModifyMyState$ and one of
rule $\ResetRule$, or one execution of rule $\StartAgain$. By
\cref{lem:free}, as long as its state is not $-1$, it executes a local rule
in every period of $5B$ rounds. Thus
\begin{equation}\label{eq:nonact}
\state_x(C)=-1\qquad\text{for every configuration $C$ with } J_{33B}\le C.
\end{equation}

\emph{Step 2: the nodes of $\Act$.} Let $x\in\Act$. After $J_{3B}$,
$\isAct(x,\A)$ is true, so $x$ does not execute rule $\ResetRule$. It does
not execute rule $\ExecuteAlgo$, since $J$ is a window. From a state that is
not $5$ or $6$, it reaches state $5$ after at most seven local rules:
$\StartAgain$, $\WantToAct$ and five executions of rule $\ModifyMyState$. By
\cref{lem:free},
\begin{equation}\label{eq:act}
\state_x(C)\in\{5,6\}\qquad\text{for every configuration $C$ with } J_{38B}\le C,
\end{equation}
and after $J_{38B}$ the state of $x$ changes at most once, from $5$ to $6$,
by an execution of rule $\toWAIT$.

\emph{Step 3: the case $\Act=\emptyset$.} By \eqref{eq:nonact}, all states
are $-1$ in every configuration $C$ with $J_{33B}\le C$. Let $C$ be a
configuration with $J_{37B}\le C$ and let $(u,v)$ be an edge. By
\cref{lem:propagation} with $\Sigma=\{-1\}$ and $r=33B$, the fields
$\reg uv.\state$, $\reg uv.\YS$, $\cp vu.\YS$ and $\cp vu.\MS$ are equal to
$-1$ in $C$. With \cref{lem:cleaning}, we obtain
\[
\reg uv(C)=\cp vu(C)=\bigl(\id_u,-1,-1,\VAR_u(C)\bigr).
\]
Thus $\pi(C)$ is quiescent and all states are $-1$. Since $\Act=\emptyset$,
$\proj C$ is terminal for $\A$. By \cref{lem:terminal}, $\pi(C)$ is terminal
for $\TA(\A)$. This proves the lemma in this case.

Now, we assume that $\Act\neq\emptyset$, and we derive a contradiction.

\emph{Step 4: copies of the states.} By \eqref{eq:nonact}, \eqref{eq:act}
and \cref{lem:propagation}(\ref{pr:2}), for every configuration $C$ with
$J_{40B}\le C$ and every edge $(x,v)$,
\begin{equation}\label{eq:copies}
\cp xv.\YS(C)=-1\ \text{ if } v\notin\Act,\qquad
\cp xv.\YS(C)\in\{5,6\}\ \text{ if } v\in\Act .
\end{equation}

\emph{Step 5: no execution of rule $\toWAIT$ after $J_{43B}$.} Let
$x\in\Act$. We split the proof into two cases.

Case (a). Node $x$ has a neighbor $y\in\Act$ with $\id_y<\id_x$. Let $C$ be
a configuration with $J_{40B}\le C$ such that $\state_x(C)=5$. By
\eqref{eq:copies} and \cref{lem:cleaning}(\ref{cl:copy}),
$\cp xy.\YS(C)\ge 5$ and $\cp xy.\id(C)=\id_y<\id_x$. Thus $\wait(x)(C)$ is
true. So $x$ does not execute rule $\toWAIT$ after $J_{40B}$, and its state
remains unchanged after $J_{40B}$.

Case (b). Every neighbor of $x$ in $\Act$ has a larger identifier than $x$.
Each such neighbor $y$ is in case (a). Its state remains unchanged after
$J_{40B}$. By \cref{lem:propagation}(\ref{pr:2}),
$\cp xy.\YS(C)=\state_y(J_{40B})$ for every configuration $C$ with
$J_{42B}\le C$.
\begin{itemize}
\item Assume that $x$ has a neighbor $y\in\Act$ such that
$\state_y(J_{40B})=6$. Then, for every configuration $C$ with
$J_{42B}\le C$, we have $\cp xy.\YS(C)=6$. So $\wait(x)(C)$ is true whenever
$\state_x(C)=5$, and $x$ does not execute rule $\toWAIT$ after $J_{42B}$.
\item Otherwise, all the neighbors of $x$ in $\Act$ are in state $5$ after
$J_{40B}$. Assume that $\state_x(J_{42B})=5$ and that $x$ does not execute
rule $\toWAIT$ between $J_{42B}$ and $J_{43B}$. By \eqref{eq:act},
$\state_x(C)=5$ for every configuration $C$ between $J_{38B}$ and $J_{43B}$.
Let $C$ be a configuration between $J_{42B}$ and $J_{43B}$. By
\cref{lem:propagation}(\ref{pr:4}), $\CBS(x)(C)$ is true. By
\cref{lem:cleaning}, $\isAct(x,\A)(C)$ is true. By \eqref{eq:copies}, the
copies $\cp xv.\YS(C)$ are equal to $-1$ or to $5$, and those equal to $5$
come with an identifier $\cp xv.\id(C)=\id_v>\id_x$. Thus $\wait(x)(C)$ is
false. So rule $\toWAIT$ is enabled at $x$ in every configuration between
$J_{42B}$ and $J_{43B}$, and rule $\ToTrue$ is not. This leads to a
contradiction with \cref{lem:totrue}. Thus $\state_x(J_{43B})=6$.
\end{itemize}
In all cases, $x$ does not execute rule $\toWAIT$ after $J_{43B}$. With
\eqref{eq:nonact} and \eqref{eq:act}, all states remain unchanged after
$J_{43B}$.

\emph{Step 6: a node is in state $6$.} Let $x_0$ be the node of $\Act$ with
the smallest identifier. It is in case (b). By Step~5, either one of its
neighbors is in state $6$ after $J_{40B}$, or $\state_{x_0}(J_{43B})=6$. Thus
the set $U$ of the nodes that are in state $6$ in $J_{43B}$ is not empty.

\emph{Step 7: contradiction.} Let $x^\ast$ be the node of $U$ with the
smallest identifier. All states remain unchanged after $J_{43B}$. Let $C$ be
a configuration between $J_{47B}$ and $J_{48B}$. By \cref{lem:propagation},
applied to every edge with $r=43B$, the copies of the states are equal to the
states in $C$, and $\CBS(u)(C)$ is true for every node $u$. We have
\begin{itemize}
\item $\CBS(x^\ast)(C)$ is true and $\state_{x^\ast}(C)=6$;
\item $\isAct(x^\ast,\A)(C)$ is true, since $x^\ast\in\Act$ by
\eqref{eq:nonact};
\item $\wait(x^\ast)(C)$ is false: the neighbors of $x^\ast$ are in state
$-1$, $5$ or $6$, those in state $6$ have a larger identifier than $x^\ast$,
and the copies of the states and of the identifiers are equal to their
current values.
\end{itemize}
Thus rule $\ExecuteAlgo$ is enabled at $x^\ast$ in every configuration
between $J_{47B}$ and $J_{48B}$, and rule $\ToTrue$ is not. This leads to a
contradiction with \cref{lem:totrue}.
\end{proof}

\subsection{Round complexity}

\begin{theorem}[Round complexity of $\TB$]\label{thm:rounds}
Let $\A$ be an algorithm in the state model. Let $\Ex$ be an execution of
algorithm $\TB(\A)$ in the link-register model under the weakly fair
distributed daemon, in which rule $\ExecuteAlgo$ is executed a finite number
$M$ of times. Then $\Ex$ reaches, after $O(\Delta(M+1))$ rounds, a
configuration $T$ such that
\begin{itemize}
\item $\pi(T)$ is a terminal configuration of algorithm $\TA(\A)$;
\item $\proj T$ is a terminal configuration of algorithm $\A$;
\item $\pi(C)=\pi(T)$ for every configuration $C$ with $T\le C$: all the
moves after $T$ are stutters.
\end{itemize}
\end{theorem}

\begin{proof}
Recall that $B=59\Delta+11$. We prove that $48B(M+1)$ rounds are enough. At
most $M$ rounds contain an execution of rule $\ExecuteAlgo$. Let
$\rho_1<\cdots<\rho_p$ be the indices of these rounds, with $p\le M$, and let
$\rho_0=0$. Since $\Ex$ has infinitely many rounds, there is a smallest index
$j\in\{0,\ldots,p\}$ such that the rounds $\rho_j+1,\ldots,\rho_j+48B$
contain no execution of rule $\ExecuteAlgo$. For $i<j$ we have
$\rho_{i+1}\le\rho_i+48B$. Thus $\rho_j\le 48Bj\le 48BM$.

The rounds $\rho_j+1,\ldots,\rho_j+48B$ form a window. Let $T$ be the last
configuration of round $\rho_j+37B$. By \cref{lem:window}, $\pi(T)$ is
terminal for $\TA(\A)$. Configuration $T$ is reached after at most
$48BM+37B\le 48B(M+1)$ rounds. By \cref{lem:terminal}, $\proj T$ is terminal
for $\A$. By \cref{lem:projection}, a transition from a configuration $C$
such that $\pi(C)$ is terminal contains only stutters, and it does not
modify $\pi(C)$. This proves the last point.
\end{proof}

\begin{corollary}\label{cor:t2-silent}
Let $\A$ be an algorithm in the state model with move complexity at most $F$.
Let $\Ex$ be an execution of algorithm $\TB(\A)$ under the weakly fair
distributed daemon{, and let $M$ be the number of executions of rule
$\ExecuteAlgo$ in $\Ex$.} Then
\begin{enumerate}
\item {$M\le(n+1)F+n$;}
\item $\Ex$ reaches, after {$O(\Delta n(F+1))$} rounds, a configuration $T$
such that $\proj T$ is a terminal configuration of algorithm $\A$ and
$\pi(C)=\pi(T)$ for every configuration $C$ with $T\le C$\modif{.}
\end{enumerate}
\end{corollary}

\begin{proof}
By \cref{lem:transfer}, \cref{cor:unfaithful,cor:M} hold for $\TB(\A)$.
{Thus at most $n$ executions of rule $\ExecuteAlgo$ are not \faithful,
and $M\le(n+1)F+n$. In particular, $M$ is finite and $M+1\le(n+1)(F+1)$,}
and \cref{thm:rounds} applies. Its proof shows that
{$48B(M+1)$ rounds are enough, and
$48B(M+1)\le 48(59\Delta+11)(n+1)(F+1)$.}
\end{proof}

After $T$, the registers, the copies, the states and the variables of $\A$
never change. If $\A$ is self-stabilizing for a predicate $P$ on its terminal
configurations, then $\proj C$ satisfies $P$ for every configuration $C$ with
$T\le C$.

\begin{remark}[Arbitrary algorithms]
The analog of \cref{thm:t1-general} holds for $\TB$. Let $S$ be a
configuration after the last execution of rule $\ExecuteAlgo$ that is not
\faithful, and let $\Ex'$ be the part of $\Ex$ that starts in $S$. If rule
$\ExecuteAlgo$ is executed infinitely often, \cref{thm:simulation} shows that
the projected execution $\proj{\Ex'}$ is an infinite execution of $\A$.
Otherwise, \cref{thm:rounds} shows that $\proj{\Ex'}$ is a finite sequence of
transitions of $\A$ that ends in a terminal configuration.
\end{remark}

\section{The transformer \texorpdfstring{\TC}{S2LR-C}: a cyclic transformer
for the link-register model}\label{sec:t3}

The transformer $\TB$ uses rule $\ToTrue$, whose guard checks that no other
rule is enabled at the node. In particular, this guard depends on all the
registers written by its neighbors. We now present a third transformer,
called $\TC$ (state to link-register, with a cyclic schedule), which does not
need this test. It works in the link-register model with strict read/write
atomicity: a move reads at most one communication register, writes at most
one communication register, and never both reads and writes a communication
register. Local computations and updates of local variables may be performed
in the same move. We write $\ACLRM=\TC(\A)$.

\subsection{Principle of the transformation}

The variables, registers, predicates and local rules of $\TC$ are those of
$\TA$ in \cref{fig:t1}. The difference is in the way the rules are scheduled
at a node. Each node $u$ writes in all its outgoing registers, reads all its
incoming registers and then checks whether one of its six local rules is
enabled. It repeats these operations forever. The writes and reads are
performed even when they do not change the content of a register or of a
copy. A local rule is executed only when its guard holds. If none of the local
rules is enabled, the node starts a new cycle without modifying its state or
its variables of algorithm $\A$.

For each node $u$ of degree $d_u=\deg(u)$, let
$v_1,\ldots,v_{d_u}$ be a fixed local enumeration of its neighbors. This
enumeration is given by the local port numbers and is not modified by
transient faults. In particular, it does not require knowledge of the
identifiers of the neighbors. The only additional variable of $u$ is a
phase counter $\phase_u$. Let
\[
  \ell_u=\max\{2,2d_u+1\}.
\]
The variable $\phase_u$ takes values in $\{0,\ldots,\ell_u-1\}$. Its initial
value is arbitrary in this domain. We denote by $\mathit{next}_u(p)$ the value
$(p+1)\bmod\ell_u$. The second phase for an isolated node is an idle phase;
it ensures that even in this case the phase counter changes at each move.

\subsection{Formal definition}

\Cref{fig:t3} defines the transformer. In the figure, $\mathcal G_j$ and
$\mathcal I_j$ denote respectively the guard and the action of the $j$th
local rule of $\TA$, for $1\le j\le 6$, in the order
$\WantToAct$, $\ModifyMyState$, $\ExecuteAlgo$, $\toWAIT$,
$\StartAgain$, $\ResetRule$. These six guards are mutually exclusive and
depend only on local variables. The notation $\mathcal I_j$ includes the
update of $\state_u$ and, for rule $\ExecuteAlgo$, the update of $\VAR_u$.
We keep the name of the corresponding local rule when it is executed in the
local phase. If the source algorithm has several enabled rules, the choice
inside $\Activate(u,\A)$ is still arbitrary.

\begin{figure}[htbp]
\figrule
\small
\noindent\textbf{$\ACLRM$ - Local Variables and Registers}

\noindent The local variables and registers of $\ASLRM$ in \cref{fig:t1},
with the additional local variable
$\phase_u\in\{0,\ldots,\ell_u-1\}$.
The register $\reg ua$ is written by $u$ and read by $a$ only.

\smallskip
\noindent\textbf{$\ACLRM$ - Predicates and functions}

\noindent The predicates and functions of \cref{fig:t1}, with
$\ell_u=\max\{2,2d_u+1\}$ and
$\mathit{next}_u(p)=(p+1)\bmod\ell_u$.

\smallskip
\noindent\textbf{$\ACLRM$ - Rules for each node $u$}
\begin{tabbing}
\hspace{2.65cm}\=\hspace{0.55cm}\=\kill
$\forall i\in\{1,\ldots,d_u\}:$\\[2pt]
\textbf{\Write($v_i$)} \textbf{::} \> $\phase_u=i-1\ \to$\\
\>\> $\reg u{v_i}:=\LV u{v_i}$ ; $\phase_u:=\mathit{next}_u(\phase_u)$\\[3pt]
\textbf{\Read($v_i$)} \textbf{::} \> $\phase_u=d_u+i-1\ \to$\\
\>\> $\CR u{v_i}:=\RR {v_i}u$ ; $\phase_u:=\mathit{next}_u(\phase_u)$\\[5pt]
$\forall j\in\{1,\ldots,6\}:$\\[2pt]
\textbf{\CycleLocal($j$)} \textbf{::} \> $\phase_u=2d_u\wedge\mathcal G_j\ \to$\\
\>\> $\mathcal I_j$ ; $\phase_u:=\mathit{next}_u(\phase_u)$\\[3pt]
\textbf{\CycleIdle} \textbf{::} \> $\phase_u=2d_u\wedge\bigwedge_{j=1}^{6}\neg\mathcal G_j\ \to$\\
\>\> $\phase_u:=\mathit{next}_u(\phase_u)$\\[3pt]
\textbf{\CycleIdle$_0$} \textbf{::} \> $d_u=0\wedge\phase_u=1\ \to\ \phase_u:=0$
\end{tabbing}
\figrule
\caption{Transformer $\TC$. Each node writes its outgoing registers,
reads its incoming registers, and checks its local rules in a cyclic order.
Every move advances the phase counter.}
\label{fig:t3}
\end{figure}

The guard of a writing rule reads only the local phase counter, and its
action writes one register. The guard of a reading rule reads only the local
phase counter, and its action reads one register and updates local variables.
The other rules read and write only local variables. In particular, there is
no rule whose guard checks the contents of several communication registers.
Thus $\TC$ respects strict read/write atomicity. Notice also that the writing
rules do not read the registers that they write.

Every node has an enabled rule in every configuration: at a communication
phase the corresponding read or write is enabled, and at the local phase
either one local rule or rule $\CycleIdle$ is enabled. The isolated-node
phase is enabled by its own rule. Hence, under the weakly fair distributed
daemon, every node performs at least one move per round, and all rounds are
finite. As for $\TB$, the executions of $\TC(\A)$ are infinite, even when
algorithm $\A$ is silent.

\section{Analysis of  \TC }\label{sec:t3-analysis}
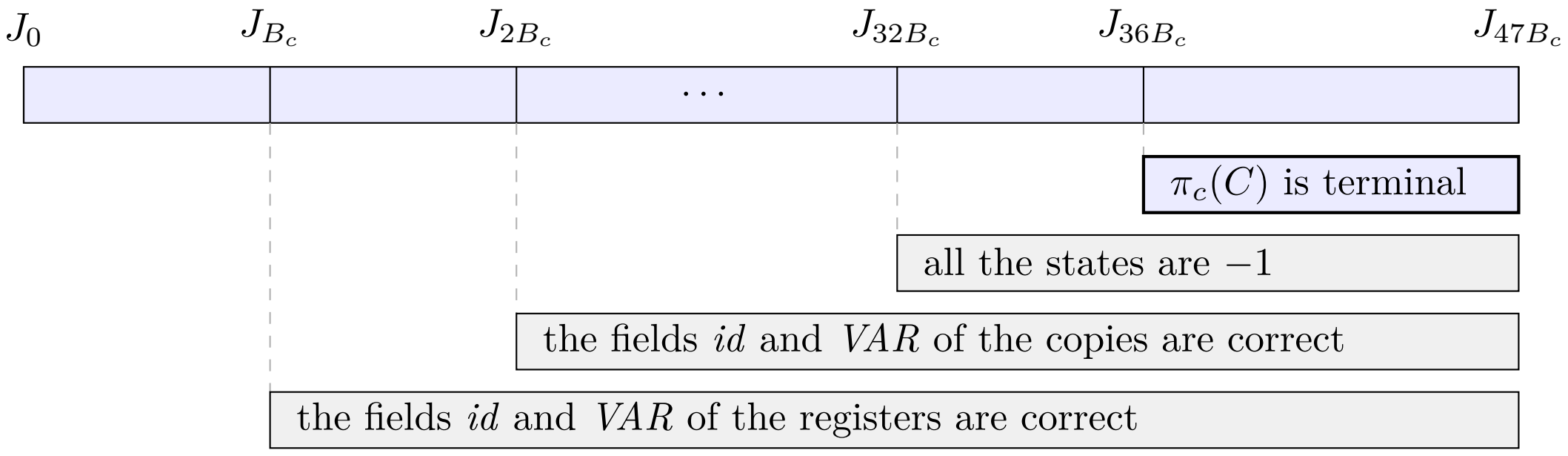
\begin{figure}[ht]
\centering
\begin{tikzpicture}[x=1cm,y=1cm,font=\small]
  \draw[fill=blue!8] (0,3.0) rectangle (13.35,3.5);
  \node at (6.1,3.25) {$\cdots$};
  \foreach \x/\t in {0/{J_0},2.2/{J_{B_c}},4.4/{J_{2B_c}},7.8/{J_{32B_c}},
      10.0/{J_{36B_c}},13.35/{J_{47B_c}}} {
    \draw (\x,3.0) -- (\x,3.5);
    \node[anchor=south] at (\x,3.55) {$\t$};
  }
  \foreach \x/\y/\t in {
      7.8/1.5/{all the states are $-1$},
      4.4/0.8/{the fields $\id$ and $\VAR$ of the copies are correct},
      2.2/0.1/{the fields $\id$ and $\VAR$ of the registers are correct}} {
    \draw[gray!60,dashed] (\x,3.0) -- (\x,\y+0.5);
    \draw[fill=black!6] (\x,\y) rectangle (13.35,\y+0.5);
    \node[anchor=west] at (\x+0.1,\y+0.25) {\t};
  }
  \draw[gray!60,dashed] (10.0,3.0) -- (10.0,2.7);
  \draw[thick,fill=blue!8] (10.0,2.2) rectangle (13.35,2.7);
  \node[anchor=west] at (10.1,2.45) {$\pi_c(C)$ is terminal};
\end{tikzpicture}
\caption{{A window $J$ with at least $47B_c$ rounds, where
$B_c=2\Delta+1$. Each bar starts at the
configuration from which the property holds, and the property then holds in
every configuration of $J$. The first two properties are those of
\cref{lem:t3-cleaning}. The last two are established in the proof of
\cref{lem:t3-window}, where $\pi_c(C)$ is terminal for algorithm $\TA(\A)$.}}
\label{fig:t3-window-thresholds}
\end{figure}

\begin{toappendix}
\label{app:t3}
\end{toappendix}
 The analysis relies on the notion of window, as for $\TB$. A window
is a sequence of consecutive rounds in which no node executes rule
$\ExecuteAlgo$. \Cref{fig:t3-window} illustrates this notion.
\Cref{fig:t3-window-thresholds} shows the properties that we establish in a
window with at least $47B_c$ rounds, where $B_c=2\Delta+1$. The projection
$\pi_c$ is defined in the next subsection, and the windows in
\cref{sec:t3-windows}. All the proofs omitted are given in \Cref{app:t3}.

\begin{figure}[ht]
\centering
\begin{tikzpicture}[x=1cm,y=1cm,font=\small]
  \foreach \a/\b in {0/1.7,2.05/3.05,3.4/5.2} {
    \draw[fill=blue!8] (\a,3.6) rectangle (\b,4.2);
  }
  \foreach \a in {1.7,3.05,5.2,13.0} {
    \fill[black!75] (\a,3.6) rectangle (\a+0.35,4.2);
  }
  \draw[thick,fill=blue!8] (5.55,3.6) rectangle (13.0,4.2);
  \node at (9.275,3.9) {window $J$};
  \draw[gray!60,dashed] (5.55,3.6) -- (0,2.3);
  \draw[gray!60,dashed] (13.0,3.6) -- (13.35,2.3);
  \foreach \a/\b/\t in {0/2.1/{round $1$},2.1/4.2/{round $2$},4.2/5.9/{$\cdots$},
      5.9/8.0/{round $r$},8.0/11.25/{$\cdots$},11.25/13.35/{round $q$}} {
    \draw[fill=blue!8] (\a,1.7) rectangle (\b,2.3);
    \node at ({(\a+\b)/2},2.0) {\t};
  }
  \foreach \x/\t in {0/{J_0},2.1/{J_1},4.2/{J_2},5.9/{J_{r-1}},8.0/{J_r},
      11.25/{J_{q-1}},13.35/{J_q}} {
    \draw (\x,1.7) -- (\x,1.55);
    \node[anchor=north] at (\x,1.55) {$\t$};
  }
  \fill[black!75] (0,0.2) rectangle (0.35,0.55);
  \node[anchor=west] at (0.5,0.375) {round that contains an execution of $\ExecuteAlgo$};
  \draw[fill=blue!8] (8.6,0.2) rectangle (9.3,0.55);
  \node[anchor=west] at (9.45,0.375) {window};
\end{tikzpicture}
\caption{{The rounds of an execution of $\TC(\A)$. A window is a
sequence of consecutive rounds in which no node executes rule $\ExecuteAlgo$.
If rule $\ExecuteAlgo$ is executed $M$ times, at most $M$ rounds contain an
execution of this rule. Bottom: a window $J$ with $q$ rounds. The
configuration $J_0$ is the first configuration of $J$, and $J_r$ is the last
configuration of its $r$th round.}}
\label{fig:t3-window}
\end{figure}

We study an execution of algorithm $\TC(\A)$ under the weakly fair
distributed daemon. The initial configuration is arbitrary. We use the
notation of \cref{sec:notation}. In particular, $\proj C$ denotes the
configuration of $\A$ obtained by retaining the variables $\VAR$ only.

\subsection{Relation with \TA}

For a configuration $C$ of $\TC(\A)$, let $\pi_c(C)$ be the configuration
of $\TA(\A)$ obtained by removing the phase counters. A move of $\TC(\A)$
is said to be a \emph{stutter} if it does not modify this projection. More
precisely, a stutter is an execution of $\CycleIdle$ or $\CycleIdle_{0}$, an
execution of $\Write(a)$ by $u$ with $\reg ua=\LV ua$, or an execution of
$\Read(a)$ by $u$ with $\CR ua=\RR au$. The other moves are said to be
\emph{effective}. The conditions distinguishing effective moves from
stutters are used only in the analysis; the nodes do not test them.

\begin{lemma}\label{lem:t3-projection}
Let $C\mapsto D$ be a transition of algorithm $\TC(\A)$. We have
\begin{itemize}
\item if all the moves of $C\mapsto D$ are stutters, then
$\pi_c(C)=\pi_c(D)$;
\item otherwise, $\pi_c(C)\mapsto\pi_c(D)$ is a transition of algorithm
$\TA(\A)$, whose moves are the effective moves of $C\mapsto D$.
\end{itemize}
\end{lemma}

\begin{proof}
An effective writing move of $u$ in $\reg ua$ satisfies
$\reg ua(C)\neq\LV ua(C)$, so the corresponding rule $\Write(a)$ is enabled
in $\pi_c(C)$. An effective reading move of $u$ satisfies
$\CR ua(C)\neq\RR au(C)$, so the corresponding rule $\Read(a)$ is enabled
in $\pi_c(C)$. A local rule has the same guard and the same action, on the
variables of $\TA$, as the corresponding rule of $\TA$.

All these guards and actions are evaluated in $C$, as in the definition of
a transition of the distributed daemon. Stutters change only the phase
counters. Since each node performs at most one move in a transition, the
nonempty set of effective moves, if any, defines a transition of $\TA(\A)$
from $\pi_c(C)$ to $\pi_c(D)$.
\end{proof}

By \cref{lem:t3-projection}, deleting the transitions that contain only
stutters gives a sequence of transitions of $\TA(\A)$. If the sequence is
finite, it can be extended to a maximal execution of $\TA(\A)$. We can
therefore apply to any finite prefix of this sequence the results of
\cref{sec:t1-analysis}. In particular, a local rule of $\TC$ is always an
effective move, and the values of the copies and of the variables used to
define a \faithful{} execution of $\ExecuteAlgo$ are unchanged by projection.

\begin{corollary}\label{cor:t3-faithful}
Let $\Ex$ be an execution of algorithm $\TC(\A)$ with initial configuration
$C_0$. We have
\begin{enumerate}
\item at most one execution of rule $\ExecuteAlgo$ per node is not faithful;
\item the number $k$ of executions of rule $\ExecuteAlgo$ that are not
\faithful{} satisfies
\[
 k\le\bigl|\{u\in V:\state_u(C_0)\in\{3,4,5,6\}\}\bigr|\le n;
\]
\item every part of $\Ex$ in which all executions of $\ExecuteAlgo$ are
\faithful{} projects to a sequence of transitions of $\A$, with the same
number of simulated moves;
\item if $\A$ has move complexity at most $F$, the number $M$ of executions
of rule $\ExecuteAlgo$ in $\Ex$ satisfies $M\le(k+1)F+k$.
\end{enumerate}
\end{corollary}

\begin{proof}
Apply \cref{cor:un\faithful,thm:simulation,cor:M} to the sequence of effective
transitions given by \cref{lem:t3-projection}, or to an extension of each
finite prefix of this sequence. The projection removes no execution of
$\ExecuteAlgo$, preserves the initial states and preserves the local view
immediately before each such execution. The bounds on finite prefixes imply
the same bounds for the whole execution.
\end{proof}

\subsection{Properties on cycles and windows}\label{sec:t3-windows}

Let $B_c=2\Delta+1$. Recall that $\Delta\ge 1$, so $\ell_u\le B_c$ for
every node $u$. Thus $B_c$ bounds the length of the cycle of a node:
$\deg(u)$ writes, $\deg(u)$ reads and one local phase. Since a node performs
at least one move per round, it goes through its whole cycle during any $B_c$
consecutive rounds (\cref{lem:t3-cycle}). The value $B_c$ plays here the role
of the value $B$ of \cref{sec:t2-analysis}.

 A window $J$ is a sequence of consecutive rounds in which
no node executes rule $\ExecuteAlgo$, as in the analysis of $\TB$. We use
$J_0$ for the first configuration of the window and $J_r$ for the last
configuration of its $r$th round. Let $\Act$ be the set of nodes enabled in
$\proj{J_0}$ for algorithm $\A$. Since only rule $\ExecuteAlgo$ modifies
$\VAR$, the set $\Act$ and all variables $\VAR$ are fixed during $J$.
Statements referring to $J_r$ below are made only when the window contains
at least $r$ rounds.

\begin{lemma}\label{lem:t3-cycle}
 {Let $u$ be a node. Let $\Ex$ be an execution of algorithm $\TC(\A)$.
During any $B_c$ consecutive rounds of $\Ex$, we have}
\begin{itemize}
\item  {$u$ executes rule $\Write(a)$ and rule $\Read(a)$ at least once,
for every $a\in N(u)$;}
\item  {$u$ visits its local phase at least once, that is, it executes a
local rule or rule $\CycleIdle$ at that phase.}
\end{itemize}
\end{lemma}

\begin{proof}
Every node performs at least one move in each round. The phase counter of
$u$ advances by one modulo $\ell_u$ at each of its moves. In any sequence of
$\ell_u$ consecutive moves of $u$, every phase is visited exactly once,
independently of the initial phase. Since $B_c\ge\ell_u$, the lemma follows.
\end{proof}

\begin{lemmarep}\label{lem:t3-cleaning}
 {Let $(u,a)$ be an edge. Let $\Ex$ be an execution of algorithm
$\TC(\A)$ which contains a window $J$.} We have
\begin{enumerate}
\item $\reg ua.\id(C)=\id_u$ and $\reg ua.\VAR(C)=\VAR_u(C)$ for every
configuration $C$ of $J$ with $J_{B_c}\le C$;
\item $\cp ua.\id(C)=\id_a$ and $\cp ua.\VAR(C)=\VAR_a(C)$ for every
configuration $C$ of $J$ with $J_{2B_c}\le C$.
\end{enumerate}
In particular, for every configuration $C$ of $J$ with $J_{2B_c}\le C$,
$\isAct(u,\A)(C)$ is true if and only if $u\in\Act$.
\end{lemmarep}

\begin{proof}
During the first $B_c$ rounds, node $u$ executes $\Write(a)$ at least once
by \cref{lem:t3-cycle}. Each such execution writes $\id_u$ and $\VAR_u$
in the corresponding fields of $\reg ua$. These values remain unchanged
during the window, which proves the first point.

During the following $B_c$ rounds, node $u$ executes $\Read(a)$ at least
once. By the first point, the fields $\id$ and $\VAR$ of $\reg au$ have
the right values at the beginning of this reading move. The reading move
copies them, and any later reading move during the window copies the same
values. This proves the second point. The last claim follows from the
fact that the local view of $u$ is now its view in $\proj C=\proj{J_0}$.
For an isolated node this equality holds throughout the window, since its
view contains no copies.
\end{proof}

{In the statement of the next lemma, no configuration is between
$J_{r'}$ and $T$ if $J$ has fewer than $r'$ rounds or if $T<J_{r'}$.}

\begin{lemmarep}\label{lem:t3-propagation}
{Let $(x,v)$ be an edge. Let $\Ex$ be an execution of algorithm
$\TC(\A)$ which contains a window $J$. Let $\Sigma$ be a set of states, let
$r\ge 0$ be an integer and let $T$ be a configuration of $J$ such that
$\state_x(C)\in\Sigma$ for every configuration $C$ between $J_r$ and $T$.
We have}
\begin{enumerate}
\item\label{t3:pr1} $\reg xv.\state(C)\in\Sigma$ for every configuration $C$
between $J_{r+B_c}$ and $T$;
\item\label{t3:pr2} $\cp vx.\YS(C)\in\Sigma$ for every configuration $C$
between $J_{r+2B_c}$ and $T$.
\end{enumerate}
If moreover $\Sigma=\{s\}$, we have
\begin{enumerate}[resume]
\item\label{t3:pr3} $\reg vx.\YS(C)=s$ for every configuration $C$ between
$J_{r+3B_c}$ and $T$;
\item\label{t3:pr4} $\cp xv.\MS(C)=s$ for every configuration $C$ between
$J_{r+4B_c}$ and $T$.
\end{enumerate}
\end{lemmarep}

\begin{proof}
By \cref{lem:t3-cycle}, if $J_{r+B_c}\le T$, node $x$ executes $\Write(v)$
between $J_r$ and $J_{r+B_c}$. It writes a value of $\state_x$ that belongs
to $\Sigma$. All later writes up to $T$ also write a value in $\Sigma$.
This proves the first point.

Similarly, between $J_{r+B_c}$ and $J_{r+2B_c}$, if this interval exists and
ends no later than $T$, node $v$ executes $\Read(x)$ at least once. By the
first point, it copies a value in $\Sigma$ in $\cp vx.\YS$. All later
executions of $\Read(x)$ up to $T$ have the same property. This proves the
second point.

Assume now that $\Sigma=\{s\}$. Between $J_{r+2B_c}$ and $J_{r+3B_c}$,
node $v$ executes $\Write(x)$ at least once whenever the whole interval
exists before $T$. By the second point, it writes $s$ in $\reg vx.\YS$,
and all later writes keep this field equal to $s$. Finally, between
$J_{r+3B_c}$ and $J_{r+4B_c}$, node $x$ executes $\Read(v)$ at least once
whenever the whole interval exists before $T$. By the third point, it
copies $s$ in $\cp xv.\MS$, and all later reads keep this field equal to
$s$ up to $T$. This proves the last two points.
\end{proof}

\begin{lemmarep}\label{lem:t3-free}
{Let $x$ be a node. Let $\Ex$ be an execution of algorithm $\TC(\A)$
which contains a window $J$ with at least $r+5B_c$ rounds, where
$r\ge 2B_c$.} Assume that one of the following conditions holds:
\begin{itemize}
\item $\state_x(J_r)\in\{0,\ldots,4,7\}$;
\item $\state_x(J_r)=-1$ and $x\in\Act$;
\item $\state_x(J_r)\in\{5,6\}$ and $x\notin\Act$.
\end{itemize}
Then $x$ executes a local rule between $J_r$ and $J_{r+5B_c}$.
\end{lemmarep}

\begin{proof}
 {We prove this lemma by contradiction. We assume that $x$ executes no
local rule between $J_r$ and $J_{r+5B_c}$. Then $\state_x(C)=\state_x(J_r)$
for every configuration $C$ between $J_r$ and $J_{r+5B_c}$.}

 {Let $C$ be a configuration between $J_{r+4B_c}$ and $J_{r+5B_c}$. By
\cref{lem:t3-propagation}(\ref{t3:pr4}), applied to every neighbor of $x$,
$\CBS(x)(C)$ is true. By \cref{lem:t3-cleaning}, $\isAct(x,\A)(C)$ is true if
and only if $x\in\Act$. Thus the guard of rule $\ModifyMyState$,
$\StartAgain$, $\WantToAct$ or $\ResetRule$, according to the case, holds at
$x$ in $C$.}

{By \cref{lem:t3-cycle}, $x$ visits its local phase in a transition
between $J_{r+4B_c}$ and $J_{r+5B_c}$. In this transition, $x$ executes the
local rule whose guard holds. This leads to a contradiction.}
\end{proof}

\begin{lemmarep}[Long windows]\label{lem:t3-window}
Let $\Ex$ be an execution of algorithm $\TC(\A)$ which contains a window $J$
with at least $47B_c$ rounds. Then, for every configuration $C$ of $J$ with
$J_{36B_c}\le C$, $\pi_c(C)$ is a terminal configuration of algorithm $\TA(\A)$.
\end{lemmarep}

\begin{proof}
Throughout the proof, all configurations and transitions belong to $J$.
Since $J$ is a window, no node executes $\ExecuteAlgo$ during $J$.
Thus $\proj C=\proj{J_0}$ for every configuration $C$ of $J$, and $\Act$
is fixed. By \cref{lem:t3-cleaning}, for every node $x$ and every configuration
$C$ with $J_{2B_c}\le C$,
\[
  \isAct(x,\A)(C)\iff x\in\Act.
\]

\emph{Step 1: the nodes outside $\Act$.}
Let $x\notin\Act$. After $J_{2B_c}$, the only local rules that $x$ can
execute are $\ModifyMyState$, $\ResetRule$ and $\StartAgain$.
If its state is in $\{0,\ldots,4\}$, it needs at most five executions of
$\ModifyMyState$ to reach state $5$, followed by one execution of
$\ResetRule$ to reach state $-1$. If its state is $5$ or $6$, only one
execution of $\ResetRule$ is needed. If its state is $7$, only one
execution of $\StartAgain$ is needed. Once in state $-1$, it cannot execute
any local rule.

For each $i\in\{0,\ldots,5\}$, set $r_i=(2+5i)B_c$.
If $\state_x(J_{r_i})\neq-1$, one of the conditions of
\cref{lem:t3-free} holds. Hence $x$ executes at least one local rule
between $J_{r_i}$ and $J_{r_i+5B_c}$.
These six intervals do not share any transition. Since at most six local
rules are needed to reach state $-1$, we obtain
\begin{equation}\label{eq:t3-nonact}
  \state_x(C)=-1\qquad\text{for every $C$ with }J_{32B_c}\le C.
\end{equation}

\emph{Step 2: the nodes of $\Act$.}
Let $x\in\Act$. After $J_{2B_c}$, node $x$ cannot execute $\ResetRule$,
since $\isAct(x,\A)$ is true. It cannot execute $\ExecuteAlgo$ either,
by the definition of a window.
From state $7$, the longest sequence of local rules needed to reach
$\{5,6\}$ consists of $\StartAgain$, $\WantToAct$ and five executions
of $\ModifyMyState$, that is, seven local rules. From any other state
outside $\{5,6\}$, at most six local rules are needed.

For each $i\in\{0,\ldots,6\}$, set $r_i=(2+5i)B_c$.
If $\state_x(J_{r_i})\notin\{5,6\}$, then \cref{lem:t3-free}
ensures that $x$ executes a local rule between $J_{r_i}$ and
$J_{r_i+5B_c}$. Once $x$ reaches $\{5,6\}$, it cannot leave this set
and the only local rule it can still execute is $\toWAIT$, at most once.
Therefore
\begin{equation}\label{eq:t3-act}
  \state_x(C)\in\{5,6\}\qquad\text{for every $C$ with }J_{37B_c}\le C.
\end{equation}
In particular, after $J_{37B_c}$, the state of each node of $\Act$ either
remains unchanged or changes once from $5$ to $6$.

\emph{Step 3: the case $\Act=\emptyset$.}
Assume that $\Act=\emptyset$. By \eqref{eq:t3-nonact}, all nodes have
state $-1$ from $J_{32B_c}$ on. Let $C$ be a configuration with
$J_{36B_c}\le C$ and let $(u,v)$ be an ordered pair of neighbors.
Apply \cref{lem:t3-propagation} with $r=32B_c$, $T=C$ and
$\Sigma=\{-1\}$, first to $(u,v)$ and then to $(v,u)$. We obtain
\begin{align*}
  \reg uv.\state(C)&=\cp vu.\YS(C)=-1,\\
  \reg uv.\YS(C)&=\cp vu.\MS(C)=-1.
\end{align*}
Together with \cref{lem:t3-cleaning}, this gives
\[
  \reg uv(C)=\cp vu(C)=(\id_u,-1,-1,\VAR_u(C)).
\]
In particular, for every ordered pair of neighbors, $\reg uv(C)=\LV uv(C)$
and $\CR vu(C)=\RR uv(C)$. No $\Write$ or $\Read$ rule of $\TA(\A)$ is
therefore enabled in $\pi_c(C)$. Moreover, all its states are $-1$, and
$\proj C=\proj{J_0}$ is terminal for $\A$ since $\Act=\emptyset$.
By \cref{lem:terminal}, $\pi_c(C)$ is terminal for $\TA(\A)$.

We now assume that $\Act\neq\emptyset$ and derive a contradiction.

\emph{Step 4: copies of the states.}
Let $(x,v)$ be an ordered pair of neighbors and let $C$ be any configuration
with $J_{39B_c}\le C$. By \eqref{eq:t3-nonact} and \eqref{eq:t3-act},
the state of $v$ belongs to $\{-1\}$ if $v\notin\Act$, and to
$\{5,6\}$ if $v\in\Act$, throughout the interval from $J_{37B_c}$ to $C$.
Applying \cref{lem:t3-propagation}(\ref{t3:pr2}) to $(v,x)$ with
$r=37B_c$, $T=C$ and the corresponding set of states yields
\begin{equation}\label{eq:t3-copies}
\begin{cases}
  \cp xv.\YS(C)=-1 & \text{if }v\notin\Act,\\
  \cp xv.\YS(C)\in\{5,6\} & \text{if }v\in\Act.
\end{cases}
\end{equation}
By \cref{lem:t3-cleaning}, we also have $\cp xv.\id(C)=\id_v$.
These properties hold for every configuration from $J_{39B_c}$ on.

\emph{Step 5: all states remain unchanged after $J_{42B_c}$.}
Let $P$ be the set of nodes of $\Act$ that have no neighbor in $\Act$
with a smaller identifier:
\[
  P=\{x\in\Act : \forall y\in N(x)\cap\Act,\ \id_x<\id_y\}.
\]
We first study the nodes of $\Act\setminus P$, and then those of $P$.

\emph{Case (a):} let $x\in\Act\setminus P$. There exists a neighbor
$y\in\Act$ with $\id_y<\id_x$. In every configuration $C$ with
$J_{39B_c}\le C$ and $\state_x(C)=5$, we have
$\cp xy.\YS(C)\ge5$ and $\cp xy.\id(C)=\id_y<\id_x$, by Step~4.
Thus $\wait(x)(C)$ is true: if the copied state is $6$, it is larger
than $\state_x(C)$; if it is $5$, the copied identifier is smaller.
Consequently, $x$ cannot execute $\toWAIT$ in any transition starting
at or after $J_{39B_c}$. Since this is the only local rule that $x$
could still execute, its state remains unchanged from $J_{39B_c}$ on.

\emph{Case (b):} let $x\in P$. Every neighbor $y\in N(x)\cap\Act$
belongs to $\Act\setminus P$, since $\id_x<\id_y$.
By Case~(a), $\state_y$ remains unchanged from $J_{39B_c}$ on.
Hence, by \cref{lem:t3-propagation}(\ref{t3:pr2}) applied to $(y,x)$
with $r=39B_c$, $T=C$ and $\Sigma=\{\state_y(J_{39B_c})\}$, for every
configuration $C$ with $J_{41B_c}\le C$,
\[
  \cp xy.\YS(C)=\state_y(J_{39B_c}).
\]
There are two possibilities.
\begin{itemize}
\item If some $y\in N(x)\cap\Act$ has $\state_y(J_{39B_c})=6$,
then $\wait(x)(C)$ is true whenever $J_{41B_c}\le C$ and
$\state_x(C)=5$. Thus $x$ cannot execute $\toWAIT$ from $J_{41B_c}$ on.
\item Otherwise, every neighbor in $N(x)\cap\Act$ has state $5$
from $J_{39B_c}$ on (this includes the case $N(x)\cap\Act=\emptyset$).
We prove that $\state_x(J_{42B_c})=6$.
Assume, by contradiction, that $\state_x(J_{42B_c})=5$.
By Step~2, $\state_x(C)=5$ for every configuration $C$ between
$J_{37B_c}$ and $J_{42B_c}$.
By \cref{lem:t3-propagation}(\ref{t3:pr4}), applied to each ordered
pair $(x,v)$ with $v\in N(x)$, $r=37B_c$, $T=J_{42B_c}$ and
$\Sigma=\{5\}$, we obtain $\cp xv.\MS(C)=5=\state_x(C)$ for every
configuration $C$ between $J_{41B_c}$ and $J_{42B_c}$.
Thus $\CBS(x)(C)$ is true throughout this interval.
Moreover, $\isAct(x,\A)(C)$ is true by \cref{lem:t3-cleaning}.
By Step~4 and the equality above for the copied states of active
neighbors, every copied state is either $-1$ or $5$; in the latter
case the copied identifier is larger than $\id_x$.
Hence $\wait(x)(C)$ is false throughout this interval.

All four conditions in the guard of $\toWAIT$ therefore hold in every
configuration between $J_{41B_c}$ and $J_{42B_c}$.
By \cref{lem:t3-cycle}, $x$ visits its local phase in a transition
$D\mapsto D'$ with $J_{41B_c}\le D<D'\le J_{42B_c}$.
Since the guards of the local rules are mutually exclusive, it executes
$\toWAIT$ in this transition. Thus $\state_x(D')=6$. By Step~2, its
state remains $6$ up to $J_{42B_c}$, a contradiction.
\end{itemize}
In both possibilities, the state of $x$ remains unchanged from $J_{42B_c}$ on.
Combining Cases~(a) and~(b) with Step~1, we conclude that, for every node $u$,
\begin{equation}\label{eq:t3-fixed-states}
  \state_u(C)=\state_u(J_{42B_c})
  \qquad\text{for every $C$ with }J_{42B_c}\le C.
\end{equation}

\emph{Step 6: at least one node is in state $6$.}
Let $x_0$ be the node of $\Act$ with the smallest identifier. Then $x_0\in P$.
By Case~(b) of Step~5, either a neighbor of $x_0$ is in state $6$ at
$J_{39B_c}$ and remains in state $6$, or $x_0$ is in state $6$ at
$J_{42B_c}$. Therefore the set
\[
  U=\{u\in V:\state_u(J_{42B_c})=6\}
\]
is nonempty. By Step~1, $U\subseteq\Act$.

\emph{Step 7: contradiction.}
Let $x^*$ be the node of $U$ with the smallest identifier.
By \eqref{eq:t3-fixed-states}, all states remain unchanged from $J_{42B_c}$ on.
For each $v\in N(x^*)$, apply \cref{lem:t3-propagation} to $(v,x^*)$
with $r=42B_c$, $T=J_{47B_c}$ and $\Sigma=\{\state_v(J_{42B_c})\}$,
and to $(x^*,v)$ with the same $r,T$ and $\Sigma=\{6\}$.
By points~\ref{t3:pr2} and~\ref{t3:pr4} of that lemma, for every
configuration $C$ between $J_{46B_c}$ and $J_{47B_c}$,
\[
  \cp {x^*}v.\YS(C)=\state_v(C),\qquad
  \cp {x^*}v.\MS(C)=6=\state_{x^*}(C).
\]
By \cref{lem:t3-cleaning}, the copied identifiers are also correct and
$\isAct(x^*,\A)(C)$ is true, since $x^*\in\Act$.
Every neighbor of $x^*$ has state $-1$, $5$ or $6$; if its state is $6$,
it belongs to $U$ and has a larger identifier than $x^*$.
Consequently, $\wait(x^*)(C)$ is false, $\CBS(x^*)(C)$ is true, and
$\state_{x^*}(C)=6$ throughout this interval.

All four conditions in the guard of $\ExecuteAlgo$ hold in every
configuration between $J_{46B_c}$ and $J_{47B_c}$.
By \cref{lem:t3-cycle}, $x^*$ visits its local phase in a transition
$D\mapsto D'$ with $J_{46B_c}\le D<D'\le J_{47B_c}$.
It must execute $\ExecuteAlgo$ in this transition, contradicting the
definition of a window.
Thus $\Act=\emptyset$, and the conclusion follows from Step~3.
\end{proof}

\subsection{Round complexity}

\begin{theorem}[Round complexity of $\TC$]\label{thm:t3-rounds}
Let $\A$ be an algorithm in the state model. Let $\Ex$ be an execution of
algorithm $\TC(\A)$ in the link-register model with strict read/write
atomicity under the weakly fair distributed daemon, in which rule
$\ExecuteAlgo$ is executed a finite number $M$ of times. Then $\Ex$ reaches,
{after $O(\Delta(M+1))$ rounds}, a configuration $T$ such that
\begin{itemize}
\item $\pi_c(T)$ is a terminal configuration of algorithm $\TA(\A)$;
\item $\proj T$ is a terminal configuration of algorithm $\A$;
\item $\pi_c(C)=\pi_c(T)$ for every configuration $C$ with $T\le C$: all the
moves after $T$ are stutters.
\end{itemize}
\end{theorem}

\begin{proof}
{Recall that $B_c=2\Delta+1$. We prove that $(47M+36)B_c$ rounds are
enough.} At most $M$ rounds of $\Ex$ contain an execution of
rule $\ExecuteAlgo$. Denote their indices by $\rho_1<\cdots<\rho_p$,
where $p\le M$, and set $\rho_0=0$. There exists a smallest $j\in\{0,\ldots,p\}$ such that rounds
{$\rho_j+1,\ldots,\rho_j+47B_c$} contain no execution of $\ExecuteAlgo$.
For every $i<j$, we have  {$\rho_{i+1}\le\rho_i+47B_c$}. Thus
 {$\rho_j\le 47B_cj\le47B_cM$}.

The rounds  {$\rho_j+1,\ldots,\rho_j+47B_c$} form a window. Let $T$ be the last
configuration of round  {$\rho_j+36B_c$}. By \cref{lem:t3-window}, $\pi_c(T)$
is terminal for $\TA(\A)$. The configuration $T$ is reached after at most
 {$(47M+36)B_c$} rounds. By \cref{lem:terminal}, $\proj T$ is terminal for $\A$.
Finally, by \cref{lem:t3-projection}, no transition starting from a
configuration whose projection $\pi_c$ is terminal can contain an effective
move. Therefore all subsequent transitions preserve $\pi_c(T)$.
\end{proof}

\begin{corollary}\label{cor:t3-silent}
Let $\A$ be an algorithm in the state model with move complexity at most
$F$. Let $\Ex$ be an execution of algorithm $\TC(\A)$ under the weakly fair
distributed daemon, and let $M$ be the number of executions of rule
$\ExecuteAlgo$ in $\Ex$. Then
\begin{enumerate}
\item  {$M\le(n+1)F+n$;}
\item $\Ex$ reaches, after {$O(\Delta n(F+1))$} rounds, a configuration
$T$ such that $\proj T$ is terminal for $\A$ and $\pi_c(C)=\pi_c(T)$ for
every configuration $C$ with $T\le C$.
\end{enumerate}
\end{corollary}

\begin{proof}
{By \cref{cor:t3-faithful}, at most $n$ executions of rule
$\ExecuteAlgo$ are not \faithful, and thus $M\le(n+1)F+n$.} In particular,
$M$ is finite and  {$M+1\le(n+1)(F+1)$}.  {The proof of
\cref{thm:t3-rounds} shows that $(47M+36)B_c$ rounds are enough, and
$(47M+36)B_c\le 47(2\Delta+1)(n+1)(F+1)$.}
\end{proof}

After $T$, the registers, the copies, the states and the variables of $\A$
remain unchanged. Only the phase counters change. If $\A$ is self-stabilizing
for a predicate $P$ on its terminal configurations, $\TC(\A)$ is
self-stabilizing for the predicate $P$ on the projected configurations.
The set of configurations whose projection $\pi_c$ is terminal for
$\TA(\A)$ is closed by \cref{lem:t3-projection}. As for $\TB$, the
transformed algorithm is silent in the sense of fixed communication
registers, not in the sense of finite executions (\cref{rem:silence}).

\begin{remark}[Arbitrary algorithms]\label{rem:t3-general}
The analog of \cref{thm:t1-general} holds for $\TC$. Indeed, by
\cref{cor:t3-faithful}, at most $n$ executions of rule $\ExecuteAlgo$ are
not \faithful. Take $S$ after the last such execution (or the initial
configuration if there is none). If $\ExecuteAlgo$ occurs infinitely often,
the projected suffix starting in $S$ is an infinite execution of $\A$.
Otherwise, by \cref{thm:t3-rounds}, the projected suffix ends in a terminal
configuration of $\A$. This is an eventual simulation result; it does not
by itself assert preservation of an arbitrary specification for non-silent
algorithms.
\end{remark}

\begin{remark}[Comparison of the two link-register transformers]
Both $\TB$ and $\TC$ have a round bound of $O(\Delta(M+1))$ under the weakly
fair distributed daemon. The former uses flags to avoid some redundant
communications and tests the absence of enabled reading rules in
$\ToTrue$. The latter performs unconditional reads and writes, and uses
only a local phase counter to schedule them. In particular, $\TC$ requires
no simultaneous reading of several registers. It has at most
$\max\{2,2\deg(u)+1\}$ additional local phase values per node $u$, and its
progress proof does not need to count the reading or writing moves between
successive executions of a refresh rule. The result concerns the worst-case
round complexity, not the total number of communication operations or the
relative practical efficiency of the two transformers.
\end{remark}


\section{Conclusion}\label{sec:conclusion}

We have presented two transformers that convert a self-stabilizing algorithm
of the state model into an algorithm for a link-register model with
read/write atomicity. The first one, $\TA$, targets the strong-link-register
model, in which a node can read its own registers, and works under the unfair
distributed daemon. One move of the original algorithm costs $O(\Delta^2)$
moves\modif{.}
The second one, $\TB$, targets the
link-register model and works under the weakly fair distributed daemon. One
move of the original algorithm costs $O(\Delta)$ rounds\modif{.}
In both cases the simulation is exact, except for at most one move
per node, which is the trace of a corrupted initial configuration of the
transformer.

The link-register model and the strong-link-register model are intermediate
steps toward the message-passing model. When a node $u$ writes in a register
shared with a node $v$, the written value becomes available to $v$. This is
similar to the reception of a message by $v$. In the strong-link-register
model, a node can check that its register $\reg uv$ is up to date. In our
transformers, the node $v$ acknowledges the values it reads, and $u$ changes
its state only when it has received the acknowledgments of all its neighbors.
The analysis shows that four consecutive exchanges of this kind are enough to
remove the effect of corrupted registers and copies.

\paragraph{Use of generative AI.}
Large language models (Claude and ChatGPT) were used during the
preparation of this report for four purposes: (i) to help update the
bibliography, (ii) to improve the English of the manuscript,
(iii) to produce referee-style reviews of draft versions, whose
comments were used to improve the text, and (iv) to assist in the
development and refinement of the proofs, in particular in deriving
some of the constants appearing in the complexity bounds.
All suggestions and mathematical arguments were checked by the authors,
and every reference was verified against its original source.
The authors take full responsibility for the correctness of the results
and proofs and for the final content of the report.


\bibliographystyle{plain}
\bibliography{refs}

\appendix

\end{document}